\documentclass[11pt]{article}

\usepackage{jheppubmod}
\usepackage{cite}

\usepackage{color}
\usepackage[usenames,dvipsnames,svgnames,table]{xcolor}
\usepackage[showonlyrefs]{mathtools}
\usepackage{psfrag}
\usepackage{array}
\usepackage{amssymb}
\usepackage{mathrsfs}
\usepackage{amsmath}
\usepackage{empheq}
\usepackage{slashed}

\usepackage{tikz}
\usetikzlibrary{trees}
\usetikzlibrary{snakes}
\usetikzlibrary{arrows}

\usepackage{graphicx}
\usepackage{epsfig}
\usepackage{epstopdf}
\usepackage{caption}
\usepackage[labelsep=quad]{subcaption}

\usepackage[punctsep]{collref}
\collectsep[]{;}

\allowdisplaybreaks

\def\beq{\begin{equation}}
\def\eeq{\end{equation}}
\newcommand{\bea}{\begin{eqnarray}}
\newcommand{\eea}{\end{eqnarray}}

\def\a{\alpha}
\def\b{\beta}
\def\d{\delta}
\def\s{\sigma}
\def\lm{\lambda}

\def\g{\gamma}
\def\G{\Gamma}
\def\ep{\epsilon}
\def\vep{\varepsilon}
\def\etah{\hat{\eta}}
\def\xih{\hat{\xi}}
\def\Jh{\hat{J}}

\def\lmb{\tilde{\lambda}}
\def\psiT{\widetilde{\psi}}

\def\zetaT{\tilde{\zeta}}
\def\chib{\tilde{\chi}}
\def\phiT{\widetilde{\phi}}
\def\phib{\bar{\phi}}

\def\FT{\widetilde{F}}
\def\epT{\tilde{\epsilon}}
\def\epb{\bar{\epsilon}}

\def\nb{\bar{n}}

\def\metric{g}
\def\gYM{\bold{e}}
\def\D{\mathcal{D}}
\def\A{\mathcal{A}}
\def\F{\mathcal{F}}
\def\Fh{\hat{\mathcal{F}}}

\def\Pr{\mathscr{P}}
\def\Prb{\widetilde{\mathscr{P}}}
\def\bPhi{\mathbf{\Phi}}
\def\bPsi{\mathbf{\Psi}}
\def\bK{\mathbf{K}}
\def\bW{\mathbf{W}}
\def\Lagr{\mathcal{L}}
\def\numb{\mathfrak{s}}
\def \TD{\mathscr{V}}
\def\TDWZ{\mathscr{V}_{\NLs}}
\def\W{W}
\def\Wb{\widetilde{W}}
\def\Id{1}  
\def\an{{\varpi}}

\def\matrUnity{1}
\def\matrRot{\mathscr{R}}
\newcommand{\lp}{\left(}
\newcommand{\rp}{\right)}
\newcommand{\nn}{\nonumber}

\def\AA{{\cal A}}
\def\BB{\mathbb{B}} 

\def\DD{{\cal D}}

\def\FF{{\cal F}}

\def\NN{{\cal N}}

\def\TT{{\cal T}}

\def\VV{{\cal V}}

\def\XX{{\cal X}}

\def\Tr{{\rm Tr}}

\def\x{{\varpi}}

\def\S{S}
\def\NLs{{NL\sigma}}

\title{$\NN=2$ supersymmetric field theories \\
	on 3-manifolds with A-type boundaries
	}

\author{Francesco Aprile,$^\dagger$}
\author{Vasilis Niarchos$^\natural$}
\affiliation{
$^\dagger$STAG Research Centre \& Physics and Astronomy \& Mathematical Sciences,\\
$~$ University of Southampton, Highfield, Southampton SO17 1BJ, UK}
\affiliation{$^\natural$Crete Center for Theoretical Physics, 
Institute of Theoretical and Computational Physics,\\
$~$ Crete Center for Quantum Complexity and Nanotechnology \\
$~$ Department of Physics, University of Crete, 71303, Greece}
\emailAdd{F.Aprile@soton.ac.uk}
\emailAdd{niarchos@physics.uoc.gr}

\preprint{ITCP-IPP-2016-04
\\ \hspace*{\fill}
CCQCN-2016-136
\\\hspace*{\fill}
CCTP-2016-05
}

\date{}

\abstract{
General half-BPS $A$-type boundary conditions are formulated for $\NN=2$ supersymmetric field theories on 
compact 3-manifolds with boundary. We observe that under suitable conditions manifolds of the real 
$A$-type admitting two complex supersymmetries (related by charge conjugation) possess, besides a 
contact structure, a natural integrable toric foliation. A boundary, or a general co-dimension-1 defect, can be 
inserted along any leaf of this preferred foliation to produce manifolds with boundary that have the topology of 
a solid torus. We show that supersymmetric field theories on such manifolds can be endowed with half-BPS 
$A$-type boundary conditions. We specify the natural curved space generalization of the $A$-type 
projection of bulk supersymmetries and analyze the resulting $A$-type boundary conditions in 
generic $3d$ non-linear sigma models and YM/CS-matter theories.
}


\begin{document}
\maketitle

\section{Introduction} \label{sec:CDFK}

The study of supersymmetric quantum field theories on rigid curved backgrounds in diverse spacetime dimensions 
has been a powerful source of new non-perturbative results in recent years. So far, a rather complete and systematic 
understanding of such results has been obtained for supersymmetric field theories on closed manifolds. 
Most notably, these theories can be engineered by taking appropriate rigid limits of certain supergravity theories. 
This framework constrains the background geometry and determines the couplings of the field theory to the 
curvature and the auxiliary background fields in the supergravity multiplet 
\cite{Festuccia:2011ws,Dumitrescu:2012ha,Closset:2012ru}. 
Partition functions and other supersymmetric observables can then be evaluated exactly 
with the powerful technique of supersymmetric localization providing a new window into non-perturbative 
physics in quantum field theory. Some of the original work in this direction in two, three, four, and five 
spacetime dimensions includes \cite{Pestun:2007rz,Kapustin:2009kz,Drukker:2012sr,Benini:2012ui,Doroud:2012xw,Kallen:2012cs,Hosomichi:2012ek,Kim:2012ava}. 

Analogous situations on manifolds with boundary, or more generally, on spaces with 
co-dimension-1 defects, are comparatively much less elaborated upon. 
There are two key aspects of this story one would like to develop systematically. 
The first aspect is related to the geometric properties of boundaries.
Given a fixed bulk supergravity background that supports supersymmetric field theories, what restrictions
should be imposed on the geometry of a co-dimension-1 surface to preserve a subset of the 
bulk supersymmetry? The second aspect is related more directly to the specific dynamic properties of the 
field theory in question, in particular, the boundary conditions that can be imposed on the defect.

Regarding the first point, it is immediately clear that since the commutator of supersymmetries squares 
to isometries on the compact manifold, the boundary should be oriented along directions parallel to these 
isometries to preserve the corresponding supersymmetries. Moreover, one can ask if supersymmetry
puts any constraints on co-dimension-1 foliations of a compact manifold. A foliation preferred by supersymmetry
could be used to decompose closed manifolds into a union of manifolds with boundary. Indeed, we will show that 
such a foliation exists in a general class of 3-manifolds.

As far as the second point is concerned, it is well known that the invariance of generic observables under bulk 
symmetries (including supersymmetries) is spoiled, in general, by boundary effects. 
A symmetry can be restored by cancelling these boundary effects. This can be achieved with the introduction 
of suitable boundary conditions and/or the introduction of appropriate boundary degrees of freedom.

In the present work we concentrate on three dimensions and develop a systematic treatment of half-BPS 
boundaries in $\NN=2$ supersymmetric field theories on compact 3-manifolds. We discuss general aspects
of the interplay between supersymmetry and the geometry of manifolds with boundary, and analyze a wide 
class of related half-BPS boundary conditions. We concentrate on the classical aspects of the problem. 
The main contributions of this work can be summarized as follows.

\subsection*{Summary of main results}

We begin in section \ref{sec:CDFK} with a concise collection of useful results on rigid 
supersymmetry in curved three-dimensional backgrounds. We follow closely the conventions of Ref.\ 
\cite{Closset:2012ru}, where it was recognized that the existence of a supersymmetry implies a tranversely 
holomorphic foliation. Subsequently, we focus on a more specific class of 
curved 3-manifolds dubbed $A$-type backgrounds. These backgrounds are introduced in 
section \ref{AtypeBacks}. By definition, they admit two complex Killing spinors related by charge conjugation,
\cite{Alday:2013lba}. 
We show, using supersymmetry, that they admit a Reeb vector and, under suitable conditions, 
a preferred integrable foliation whose vector distribution is defined only in terms of Killing spinors bilinears. 
The Reeb vector belongs to the foliation, thus the algebra of supersymmetry is preserved. 
Geometrically, global properties and degenerations of this foliation are classified as regular, 
quasi-regular, and irregular, as reviewed in \cite{Martelli:2015kuk}. The manifolds covered by this 
analysis include well-known examples of Seifert manifolds, like for instance the round and squashed 3-spheres, and geometries of the 
$\mathbb{S}^2\times \mathbb{S}^1$ type. 

A boundary can be introduced along the generic leaf of these foliations. 
Technically, our construction of the foliation in terms of vector fields does not require the use of coordinates, 
which may be problematic if the coordinates are not globally well defined. We argue that the topology of the leaves 
is that of a torus. Hence, the manifold decomposition, that follows from supersymmetry 
considerations, selects spaces with boundary which are topologically solid tori. The main goal of the
paper is to formulate $\NN=2$ supersymmetric field theories with half-BPS boundary conditions on such spaces.

In section \ref{sec_SUSY_CDFK} we show that the geometry of the $A$-type backgrounds 
admits a natural half-BPS projection on the bulk supersymmetries that generalizes in curved 
space the well-known $A$-type projection familiar from studies of $2d$ $\NN=(2,2)$, \cite{Hori:2000ck},
and $3d$ $\NN=2$ theories in flat space, \cite{Okazaki:2013kaa}.
Unlike the case of flat space where one projects constant spinors, in curved spaces one has to project spinors that
are in general non-trivial functions of the spacetime coordinates. We propose a `canonical' way to implement
a generalized $A$-type projection in curved space, that reduces to the familiar $A$-type projections in flat space. 
To the best of our knowledge this formulation is new. A similar generic formulation for 
$B$-type projections in curved space is left to future work.

The generalized $A$-type projection can be employed to formulate corresponding $A$-type boundary conditions in
$\NN=2$ supersymmetric field theories, that preserve half of the bulk supersymmetry. In sections 
\ref{sec_Lagrangians_N=2}-\ref{SUSYbcsII} we present these boundary conditions for arbitrary non-linear sigma 
models and YM/CS-matter theories. In both cases, we relate the boundary condition to the geometry of certain 
$2$-forms defined on the space of field configurations at the boundary. These $2$-forms are also relevant in the 
analysis of the on-shell boundary value problem, that we review in section \ref{preview}.

Section \ref{Bound_Sec_1} studies the instructive case of non-linear sigma models
with generic K\"ahler potential and superpotential. 
The boundary conditions describe Lagrangian submanifolds of the K\"ahler form in target space.
The effect of the curvature and the presence of couplings to the background fields, generalize the more
familiar analysis in flat space \cite{Hori:2000ck} \cite{Herbst:2008jq}. 

The case of general (non-abelian) YM/CS-matter theories is discussed in section \ref{SUSYbcsII}. 
We find boundary conditions that include the curved space generalization of holomorphic Neumann boundary 
conditions for Yang-Mills gauge fields and matter fields, and holomorphic Dirichlet boundary conditions for 
the gauge fields in CS theories. 

A summary of useful formulae, and an exposition of technical details for results used in the main text 
are relegated to two appendices at the end of the paper.

\subsection*{Prospects}

We conclude this short introduction with a few remarks on some of the interesting open questions raised in this
work and the prospects of further related developments. 

Our main motivation for the study of the classical problem in this paper is the eventual formulation of general 
half-BPS co-dimension-1 defects in $3d$ $\NN=2$ supersymmetric quantum field theories on curved spaces, and 
the non-perturbative computation of observables associated with these defects. 

The observables we are interested in include the partition function of $\NN=2$ supersymmetric gauge theories 
on curved backgrounds with boundary. With $A$-type boundary conditions these partition functions are
computing a class of supersymmetric wavefunctions. It would be interesting to explore the dependence of 
these observables on the moduli of the defects, i.e.\ the moduli of the boundary conditions we formulate, 
generalizing the bulk analysis of Ref. \cite{Closset:2013vra}. A preliminary computation of partition functions on 
manifolds with boundary in three dimensions using localization techniques has been performed in special cases 
in \cite{Sugishita:2013jca,Yoshida:2014ssa}. The results in the present paper can be used to extend known results in 
this direction. 

Moreover, one can also attempt to use the information of supersymmetric wavefunctions to study the structure of 
observables on closed manifolds that do not involve co-dimension-1 defects.
Hints of such a possibility come from a variety of previous results:
the holomorphic block decomposition of $3d$ partition functions \cite{Beem:2012mb,Yoshida:2014ssa}, 
and the analogous phenomenon in different dimensions \cite{Benini:2012ui,Nieri:2015yia,Qiu:2014oqa},
the recent progress in D-brane amplitudes in $2d$ $\NN=(2,2)$ theories 
\cite{Hori:2013ika,Honda:2013uca,Gomis:2012wy}, and $tt^\star$ arguments in flat toroidal backgrounds in 
three, and four spacetime dimensions \cite{Cecotti:2013mba}.

Boundary conditions also introduce another tool to probe dualities between quantum field theories.  
If two theories are dual at the quantum level, we expect corresponding boundary 
conditions on each side to be mapped to each other in a non-trivial way. For instance, in the case of mirror 
symmetry, the duality between boundary conditions can be understood in the mathematical framework of 
{\it symplectic duality} \cite{Bullimore:2016nji}. $3d$ Seiberg duality also acts non-trivially on boundary
conditions.
We refer the reader to Ref.\ \cite{Armoni:2015jsa} for a recent discussion of the relation between $3d$ 
Seiberg dualities and $2d$ level-rank dualities in this context. Similar problems 
with Wilson loops were investigated in \cite{Kapustin:2013hpk},\cite{Assel:2015oxa}.
Finally, an intriguing interpretation of co-dimension-1 defects relates the expectation value of 
these operators to the entanglement structure of the field theory \cite{Hung:2014npa}.

Another arena of potential applications of such computations is M-theory. The study of boundary conditions 
in the ABJM theory \cite{Aharony:2008ug}, which is an $\NN=6$ Chern-Simons-matter theory, is expected to 
yield information about the physics of M2, M5-branes and their interactions. For instance, it is anticipated that 
the low-energy theory at the orthogonal intersection of M2 and M5-branes in $\mathbb{C}^4\times \mathbb{Z}_k$ 
is a $2d$ theory with $\NN=(4,2)$ (or in special cases $\NN=(4,4)$) supersymmetry. 
The non-abelian quantum properties of this theory are still illusive. A recent bare Lagrangian formulation of this 
theory in terms of boundary degrees of freedom motivated by D-brane physics in type IIB Hanany-Witten setups
was proposed recently in \cite{Niarchos:2015lla}. For a study of half-BPS boundary conditions 
in ABJM theory see \cite{Berman:2009xd}, \cite{Okazaki:2015fiq}.   

Finally, there are several aspects of the general theory of supersymmetric boundaries in three dimensions that are 
not discussed in this paper. One of these aspects is the general curved space analog of B-type boundary 
conditions in $2d$ $\NN=(2,2)$ theories. Another aspect that is worth exploring further is the 
formulation of half-BPS boundaries using explicit boundary degrees of freedom and boundary 
actions \cite{DiPietro:2015zia}. The analysis of supersymmetric boundaries in $2d$ $\NN=(2,2)$ theories 
in \cite{Hori:2013ika} was performed in this manner.

\section{Review of rigid supersymmetry on curved 3-manifolds} \label{sec:CDFK}

In the modern approach to rigid supersymmetry on curved spaces, the metric tensor $g_{\mu\nu}$ 
(or any other background field) is embedded into a certain supergravity multiplet, and the field theory 
is obtained by taking the rigid limit of Festuccia-Seiberg \cite{Festuccia:2011ws} (FS). With a $U(1)_R$ 
symmetry, the supergravity of interest in $4d$ is the ``new minimal supergravity" of \cite{Ferrara:1988qxa}, 
and the supergravity multiplet contains an $R$-symmetry gauge field $A_\mu^{(R)}$, a conserved vector 
$V^\mu$ and the two gravitini $\Psi_{\mu\a}$, $\widetilde{\Psi}_{\mu\dot{\a}}$. Following FS, the rigid field 
theory of chiral and vector superfields on the curved space, is obtained from the action of off-shell 
supergravity coupled to chiral and vector fields, by freezing the bosonic components of the supergravity 
multiplet to a configuration in which $\d\Psi_{\mu}=\d\widetilde{\Psi}_\mu=0$. The advantage of this 
formulation is that the whole procedure can be carried out without the need of an explicit solution to the 
equations $\d\Psi_{\mu}=\d\widetilde{\Psi}_\mu=0$. 

In $3d$ it is possible to perform a twisted 
dimensional reduction of the $4d$ rigid theories to infer a consistent new minimal $3d$ algebra 
\cite{Closset:2012ru}. At the end of this process, the background fields are, the metric $g_{\mu\nu}$, an
$R$-symmetry gauge field $A_\mu^{(R)}$, a conserved vector $V^\mu$ (as in $4d$), and an extra scalar field $H$. 
The conditions $\d\Psi_\mu=\d\widetilde{\Psi}_\mu=0$ reduce to the following two Killing spinor equations
\bea
 \big( \nabla_\mu - i A^{\lp R\rp}_\mu \big) \zeta &=& -\frac{1}{2} H \g_{\mu}\zeta +\frac{i}{2} V_{\mu}\zeta - \frac{1}{2}\vep_{\mu\nu\rho}V^{\nu}\g^{\rho}\zeta\nn\\
 										  &=&  -\frac{1}{2} \g_{\mu}( H \zeta -i  V_\nu\g^\nu \zeta)~, \label{eq_zeta}\\
 \big( \nabla_\mu + i A^{(R)}_\mu \big) \zetaT   &=& -\frac{1}{2} H \g_{\mu}\zetaT - \frac{i}{2} V_{\mu}\zetaT + \frac{1}{2}\vep_{\mu\nu\rho}V^{\nu}\g^{\rho}\zetaT\nn\\
 										&=&  -\frac{1}{2} \g_\mu ( H\zetaT + i  V_\nu\g^\nu \zetaT)~. \label{eq_zetaT}
\eea
The two Weyl spinors $\zeta$ and $\zetaT$ have  R-charges $+1$ and $-1$ respectively. 

In practice, given a choice of the background metric, the other background fields can be adjusted to obtain 
at least one solution of the Killing spinor equations. On the other hand, assuming that at least one Killing 
spinor exists as a solution of the equations \eqref{eq_zeta}, \eqref{eq_zetaT}, it is possible to deduce what 
geometric structure the manifold needs to possess. In $3d$, this analysis was first carried out in 
\cite{Klare:2012gn, Closset:2012ru, Closset:2013vra}. In sec.~\ref{The_geometry_M3} we will review in some 
detail the relevant geometry since it will play an important role in our problem. In fact, in order to set 
up supersymmetric boundary conditions, it will be useful to improve slightly the way in which the 
relevant geometric structure is characterized. The new material is presented in section \ref{AtypeBacks}.
Experts familiar with rigid supersymmetry on spaces without boundary (e.g.\ the work in 
\cite{Klare:2012gn, Closset:2012ru, Closset:2013vra}) may skip to section \ref{AtypeBacks}. 
We follow closely the notation of Ref.\ \cite{Closset:2012ru}.

In our presentation, it will be convenient to make an explicit distinction between commuting and anti-commuting Killing 
spinors. In particular, we will denote the commuting spinors with $\zeta$ and $\zetaT$, and the anti-commuting 
spinors with $\ep$ and $\epT$. Both sets of Killing spinors satisfy the same equations. The anti-commuting 
spinors $\ep$ and $\epT$, will provide the parameters of the supersymmetry transformations of the field theory. 
The commuting spinors, $\zeta$ and $\zetaT$, will be used to explore the geometry of the manifold.

\subsection{Geometry of $\mathcal{M}_3$}\label{The_geometry_M3}

The existence of Killing spinor solutions, $\zeta$ and $\zetaT$, strongly constrains the geometric structure of the 
background fields. We will not repeat the general analysis here, but we recall two important results of 
\cite{Closset:2012ru}, which will be useful for later purposes. 
The first states that a solution of \eqref{eq_zeta}, or \eqref{eq_zetaT}, 
when it exists, is nowhere vanishing.\footnote{This property will be crucial for the consistency of the {\it canonical 
formalism} that we set up in sec.\ \ref{Canonical_Formalism}. In general situations, depending on the specifics of
the Killing spinor equations, a non-trivial solution may or may not admit zeros \cite{Dumitrescu:2012ha}.} 
The second result states that given {\it one} Killing spinor, say $\zeta$ for concreteness, it is possible to cover the 
manifold with a {\it transversely holomorphic foliation} (THF), and write the metric in the following form
\beq\label{metricCDFK_1}
ds^2= \metric_{\mu\nu} dx^\mu dx^\nu=\eta^2 + c(\tau,z,\bar{z})^2 dz d\bar{z},\qquad 
\eta=d\tau + \lp h(\tau,z,\bar{z}) dz + c.c.\rp 
~.
\eeq
By definition of the THF, the adapted coordinate $\tau$ is real, whereas $\{z,\bar{z}\}$ are complex. 
The leaves of the foliation are the submanifolds $z=const.$, and two patches are related by transitions 
functions, $f$ and $h$, such that $z'=f(z)$ with $f$ holomorphic, and $\tau'=h(\tau,z,\bar{z})$ with $g$ real. 
In particular, $g$ can be put in the form $h(\tau,z,\bar{z})=\tau+t(z,\bar{z})$.

The origin of the transversely holomorphic foliation is an integrability constraint.  
The one-form $\eta=\eta_\mu dx^\mu$ can be represented as the spinor bilinear
\beq\label{contactS_CDFK}
\eta_\mu=\frac{1}{|\zeta|^2}\, \zeta^c\g_\mu\zeta,\qquad |\zeta|^2=\zeta^c\zeta
~,
\eeq
and the following field can be defined,\footnote{The triple $(\eta_\mu,\xi^\mu, J^{\mu}_{\ \nu})$, 
with $\eta_\mu$, $\xi^\mu$, and $J^{\mu}_{\ \nu}$ 
such that $\eta_\mu\xi^\mu=1$ and $J^2=-\Id +\xi\,\eta$, is called an {\it almost contact structure} (ACS). 
This definition only requires that $\eta_\mu$, $\xi^\mu$, and $J^{\mu}_{\ \nu}$, satisfy algebraic constraints. 
It does not require the manifold to have a metric. For Riemannian manifolds, a metric $g_{\mu\nu}$ is said to be 
{\it compatible} with the ACS if $\xi^\mu=g^{\mu\nu}\eta_\mu$. The ACS is then promoted to an {\it almost contact 
metric structure} (ACMS). Similarly to the definition of a complex structure, the difference between an {\it almost} 
and a {\it contact structure}, is a differential constraint. However, this constraint is not \eqref{CDFK_int_constr} but: 
$d\eta\lp\xi,\cdot\rp=0$ for the {\it contact structure}, and $d\eta\lp\cdot,\cdot\rp=g\lp J\cdot,\cdot\rp $ for the 
{\it contact metric structure} \cite{BlairBook}. It is perhaps useful to mention that the condition for a contact metric 
structure resembles the one for K\"ahler manifolds in even dimensions \cite{daSilva}.}
\beq
\xi^\mu=g^{\mu\nu}\eta_\mu~,\qquad J^\mu_{\ \nu}=\vep^{\mu}_{\ \nu\rho}\xi^\rho~.
\eeq
The spinor $\zeta^c$ is the charge conjugate to $\zeta$.
Notice that from the properties of $\xi^\mu$ and $J^\mu_{\ \nu}$ it also follows that the Killing spinor equation of $\zeta$, \eqref{eq_zeta},
is invariant under the shift symmetry,
\beq\label{shift_inv}
\begin{array}{rcl}
V^\mu&\rightarrow\quad& V^\mu + X^\mu + k\, \xi^\mu~, \\
H& \rightarrow\quad & H + i k~,\\
\end{array}
\eeq
where the scalar $k$ and the vector field $X^\mu$ are such that $ J^{\mu}_{\ \nu}X^\nu=i X^\mu$ and $\nabla_\mu(X^\mu + k \xi^\mu)=0$. 
After gauge fixing the shift invariance, the Killing spinor equation \eqref{eq_zeta}, implies the constraint 
\beq\label{CDFK_int_constr}
J^\mu_{\ \nu}\lp \mathcal{L}_{\xi} J\rp^{\nu}_{\ \rho}=0
~.
\eeq 
Given the condition 
\eqref{CDFK_int_constr}, the authors of Ref.\ \cite{Closset:2012ru} showed that it is possible to find the adapted 
coordinates $\{\tau,z,\bar{z}\}$ introduced in \eqref{metricCDFK_1}. This is the THF associated to $\zeta$. 
On equal footing, there exists the THF associated with $\zetaT$, which is defined as in \eqref{contactS_CDFK} 
with the substitution $\zeta\rightarrow\zetaT$, i.e. $\tilde{\eta}_\mu=(\zetaT^c\g_\mu\zetaT)\,{|\zetaT|}^{-2}$,
and  the Killing spinor equation of $\zetaT$, \eqref{eq_zeta}, remains invariant under a shift similar to \eqref{shift_inv}. 

Manifolds that admit {\it two} complex supercharges of opposite $R$-charge have additional properties 
compared to the THF. They have a nowhere vanishing Killing vector $K^\mu$, and a {\it contact 
structure}. 
The Killing vector is represented as
\beq\label{Killing_rep}
K^\mu =\zetaT\g^\mu\zeta
~.
\eeq
It solves the equation 
\beq\label{eq_killing}
\nabla_\mu K_\nu=iH\, \vep_{\mu\nu\rho}K^\rho + \vep_{\mu\nu\rho} V^\rho\, \zetaT\zeta
~,
\eeq 
from which $\nabla_{\{\mu}K_{\nu\}}=0$ follows. The norm of $K^\mu$ is 
$K^\mu K_\mu=(\zetaT\zeta)^2\equiv \Omega^2$, and the function $\Omega$ is such that 
\beq
K^\mu\partial_\mu(\zetaT\zeta) 
= -K^\mu \varepsilon_{\mu \a\b} V^\a K^\b =0
~. 
\eeq
Notice that the Killing spinor equations are linear, therefore $\zeta$ and $\lambda\zeta$, with $\lambda$ an 
arbitrary complex number, are both solutions. Similarly for $\zetaT$. However, the relation $\zetaT\zeta=\Omega$ 
breaks the arbitrariness in the normalization of $\zeta$ and $\zetaT$, and only the symmetry 
$\zeta\rightarrow \lambda\zeta$ with $\zetaT\rightarrow\lambda^{-1}\zetaT$ remains. Eq.\ \eqref{Killing_rep} is also 
invariant under this scaling.

When the Killing vector is real, the geometry can be further characterized by the orbits of $K^\mu$. Two cases can 
be distinguished: either the orbits of $K^\mu$ are periodic, or they do not close. The first case consists of manifolds 
with the topology of an $\mathbb{S}^1$-bundle over a $2d$ Riemann surface. In the second case, it can be proved 
that there exists another independent Killing vector, transverse to $K^\mu$, and that the isometry group of 
$\mathcal{M}_3$ is at least $U(1)\times U(1)$ \cite{Isenberg}. 
 
The {contact structure} $(\etah_\mu,\xih^\mu, \Jh^{\mu}_{\ \nu})$, is defined by the fields
\beq\label{contact_structure}
\etah_\mu=\frac{1}{\Omega^2} K_\mu,\qquad 
\xih^\mu=K^\mu,\qquad \Jh^{\mu}_{\ \nu}=\frac{1}{\Omega} \vep^{\mu}_{\ \nu\rho} K^\rho\ 
~,
\eeq
subject to the relations: $\etah_\mu\xih^\mu=1$, $(d\etah)_{\mu\nu}\xih^\mu=0$.  
The latter condition can be checked by means of the Killing spinor equations \eqref{eq_zeta} and \eqref{eq_zetaT}. 
In particular, $(d\etah)_{\mu\nu}\xih^\mu=0$ implies through Darboux's theorem \cite{daSilva} the existence of local 
coordinates $(\psi, x_1,y_1)$ such that
\beq\label{Darboux}
\etah=\frac{K_\mu}{\Omega^2}=d\psi + x_1 dy_1\ ,\qquad \xih^\mu\partial_\mu=\partial_\psi\ .
\eeq
As a result, the Killing vector $K^\mu$ is aligned along $\partial_\psi$. $\mathcal{M}_3$ endowed with such 
contact structure is a {\it contact manifold}, and the vector $\xi^\mu=K^\mu$ is called {\it Reeb vector}. An equivalent 
characterization of a contact manifold is the condition that $\etah\wedge d\etah\neq 0$. The coordinates 
$(\psi, x_1,y_1)$ are called {\it canonical} since the condition $\etah\wedge d\etah\neq 0$ becomes trivial. 
The contact structure defined in \eqref{contact_structure}, shares the same algebraic properties of the triple 
$(\eta_\mu,\xi^\mu, J^{\mu}_{\ \nu})$ defined in \eqref{CDFK_int_constr}. These are $\eta_\mu\xi^\mu=1$ and 
$J^2=-\mathbb{I} +\xi\,\eta$, but in addition, an explicit calculation shows that the tensor $\Jh^{\mu}_{\ \nu}$ 
satisfies a stronger (integrability) constraint, $\mathcal{L}_{\xih} \Jh=0$. \\

In section \ref{AtypeBacks} we will supplement the above results on 3-manifold geometry 
with a further new refinement that facilitates the introduction of boundaries preserving a subset of the
bulk supersymmetries.

\vspace{0.3cm}

\subsection{Supersymmetric multiplets and transformations}

Rigid supersymmetric field theories exist on any curved background $\mathcal{M}_3$, equipped with the two 
Killing spinors $\ep$ and $\epT$. Their Lagrangians are obtained by exploiting the {\it multiplet calculus} of $4d$ 
new minimal supergravity \cite{Ferrara:1988qxa} and its $3d$ version (see appendix of \cite{Closset:2012ru}). 

By multiplet calculus we mean the collection of all the supersymmetry transformations of the components of a generic 
multiplet $\mathcal{S}$. The total number of independent degrees of freedom in $\mathcal{S}$ is $16$ bosonic 
plus $16$ fermionic. They are organized as follows:
\beq
\mathcal{S}=\{ C,\chi_a,\tilde{\chi}_\a, M,\tilde{M},a_\mu,\s,\lm_\a,\lmb_\a, D \}
~.
\eeq  
The R-charges are $(0,-1,+1,-2,+2,0,0,+1,-1,0)$ relative to the bottom component $C$.
The supersymmetry transformation rules $\d_\ep\mathcal{S} + \tilde{\d}_{\epT}\mathcal{S}$ are summarized in 
appendix \ref{conventions}.
The set of all these transformations realize an algebra on the space of fields. Denoting with $\varphi_{(r,z)}$ a field 
of arbitrary spin, R-charge $r$, and central charge $z$, the supersymmetric algebra is represented by
\beq\label{Algrebra_fields_susy}
[\d_\ep ,\tilde{\d}_{\epT} ]\, \varphi_{(r,z)}= -2 i \lp \mathcal{L}_K + \ep\epT \lp z- r H\rp \rp  \varphi_{(r,z)},\qquad 
[ \d,\d ]=0,\qquad [\tilde{\d},\tilde{\d}] =0
~,
\eeq
The symbol $\mathcal{L}_K$ is defined in \cite{Closset:2012ru} as a modified Lie derivative along $K$
\beq\label{CDFK_Lie_derivative}
\mathcal{L}_K \varphi_{(r,z)}
=\left[ \mathrm{Lie}_K - i r K^\mu  \lp A_\mu -\frac{1}{2} V_\mu \rp - iz\, K^\mu \mathcal{C}_\mu\right] \varphi_{(r,z)}
~.
\eeq
The covariant derivative associated to $\mathcal{L}_K$ will be denoted as
\beq
\D_{\mu}\varphi_{(r,z)} = \left[ \nabla_\mu - i r \lp A_\mu -\frac{1}{2} V_\mu \rp - i z\, \mathcal{C}_\mu\right] 
\varphi_{(r,z)}
~.
\eeq
Here the background gauge field $\mathcal{C}_\mu$ is related to the background conserved vector $V^\mu$ by the relation 
$V_\mu=-i\varepsilon^{\mu\nu\rho}\partial_\nu \mathcal{C}_\rho$. 
The gauge field $A_\mu$ is not $A_\mu^{(R)}$, but the two are related by a redefinition
\beq\label{def_A_Rsym}
A_\mu^{(R)}\equiv A_\mu -\frac{3}{2} V_\mu
~.
\eeq
The combination $A_\mu-\frac{1}{2} V_\mu$ is not invariant under the shift symmetry \eqref{shift_inv}, 
but $A^{(R)}_\mu$ is. Accordingly, it is convenient to express $(\mathcal{L}_K + \ep\epT ( z- r H)) $ as 
\beq
\mathrm{Lie}_K - i r K^\mu  A_\mu^{(R)} -ir ( K^\mu V_\mu - \ep\epT\,iH)- iz\, 
K^\mu \mathcal{C}_\mu+ \ep\epT \, z
~.
\eeq 
In what follows, we will mostly use $\D_\mu$, as defined above, since we adopt the notation of Ref.\ 
\cite{Closset:2012ru}. Sometimes, however, it will be convenient to consider $A^{(R)}$ in the covariant derivative. 
When this happens we will be very explicit.\\ 

For the benefit of the reader we list here two standard short multiplets $\mathcal{S}$ that will play a 
dominant role in the main discussion. The shortening of the multiplets is obtained by imposing restrictions on 
its components.

\subsubsection{Chiral and the anti-chiral multiplets} 
Chiral (anti-chiral) multiplets are obtained by imposing the conditions $\tilde{\chi}_{\a}=0$ (${\chi}_{\a}=0$).  
This implies that not all components of the generic multiplet are independent. 
A chiral multiplet $\Phi$, with independent components $\{ \phi,\psi_\a,F\}$ is organized as follows,
\beq
\begin{array}{ccl}
\mathcal{S}\Big|_{ \tilde{\chi}=0}\equiv \Phi& = & \{ \phi,-i\sqrt{2}\psi_\a, 0,-i2 F, 0,\\ 
			&  &	\quad -i \D_\mu\phi , (z-rH)\phi, 0, 0, \frac{r}{4} (R-2V^2-2H^2)\phi- z H\phi \}.\rule{0pt}{.5cm}

\end{array}
\eeq
In the above formula, $R[\phi]=r$ is the $R$-charge of $\phi$, and $z$ is the central charge. 
The transformation rules of $\{ \phi,\psi,F\}$ are
\beq\label{susyrules_chiral}
\begin{array}{ccl}
\d \phi & = & \sqrt{2}\, \ep\psi ~,\\
\d \psi &= & \sqrt{2} \ep F - i\sqrt{2}   \lp z- r H \rp \epT\, \phi - i\sqrt{2}\g^\mu \epT\, \D_\mu\phi ~, \rule{0pt}{.5cm} \\
\d F & = & i\sqrt{2} \lp z- (r-2) H \rp \epT\psi - i\sqrt{2}\, \D_\mu (\epT\g^\mu\psi)~. \rule{0pt}{.5cm}
\end{array}
\eeq
The shorthand notation for $\Phi$ will be $\Phi=\{ \phi,\psi,F\}$. 
The case of the anti-chiral multiplet $\tilde{\Phi}$ is analogous. The independent components are 
$\{\phiT,\psiT_\a,\FT\}$ and the supersymmetric transformation rules are
\beq\label{susyrules_antichiral}
\begin{array}{ccl}
\d \phiT & = & -\sqrt{2}\, \epT\psiT~, \rule{0pt}{.5cm}\\
\d \psiT &= & \sqrt{2}  \epT \FT + i\sqrt{2} \lp \tilde{z}-\tilde{r} H\rp \ep\,\phiT + i \sqrt{2} \g^\mu\ep\, \D_\mu\phiT~,
 \rule{0pt}{.5cm} \\
\d \FT & = &  i  \sqrt{2}  \lp \tilde{z} - (\tilde{r} -2) H \rp  \ep\,\psiT - i  \sqrt{2}\,  \D_\mu (\ep\,\g^\mu \psiT)~, 
  \rule{0pt}{.5cm}
\end{array}
\eeq
where the $R$-charge of $\phiT$ is $R[\phiT]=-\tilde{r}$ and its central charge is $-\tilde{z}$. 

\subsubsection{Real and gauge multiplets} 
\label{realmultiplets}

A real multiplet $\Sigma$ arises by imposing on $\mathcal{S}$ the conditions $M=\tilde{M}=0$, and $r=z=0$. 
The subset of independent components can be defined by 
$\{ C^{(\Sigma)}, \chi^{(\Sigma)}_\a,\tilde{\chi}^{(\Sigma)}_\a,j_\mu, {\s}^{(\Sigma)} \}$, and $\Sigma$ is organized 
as follows
\beq
\label{Sigma}
\begin{array}{ccl}
\Sigma&=&\{C^{(\Sigma)}, \chi^{(\Sigma)}_\a, \tilde{\chi}^{(\Sigma)}_\a, 0, 0, \\ 
	& &\rule{1.5cm}{0pt} - j_\mu-V_\mu C^{(\Sigma)},  -{\s}^{(\Sigma)}, \rule{0pt}{.5cm} \\
	& &\rule{1.5cm}{0pt} -\frac{i}{2} H \tilde{\chi}^{(\Sigma)}_\a + i \g^{\mu\ \b}_{\ \a}( \nabla_\mu+i V_\mu)\tilde{\chi}^{(\Sigma)}_\b\,,
								+\frac{i}{2} H \chi^{(\Sigma)}_\a - i \g^{\mu\ \b}_{\ \a}(\nabla_\mu -i V_\mu)\chi^{(\Sigma)}_\b \,, \rule{0pt}{.6cm} \\
	& &\rule{1.5cm}{0pt} - V^\mu j_\mu - H {\s}^{(\Sigma)}- (\nabla^2 + V^2) C^{(\Sigma)} \}~.  \rule{0pt}{.6cm}
\end{array}
\eeq
The vector field $j_\mu$ is a conserved current, $\nabla_\mu j^\mu=0$. The supersymmetric transformations rules 
are
\beq
\label{transformreal}
\begin{array}{ccl}
\d C^{(\Sigma)} &=&i \ep \chi^{(\Sigma)}+ i \epT\tilde{\chi}^{(\Sigma)}~, \\
\d\chi^{(\Sigma)}&=&   \epT \,\s^{(\Sigma)}+ i \g^\mu\,\epT\, (j_\mu + i \partial_\mu C^{(\Sigma)}+ V_\mu C^{(\Sigma)})~, \rule{0pt}{.5cm}\\
\d\tilde{\chi}^{(\Sigma)} &=&\ep\,\s^{(\Sigma)} - i \g^\mu\, \ep\, (j_\mu-i\partial_\mu C^{(\Sigma)} + V_\mu C^{(\Sigma)})~, \rule{0pt}{.5cm}\\
\d j_\mu &=& i \varepsilon_{\mu\nu\rho}\nabla^\nu(\ep\g^\rho\chi^{(\Sigma)}- \epT\g^\rho\tilde{\chi}^{(\Sigma)})~, \rule{0pt}{.5cm}\\
\d\s^{(\Sigma)}&=& -i \nabla_\mu( \ep\g^\rho\chi^{(\Sigma)}+ \epT\g^\rho\tilde{\chi}^{(\Sigma)} )+ 2iH(\ep\chi^{(\Sigma)}+\epT\tilde{\chi}^{(\Sigma)}) - V_\mu( \ep\g^\rho\chi^{(\Sigma)}- \epT\g^\rho\tilde{\chi}^{(\Sigma)})~. \rule{0pt}{.5cm}
\end{array}
\eeq

An abelian gauge multiplet $\mathcal{V}$ is a generic multiplet $\mathcal{S}$ subject to the gauge freedom 
$\d\mathcal{V}=\Lambda+\tilde{\Lambda}$, where $\Lambda$ is a chiral multiplet. After the standard procedure
of Wess-Zumino gauge fixing the independent fields reduce to $\{\A_\mu,\s,\lm_{\a},\lmb_{\a},D\}$. 
Notice that an abelian gauge multiplet becomes a real multiplet under the identification:
\beq
\begin{array}{lclcl}
C^{(\Sigma)}=\s~,&  \rule{1cm}{0pt} & \chi^{(\Sigma)}_\a = i \lmb_\a~, &\rule{1cm}{0pt} &\tilde{\chi}^{(\Sigma)}_\a= - i \lm_a~, \\
j_\mu =-\frac{i}{2} \vep_{\mu\nu\rho} f^{\nu\rho}~, & &\s^{(\Sigma)}= D+\s H ~,& &\rule{0pt}{.5cm} \\
\end{array}
\eeq
where $f^{\nu\rho}$ is the field strength of $\A_\mu$.  This parametrization will be particularly useful in later sections.

In the case of non-abelian gauge multiplets the supersymmetry transformation rules have extra terms compared to \eqref{transformreal}. The complete set of transformation rules in the non-abelian case is 
\beq
\begin{array}{ccl}
\d\s&=& - \ep \lmb + \epT \lm  ~,\\
\d\lm&=&+ i \ep (D+\sigma H) -\frac{i}{2} \varepsilon^{\mu\nu\rho} \gamma_\rho \ep\, \FF_{\mu\nu} 
-\gamma^\mu \ep \left( i D_\mu \sigma - V_\mu \sigma \right)  ~,\rule{0pt}{.5cm}\\
\d\lmb&=& -i \epT (D+\sigma H) - \frac{i}{2} \varepsilon^{\mu\nu\rho} \gamma_\rho \epT \, \FF_{\mu\nu}
+\gamma^\mu \epT \left( i D_\mu \sigma + V_\mu \sigma \right)~, \rule{0pt}{.5cm}\\
\d\A_\mu&=& -i ( \ep \gamma_\mu \lmb +\epT \gamma_\mu \lm ) ~, \rule{0pt}{.5cm}\\
\d D&=& D_\mu ( \ep \gamma^\mu \lmb - \epT \gamma^\mu \lm )  
-i V_\mu ( \ep \g^\mu \lmb + \epT \g^\mu \lm ) - H ( \ep \lmb - \epT \lm )
- [ \lmb \ep + \epT \lm, \sigma ] ~.\rule{0pt}{.5cm}
\end{array}
\eeq 
$\FF_{\mu\nu}$ is the field strength of $\AA_\mu$, and $D_\mu$ is the non-abelian gauge covariant derivative
\eqref{gaugecov}.

\subsubsection{Curved D- and F-terms}
So far we have not specified whether $\mathcal{S}$ is an elementary or a composite multiplet. The 
supersymmetric transformations are, of course, valid regardless of this distinction. 
Once elementary multiplets are defined, any composite multiplet $\mathcal{K}$ of the form 
$\mathcal{K}=(K,\chi^{(K)}, \tilde{\chi}^{(K)}, M^{(K)},\ldots )$ is generated by the multiplet calculus. 
In practice, given the definition of the bottom component $K$, as a function of the elementary fields $C^I$, the 
other components in the multiplet are obtained in a step-by-step procedure: varying $K(C^I)$ with the use of $\d C^I$ 
one reads off the definitions of $\chi^{(K)}$ and $\tilde{\chi}^{(K)}$, and so on. From the composite multiplets it is then 
possible to construct kinetic terms for the elementary fields and thus generic supersymmetric Lagrangians whose 
variation is a total derivative. 

Such Lagrangians can be understood as follows.
Given a generic multiplet $\mathcal{S}$ with $r=0$ and $z=0$, its $D$ component almost transforms as a 
total derivative. Terms that are not total derivatives are proportional to background fields, and the flat space 
result is recovered when these vanish. In curved space the correct combination transforming into a total derivative is
\cite{Closset:2012ru}
\beq\label{Def_SUSY_DLagrangian}
\begin{array}{ccccl}
\mathrm{curved\ D{\rm -}term} :&\rule{.5cm}{0pt} & \mathscr{L}_D&=& -\frac{1}{2}( D - a_\mu V^\mu - \s H)~,\\
			& & \d \mathscr{L}_D&=&  -\frac{1}{2} \nabla_{\mu} (\ep\g^\mu\lmb - \epT\g^\mu\lm - V^\mu \ep\chi 
			+ V^\mu \epT\chib )~.\rule{0pt}{.5cm}
\end{array}
\eeq
The result for the $F$ (or $\tilde{F}$) component of a chiral $\Phi$ (or anti-chiral $\tilde{\Phi}$) multiplet of 
R-charge $r=2$ (or $r=-2$) and central charge $z=0$ is the same as that in flat space. The F-term is 
\beq\label{Def_SUSY_FLagrangian}
\mathrm{curved\ F{\rm -}term} :\rule{.5cm}{0pt}  \mathscr{L}_F= F+ \tilde{F}~,
			\rule{.7cm}{0pt} \d \mathscr{L}_F= -2i\, \nabla_\mu( \epT\g^\mu\psi + \ep\g^\mu\psiT )~. \rule{0pt}{.5cm}
\eeq

\section{Manifold decomposition for curved $A$-type backgrounds} \label{AtypeBacks}

In this paper we focus on a class of background geometries introduced in \cite{Alday:2013lba}, that we call 
``$A$-type".\footnote{In \cite{Alday:2013lba} the partition function of $\mathcal{N}=2$ Chern-Simons theories 
on generic $A$-type backgrounds was computed explicitly using supersymmetric localization techniques
similar to \cite{Kapustin:2009kz}.} 
By definition, these backgrounds admit two supercharges related by charge conjugation. 
The charge conjugate spinors, $\zeta^c\equiv +i\g^2\zeta^\star$ and 
$\zetaT^c\equiv +i\g^2\zetaT^\star$,\footnote{Our $\g$ matrix conventions are summarized in appendix 
\ref{conventions}. In deriving the formulae \eqref{eq_zetachargeC} and \eqref{eq_zetaTchargeC} we made 
use of the relation $\g^{\mu\star}=-\g^2\g^\mu\g^2$. $^\star$ denotes the standard complex conjugation.} 
solve the equations
\bea
\label{eq_zetachargeC}
\big( \nabla_\mu + i A^{\lp R\rp}_\mu \big) \zeta^c &=&  +\frac{1}{2} \g_{\mu}( {H}^\star  \zeta^c 
-i  {V}^\star_\nu\g^\nu \zeta^c)~,\\
\label{eq_zetaTchargeC}
\big( \nabla_\mu - i A^{(R)}_\mu \big) \zetaT^c &=&  +\frac{1}{2} \g_\mu ( {H}^\star  \zetaT^c 
+ i  {V}^\star_\nu\g^\nu  \zetaT^c)~. \label{eq_zetac_eqs}
\eea
In general, given a Killing spinor, say $\zeta$, its complex conjugate $\zeta^c$ is an independent spinor that 
does not solve any of the Killing spinor equations \eqref{eq_zeta} and \eqref{eq_zetaT}. However, if the 
background fields $A_\mu^{(R)}$ and $V_\mu$ are real, and $H$ is purely imaginary, then $\zeta^c$ solves 
the same Killing spinor equation as $\zetaT$. Therefore, for an $A$-type background, $\zeta$ and $\zeta^c$ 
are the two Killing spinors of opposite R-charge. 

Now we are going to show that it is possible to understand any $A$-type background in terms of a 
{\it supersymmetric} foliation in which the leaves are topologically tori. As a mathematical statement about 
irreducible orientable closed $3$-manifolds, it is certainly well known in the literature that such a toric foliation 
exists, however we will use supersymmetry and the Killing spinors $\zeta$ and $\zetaT$ to re-derive this result. 
Very explicitly, the geometry of the foliation will be characterized by a distribution of orthogonal vector fields built 
out of the Killing spinors. One of these vectors will be the Killing vector $K^\mu$, and we will construct another 
vector $N^\mu$ that: 1) is orthogonal to $K^\mu$, and 2) can be used to define a proper orthogonal submanifold. 

The use of vector fields, instead of the adapted coordinates of the THF, will be essential in the formulation
of boundary conditions preserving a subset of the bulk supersymmetry. With such a foliation in place, 
we will be able to decompose the compact manifolds by placing a boundary (or a co-dimension-1 defect) along  
any leaf of the foliation. Our main purpose will be to formulate rigid supersymmetric fields theories on 
the resulting spaces with boundary that are topologically solid tori. 
Since the metric is part of a supergravity multiplet, the decomposition of the
manifolds should be combined with certain extra conditions on the remaining background fields. We will discuss 
concretely how the manifold decomposition is carried out in the rest of this section. In the final 
subsection \ref{SUSY_EXAMPLES}, we revisit some of the well-known examples of compact $3d$ manifolds, 
and re-discuss them from the perspective of this decomposition.

\subsection{Supersymmetric foliation}\label{SUSY_FOLIATION}

\subsection*{Normal vector}

Let us consider how the existence of the Killing spinors $\zeta$ and $\zeta^c$ determines the geometry of $A$-type 
manifolds. By fixing the normalization of $\zetaT$ to be $\zetaT=\zeta^c$, 
we show that supersymmetry provides a ``refinement" of the THF in which a special orthogonal direction to 
$K^\mu$ is selected out. 

The starting point of our treatment is based on the use of a Fierz identity for commuting spinors
that allows us to show that the real vector $N^\mu$, defined as
\beq\label{Def_N}
N^\mu= (\zeta^\star\g^\mu\zeta ) - ( \zetaT^\star\g^\mu\zetaT)= (\zeta^\star\g^\mu\zeta ) + c.c.
~,
\eeq
is orthogonal to $K^\mu$, i.e.\ $K_\mu N^\mu=0$. The same result about $N_\mu$ can be obtained by noticing that
\beq\label{Alternative_N}
\zeta^\star\g_\mu\zeta=+i\zetaT\g_2\g_\mu\zeta= +i g_{\nu\mu} e^\nu_2 ( \zetaT\zeta) -\vep_{\nu\mu\rho} 
K^\rho e^{\nu}_2\ ,
\eeq
where $e^\mu_2$ is an unspecified vielbein. Hence, the real part of \eqref{Alternative_N} gives 
$N^\mu=2\varepsilon^{\mu\nu\rho}e_\nu^2 K_\rho$, which is manifestly orthogonal to $K^\mu$. 
The tangent space $T\mathcal{M}_3$ can then be spanned by the following orthogonal vectors: 
$K^\mu, N^\mu$, and $\tilde{K}^\mu\equiv \varepsilon^{\mu\nu\rho}N_\nu K_\rho$. By construction, we also have 
\beq
N^\mu e^2_\mu=0
~,
\qquad
\label{Alternative_Ktilde}
\tilde{K}^\mu=2 \varepsilon^{\mu\nu\rho}\vep_{\nu\a\b}e^\a_2 K^\b K_\rho
= - 2e^\mu_2 ||K||^2 + 2K^\mu \lp K\cdot e^2\rp\ .
\eeq
It is, therefore, convenient to choose a reference frame, $\{ e_1,e_2,e_3\}$, such that 
$K^\mu=e_3^\mu$ and $\lp K\cdot e^2\rp=0$. For such a frame we deduce from \eqref{Alternative_N} that 
$e_2^\mu \propto \vep^{\mu}_{\ \nu \rho} N^\nu e_3^\rho$, and $N^\nu\propto e_1^\nu$. By consistency, we have to prove that the inverse metric $g^{\mu\nu}$ can be written in terms of the bilinears 
$K^\mu K^\nu$, $N^\mu N^\nu$ and $\tilde{K}^\mu \tilde{K}^\nu$. Indeed, from the Fierz identity applied to 
$K^\mu K^\nu$, and from the very definition of $\tilde{K}^\mu \tilde{K}^\nu$, we obtain the relation
\beq\label{CDFKmetric_2}
g^{\mu\nu} = \frac{1}{||K||^2} K^\mu K^\nu + \frac{1}{||N||^2}  N^\mu N^\nu  
+  \frac{1}{||N||^2||K||^2}\tilde{K}^\mu \tilde{K}^\nu\ .
\eeq
Our adapted dreibein fields are 
\beq\label{frame_fields}
e_1^{\mu}= \frac{N^\mu}{||N||} \equiv n^\mu, \quad 
e_2^\mu= \frac{\tilde{K}^\mu}{||\tilde{K}||} \equiv \tilde{k}^\mu,\quad 
e_3^\mu=\frac{K^\mu}{||K||} \equiv k^\mu 
~,
\eeq 
with $||\tilde{K}||=||N||\,||K||$. The norms of $K^\mu$ and $N^\mu$ are 
\beq\label{norms_spinors_frame_1}
K^\mu K_\mu=\Omega^2,\qquad N^\mu N_\mu=4 \lp \zeta^\star\zeta\rp^2 + 4 \Omega^2
~.
\eeq

The generic form of the metric on $\mathcal{M}_3$ was given in the previous section
\beq\label{metricCDFK_2}
ds^2= \metric_{\mu\nu} dx^\mu dx^\nu=\eta^2 + c(\tau,z,\bar{z})^2 dz d\bar{z},\qquad 
\eta=d\tau + \lp h(\tau,z,\bar{z}) dz + c.c.\rp ~.
\eeq
We can now compare it with \eqref{CDFKmetric_2}. 
For $A$-type manifolds, the knowledge of \eqref{Darboux} implies that $K_\mu=\Omega^2\,\etah$ 
and $e^3=\Omega\lp d\psi +x_1 dy_1\rp$. On the other hand, from \eqref{contactS_CDFK} and 
$\zetaT=\zeta^c$, we get $\eta=\Omega\,\etah=e^3$. Then, we can make use of the coordinates 
$\{z,\bar{z}\}$, instead of $\{x_1,y_1\}$, by implementing the contact structure condition on the function $h$. 
As a result, $h$ is $\psi$-independent, and since $\partial_\psi\Omega=K^\mu\partial_\mu\Omega=0$, the function 
$\Omega$ is also $\psi$-independent. The metric \eqref{metricCDFK_2} takes the final form \cite{Closset:2012ru}
\beq
ds^2=\Omega\lp z,\bar{z}\rp^2\lp d\psi + h(z,\bar{z})dz + c.c. \rp^2 + c(\psi,z,\bar{z})^2 dz d\bar{z}
~.
\eeq
Because of \eqref{CDFKmetric_2}, we also know that there exists a real parametrization of the plane $dz d\bar{z}$ 
in terms of the vectors, $N^\mu$ and $\tilde{K}^{\mu}$. In terms of the contact structure, $\etah$, the vectors 
$N^\mu$ and $\tilde{K}^{\mu}$ are understood as the distribution $\mathcal{H}=\mathrm{ker}\,\etah$. 
Recalling Frobenius' theorem, the defining property of a contact manifold, namely $\etah\wedge d\etah\neq 0$, 
implies that the distribution $\mathcal{H}=\mathrm{ker}\,\etah$ is not integrable. Instead, we will now study under 
what conditions the distribution generated by $K^\mu$ and $\tilde{K}^\mu$ is integrable. This will provide a regular 
foliation of the $A$-type manifold.

\subsection*{Integrability condition}\label{integrability}

Frobenius' theorem guarantees that the distribution $\mathcal{E}$, generated by $K$ and ${\tilde K}$, is integrable 
if the commutator $[K,\tilde{K}]$ belongs to $\mathcal{E}$ \cite{LeeBook}.\footnote{The normalizations of $K$ and 
$\tilde{K}$ are not important in the argument. Even though they contribute to the commutator, through the terms 
$\tilde{K}^\mu (K^\a\partial_\a\frac{1}{||\tilde{K}||})$ and ${K}^\mu (\tilde{K}^\a\partial_\a\frac{1}{||{K}||})$, these 
contributions belong to $\mathcal{E}$. Thus the statement of Frobenius' theorem remains unchanged.} 
This commutator is equal to
\bea\label{commutator_1}
[K,\tilde{K}]^\mu &\equiv& K^\a \nabla_\a \tilde{K}^\mu - \tilde{K}^\a \nabla_\a K^\mu \nn\\
                          &     =     & K^\a\vep^{\mu\nu\rho}\lp\nabla_\a N_\nu\rp K_\rho + K^\a\vep^{\mu\nu\rho} N_\nu\lp \nabla_\a K_\rho\rp - \vep^{\a\nu\rho}\ N_\nu K_\rho \nabla_\a K^\mu
~.
\eea
The second and third terms in \eqref{commutator_1} can be manipulated by using the equation of $K_\nu$, given 
in \eqref{eq_killing}. We obtain the expression
\beq
+K^\a\vep^{\mu\nu\rho} N_\nu\lp \nabla_\a K_\rho\rp- \vep^{\a\nu\rho}\ N_\nu K_\rho \nabla_\a K^\mu
=- \lp K\cdot V\rp N^\mu \, \zetaT\zeta - iH\, ||K||^2 N^\mu
~.
\eeq
A small complication arises in the calculation of $\nabla_\a N_\nu$. By definition 
$\nabla_\a N_\nu=D_\a\zeta^\star\g_\nu\zeta+\zeta^\star\g_\nu D_\a\zeta+c.c.$, but  $D_\a \zeta^\star$ 
is not just $-i\g^2D_\a\zetaT$. It is given by the more involved expression
\beq\label{express_Dzetastar}
D_\a\zeta^\star=-i\g^2 D_\a\zetaT-\frac{1}{4}\, \vep^{abc} \omega_{\a\, ab} \g^2\lp \g_c^\star +\g_c \rp\zetaT 
~.
\eeq
Substituting \eqref{express_Dzetastar} into $\nabla_\a N_\nu$, we get several contributions
\beq
\nabla_\a N_\nu= +i \lp D_\a \zetaT\rp\g^2\g_\nu\zeta +i \zetaT\,\g^2\g_\nu D_\a \zeta 
+\frac{1}{4}\omega_{\a\, ab}\vep^{abc} \zetaT\,\lp\g^2\g_c\g_\nu-\g_c\g^2\g_\nu\rp\zeta\ +\ cc.
\eeq
In the adapted frame \eqref{frame_fields}, and after some algebra, we can show that
\beq
\vep^{\mu\nu\rho}K_\rho\, \lp K^\a \nabla_\a N_\nu\rp=-\frac{1}{2} \lp K^\a \omega_{\a\, ab}\vep^{abc} K_c  \rp N^\mu
~.
\eeq
Then, the commutator becomes,
\beq\label{commutator_2}
[K,\tilde{K}]^\mu=\Big[  - \lp K\cdot V\rp  \, \zetaT\zeta - iH\, ||K||^2- \frac{1}{2} 
\lp K^\a \omega_{\a\, ab}\vep^{abc} K_c  \rp \Big]\,N^\mu\ .
\eeq

The vector $N^\mu$ does not belong to the distribution $\mathcal{E}$, thus the distribution is integrable iff the 
commutator $[K,\tilde{K}]$ vanishes. The explicit expression given in \eqref{commutator_2} can be rearranged by 
noticing that 
\beq
 \frac{\zetaT\zeta}{||K||^2}\lp K\cdot V\rp  + iH\, =\frac{1}{2}\vep^{\s\mu\nu}K_\s\nabla_\mu\etah_\nu
 =\frac{1}{2} \vep^{\s\mu\nu}e^3_\s \nabla_\mu e^3_\nu
 ~.
\eeq
The final form of the constraint coming from $[K,\tilde{K}]=0$ is
\beq\label{Frobenius_constr}
e^3_\s\, \vep^{\s\mu\nu} \Big[ e^\a_3 \lp\omega_{\a\, ab} e^a_\mu e^b_\nu\rp+\nabla_\mu e^3_\nu \Big]=
e^3_\s\, \vep^{\s\mu\nu} \Big[ e_3^\a\, \omega_{\a\, ab}\, e^a_\mu - \omega_{\mu\, 3 b}\Big] e^b_\nu\, = 0
~.
\eeq

Given an A-type manifold such that \eqref{Frobenius_constr} is satisfied, Frobenius' theorem \cite{LeeBook} then
implies the integrability of the distribution $\mathcal{E}$ and the existence of the foliation. As an example, all 
metrics of the type 
\beq\label{metric_real_param}
ds^2=\Omega^2\lp \theta,\varphi\rp \lp d\psi + F_\theta d\theta + F_\varphi d\varphi \rp^2 + g_{\theta\theta}^2 d\theta^2
+ g_{\varphi\varphi}^2 d\varphi^2
~,
\eeq
with $F_{\theta},F_{\varphi}$, $\psi$-independent, and $g_{\theta\theta}$, $g_{\varphi\varphi}$ generic, satisfy the 
constraint \eqref{Frobenius_constr}. Locally, where $N_{\mu}dx^\mu=d\theta$, the $2d$ submanifolds 
$\theta=const.$ define the foliation generated by $\mathcal{E}$. The leaves of the foliation will be denoted by 
$\mathcal{M}'_2$.\footnote{To be pedantic we should also specify a reference point $\theta_0\in\mathcal{M}_3$ for 
any leaf. This is usually implied.} We refer to $\mathcal{M}'_2$ as a {\it supersymmetric} leaf of $\mathcal{M}_3$. 
This terminology follows from the observation that the algebra of supersymmetry 
\beq
[\d_\ep ,\tilde{\d}_{\epT} ]\, \varphi_{(r,z)}= -2 i \lp \mathcal{L}_K + \ep\epT \lp z- r H\rp \rp  \varphi_{(r,z)}
\eeq
involves the Killing vector $K^\mu$ in the Lie derivative $\mathcal{L}_K$. In the simplest case, since 
the commutator of two transformations $\d_\ep$ and $\d_{\epT}$ squares to a translation along the orbit of 
the Killing vector $K^\mu$, $\mathcal{M}'_2$ preserves supersymmetry because $K^\mu$ belongs to 
$T\mathcal{M}'_2$.

\subsection{Topology and manifold decomposition}\label{sec_top_and_deco}

We can show with a simple argument that the topology of  $\mathcal{M}_2'$ cannot be genus zero, i.e.\ 
the leaves of the supersymmetric foliations are not spheres. The reasoning goes as follows. $\mathcal{M}_2'$ 
contains the orbits of the Killing vector $K^\mu$, and $K^\mu$ is nowhere vanishing because, as we mentioned 
in section \ref{The_geometry_M3}, the Killing spinors are nowhere vanishing. If $\mathcal{M}_2'$ was a sphere, 
$K^\mu$ would correspond to the $U(1)$ isometry of the sphere, which is unique. However, this cannot be the 
case since the $U(1)$ isometry of the sphere vanishes at the north and south poles. 

The topology of $\mathcal{M}_2'$ is a torus. We showed that it cannot be genus zero, but also it 
cannot be a higher genus surface either, because a $2d$ Riemann surface of genus $g>1$ would not have 
a Killing vector. Thus, an A-type background is topologically a torus fibered over a closed interval. 
The example of the round three-sphere is very instructive: $\mathbb{S}^3$ does admit a genus zero 
topological (Heegard) decomposition as the union of two 3-balls \cite{JennyHeegard}, 
however the supersymmetry that we are considering rules out this possibility and 
allows only for $g=1$ decompositions. A similar phenomenon has been noticed in $4d$. Ref.\ \cite{Dumitrescu:2012ha} 
showed that a $4d$ supersymmetric manifold for which $[K,\bar{K}]=0$ is topologically $\Sigma\times {\mathbb T}^2$, 
where $\Sigma$ is a $2d$ Riemann surface.

The results we have obtained so far can be summarized by the statement that any supersymmetric compact space 
$\mathcal{M}_3$ of $A$-type admits a toric foliation. We now pick {\it one} leaf $\mathcal{M}_2'$ of the toric 
foliation, and slice $\mathcal{M}_3$ along its volume. As a result, we obtain two manifolds $\mathcal{T}_1$ and 
$\mathcal{T}_2$, which share a common boundary, the leaf $\mathcal{M}_2'$, such that 
$\mathcal{T}_1\#\mathcal{T}_2\cong\mathcal{M}_3$. Generically, we will refer to $\mathcal{T}$ as a solid torus, 
borrowing the terminology from surgery theory. The solid torus is the analog of the hemisphere in $2d$, and 
the tip of the hemisphere corresponds here to the shrinking of one of the two boundary cycles. Following the 
analogy with the lower dimensional case, another interesting $3d$ manifold is represented by the ``cylinder", 
which topologically would be a torus fibered on the interval with both a left and a right boundary.
We note the obvious fact that when a boundary is inserted the homotopy properties of the manifold change.

Since the metric belongs to the supergravity multiplet, whose components include the $R$-symmetry gauge field 
$A_\mu^{(R)}$, the vector field $V_\mu$, and the scalar $H$, any manifold decomposition should be consistent 
with the profile of these background fields. Being a scalar field, $H$ is not constrained by the manifold decomposition. 
However, a condition on $V^\mu$ follows from the fact that $V^\mu$ is a conserved vector, and therefore we should 
require $n_\mu V^\mu \equiv V^\perp=0$ at the boundary. As a further simplifying, but not necessary, assumption in 
some of the examples that will be analysed below we will also consider $n^\mu A^{(R)}_\mu=0$.

\subsection{Clifford algebra and bilinears at the boundary}

The frame fields $k^\mu$, $\tilde{k}^\mu$ and $n^\mu$, split the algebra of the $\g$ matrices into a $2d$ 
``parallel" Clifford algebra, which lives on $\mathcal{M}'_2$, and an orthogonal matrix $\g^\perp=n_\mu\g^\mu$. 
As a consequence, all possible spinor bilinears obtained from $\zeta$ and $\zetaT$ are classified in terms of 
scalars and tensors on $T\mathcal{M}_2$. One obvious example is $K^\mu=\zetaT\g^\mu\zeta$, which is a vector 
on $T\mathcal{M}_2$, and has no scalar component because $n_\mu K^\mu=0$. 

It will be useful for later purposes to have the explicit decomposition for all spinor bilinears.
Since we have an expression for $n_\mu$ in terms of the Killing spinors, we can use Fierz identities 
to bring the bilinears in a simple form. It is enough to consider a generic bilinear with at most two $\g$ matrices; 
higher order bilinears would not be independent, because of the identity 
$\g^\mu\g^\nu=g^{\mu\nu} + i\, \varepsilon^{\mu\nu\rho}\g_\rho$. For notational convenience we use the indices 
$\nu_\parallel$ for directions parallel to $\mathcal{M}_2'$. 

The bilinears of interest are\footnote{For generic commuting spinors $\psi$ and $\chi$, we have 
$\psi\g^\mu\chi=\chi\g^\mu\psi$, thus $\psi\g^\mu\psi\neq 0$, and $\psi\chi=-\chi\psi$.}
\beq\label{Bilinears_1}
\begin{array}{lcl}
\zeta\g^\perp\zeta&=&+\frac{2\,\Omega}{||N||} (\zetaT\zeta^\star)^\star,\\
\zeta\g^\perp\g^{\nu_\parallel}\zeta&=&+\frac{ 2 }{||N||}(\zetaT\zeta^\star)^\star K^{\nu_\parallel},   \rule{0pt}{.6cm}\\
\zetaT\g^\perp\g^{\nu_\parallel}\zeta&=& -\frac{i}{||N||} \tilde{K}^{\nu_\parallel},  \rule{0pt}{.6cm}
\end{array}
\qquad
\begin{array}{lcl}
\zetaT\g^\perp\zetaT&=&-\frac{2\,\Omega}{||N||} (\zetaT\zeta^\star),\\
\zetaT\g^\perp\g^{\nu_\parallel}\zetaT&=&+\frac{ 2 }{||N||}(\zetaT\zeta^\star) K^{\nu_\parallel},    \rule{0pt}{.6cm}\\
\zeta\g^\perp\g^{\nu_\parallel}\zetaT&=& -\frac{i}{||N||} \tilde{K}^{\nu_\parallel}.  \rule{0pt}{.6cm}
\end{array}
\eeq
As a technical remark, we observe that the right column of \eqref{Bilinears_1} can be obtained by complex 
conjugation of the left column using $\zetaT=+i\g^2\zeta^\star$, and $\g^{\mu\star}=-\g^2\g^\mu\g^2$. 
The norm of the normal vector $N^\mu= (\zeta^\star\g^\mu\zeta ) +c.c.$ was given in \eqref{norms_spinors_frame_1}. 
However, by using the symmetries of the commuting spinors, we can also write 
$N^\mu N_\mu=N^\mu[(\zeta\g_\mu\zeta^\star ) + c.c.]$, and from the Fierz identity we obtain
\beq
\big[\zeta^\star\zetaT\big]\big[\zeta \zetaT^\star\big]=\frac{1}{4}||N||^2~.
\eeq
Therefore,
\beq
\lp\zeta\g^\perp\zeta\rp \lp \zetaT\g^\perp\zetaT\rp =- \frac{4\Omega^2}{||N||^2}\lp \zetaT\zeta^\star\rp^\star\lp \zetaT\zeta^\star\rp=-\Omega^2~.
\eeq
We conclude that the only new geometric information needed, in order to parametrize the bilinears \eqref{Bilinears_1}, 
is a phase 
\beq\label{Bilinears_2}
\begin{array}{lcl}
\zeta\g^\perp\zeta&\equiv&\Omega\, e^{i\x}~,\\
\zeta\g^\perp\g^{\nu_\parallel}\zeta&=&\Omega\, e^{i\x}\, k^{\nu_\parallel}~,   \rule{0pt}{.6cm} \\
\zetaT\g^\perp\g^{\nu_\parallel}\zeta&=& -i\Omega\, \tilde{k}^{\nu_\parallel}~,  \rule{0pt}{.6cm}
\end{array}
\qquad
\begin{array}{lcl}
\zetaT\g^\perp\zetaT&=&-\Omega\, e^{-i\x}~,\\
\zetaT\g^\perp\g^{\nu_\parallel}\zetaT&=&\Omega\, e^{-i\x}\, k^{\nu_\parallel}~,   \rule{0pt}{.6cm} \\
\zeta\g^\perp\g^{\nu_\parallel}\zetaT&=& -i\Omega\, \tilde{k}^{\nu_\parallel}~.  \rule{0pt}{.6cm}
\end{array}
\eeq
The phase $\x$ can be calculated explicitly, given the Killing spinor $\zeta$ and the norm of the Killing vector. 
In general, we expect $\x$ to be coordinate dependent: $\x=\x(\psi,z,\bar{z})$ if we are using the THF parametrization. 
We will present examples in section~\ref{SUSY_EXAMPLES}.

Finally, we can ask how the bilinear $\zeta\g_\mu\zeta$ decomposes in the basis $\{k_\mu, n_\mu, \tilde{k}_\mu\}$. 
The answer is again obtained by using Fierz identities and reads 
\beq
U_\mu\equiv\frac{\zeta\g_\mu\zeta}{\Omega}=  e^{i\x} (n_\mu - i \tilde{k}_\mu) ~.
\eeq

Consequently, we also find that the metric can be written equivalently as
\beq
ds^2= k_\mu k_\nu - U_\mu \widetilde{U}_\nu~,\qquad \widetilde{U}_\mu\equiv\frac{\zetaT\g_\mu\zetaT}{\Omega}
=-U^\star~.
\eeq

\subsection{Twisting and phases}\label{sec_twisting_phases}

So far we have discussed several of the characteristic properties of $A$-type backgrounds. 
The appearance of the phase $\x$ is one of the properties that will play an important role in the subsequent analysis
and as such it deserves some further elaboration. 
 
A constant shift of $\x$ can be understood as part of the $U(1)$ invariance that is built into the relations 
$\Omega=\zeta\zetaT$ and $\zetaT=\zeta^c$, as we discussed in sec.~\ref{The_geometry_M3}. 
The coordinate dependent part of $\x(\psi,z,\bar{z})$ is due to the non-trivial profile of the background fields
and is closely related to the explicit solution of the Killing spinor equations.
The choice of the frame fields, and therefore the definition of the curved $\g$ matrices becomes important when we
discuss the Killing spinor equation. We fix possible ambiguities in the choice of vielbein by working in the preferred
frame $\{ n_\mu, k_\mu, \tilde{k}_\mu\}$.  The relation between the coordinate dependent phase, $\x(\psi,z,\bar{z})$, 
and the Killing spinor, that we discuss here is made in this frame.\footnote{The reader familiar with the 
$\mathbb{S}^3$ geometry may notice that by using the Maurer-Cartan forms two out of four Killing spinors of the 
$\mathbb{S}^3$ are constant (see for example \cite{Marino:2011nm}). As we emphasize in the next section, 
$U^\mu$ is always well defined and so is $\x$. It can be explicitly checked that the phase $\x=\psi/2$ will show 
up in $U^\mu$, even in the Maurer-Cartan formulation. In the frame $\{n_\mu,k_\mu,\tilde{k}_\mu\}$ the phase 
will appear in the Killing spinors.}

We make the following observation. Given a metric $g_{\mu\nu}$ with corresponding background fields and a 
generic non-trivial $\x(\psi,z,\bar{z})$ we can consider a $U(1)_R$ gauge transformation that sets it everywhere 
to zero. As a result of this operation, the new background $R$-symmetry, in which the phase is constant, is
\beq
A^{(R)}_{new}=A_{old}^{(R)} + d\Lambda
\eeq
where $d\Lambda$ is a flat connection (in the simplest case a non-zero constant).  

Globally, the addition of a non-trivial flat connection can lead to interesting phenomena. Even though we expect 
the details of the manifold to become important at this point, we know for sure that the leaves of the supersymmetric 
foliations are tori, and therefore we can make the following general comments:
\begin{itemize}
\item
When $\pi_1(\mathcal{M}_3)$ is trivial,  
the two cycles of a generic leaf $\mathcal{M}_2'$ will shrink in the bulk, identifying the location of the north and south
pole. Then, if $A_{old}^{(R)}$ was topologically trivial, $A^{(R)}_{new}$ is inserting a singularity, effectively changing 
the topology. For example, it inserts punctures at the north/south pole. 
\item
When $\pi_1(\mathcal{M}_3)$ is non-trivial, e.g.\ $\pi_1(\mathcal{M}_3)=\mathbb{Z}$, the new flat connection will
generically decompose into a combination of an holonomy and a singularity (if both are non vanishing).
\item
When the manifold has a {\it toric contact structure}, the Killing vector $K=\partial_\psi$ is a combination of the 
vectors $\partial_{\phi_1}$ and  $\partial_{\phi_2}$, where $\phi_1$ and $\phi_2$ are $2\pi$-periodic coordinates 
on the leaves. The effect of a constant $A^{(R)}_\psi$ will result to the insertion of a vortex loop at the north and south pole, together with an holonomy along the corresponding non-shrinking cycles.  
\end{itemize}

From the point of view of the Killing spinor equations, the addition of a flat connection, from $A^{(R)}_{old}$ to 
$A_{new}^{(R)}$, is twisting the original solution.\footnote{Sometimes, even $A^{(R)}_{old}$ can be thought as a 
twisting of a theory with no $A^{(R)}$ \cite{Gomis:2012wy}. Here we are saying something slightly different, in 
particular we identify $A^{(R)}_{new}-A_{old}^{(R)}$ as a gauge transformation.} Indeed, assume that in the 
old background the spinor is of the type $e^{i\x}\eta_0$, with $\eta_0$ a constant spinor and $\x=\x(\psi,z,\bar{z})$.
Then, in the new background $\eta_0$ is the spinor and the Killing spinor equation becomes 
$\partial_\mu \eta_0=0$. 

Returning to the coordinate system \eqref{metric_real_param} we further observe that 
the $\psi$ dependence of the Killing spinor is always constrained to be a phase. This is due to the fact that 
$k=\partial_\psi$, and the fact $k^\mu\partial_\mu(\zetaT\zeta)=0$ that follows from the Killing spinor equations. 
The generic ansatz for a solution of the Killing spinor equations is then
$$\zeta=e^{i f(\psi)}\zeta_0 (\theta,\varphi)~,\qquad \zetaT=e^{-i f(\psi)} \zetaT_0(\theta,\varphi)~.$$  
According to this ansatz, for generic $f$ neither $\zeta$ nor $\zetaT$ are scalars under translations along the Killing 
vector, however, the $\psi$ dependence can always be solved by considering a gauge transformation of $A_{old}^{(R)}$ 
such that  $k^\mu\partial_\mu\zeta=k^\mu\partial_\mu\zetaT= 0$. By using the integrability condition 
\eqref{commutator_2} we can prove that
\beq\label{twisting_relation_back}
k^\mu A_{\mu\, new}^{(R)} = -iH - k^\mu V_\mu 
~.
\eeq
To prove this equation contract the Killing spinor equation of $\zeta$ with $k^\mu$ and $\zetaT$. 
The same result follows by considering the Killing spinor equation for $\zetaT$.
We come back to this relation in sections~\ref{Bound_Sec_1} and~\ref{SUSYbcsII}, where it will used as an input to 
solve for boundary conditions preserving a subset of the bulk supersymmetry. 

\vspace{0.3cm}
As a final remark, we would like to emphasize the following fact. {\it The field theories defined by the rigid limit of a  
supergravity theory depend explicitly on the background fields. As a result, for a given field theory in flat space, 
the corresponding rigid theories coupled to 
$\{ g_{\mu\nu}, A^{(R)}_{old}, V\} $ and to $\{ g_{\mu\nu},A^{(R)}_{new}, V\} $, are generally different theories with
different Lagrangians.} 
It would be interesting to understand via localization how the partition functions of the two theories are related. 
We leave this question to future work.

\subsection{Examples: spheres and their squashings}\label{SUSY_EXAMPLES}

Important examples of $A$-type backgrounds include: the round three-sphere $\mathbb{S}^3$, the ellipsoid 
$\mathbb{S}^3_b$, the $SU(2)\times U(1)$ squashed spheres of \cite{Hama:2011ea}, and geometries of the 
type $\mathbb{S}^2\times \mathbb{S}^1$. Round and squashed spheres were the first manifolds on which 
the use of supersymmetric localization made possible the exact computation of the partition function of 
$\mathcal{N}=2$ theories \cite{Kapustin:2009kz, Hama:2010av,Imamura:2011wg,Alday:2013lba}. 
Our main interest here will be to calculate the triple of vectors $\{n_\mu,k_\mu,\tilde{k}_\mu\}$ for the round sphere 
and its deformations. We will also mention the case of $\mathbb{S}^2\times \mathbb{S}^1$ which admits both an 
$A$-type and a different ``non-real" structure. In the context of squashed spheres, the distinction between these 
two structures has been also emphasized in \cite{Martelli:2013aqa}.

\subsubsection{Ellipsoid}

Our first example is $\mathbb{S}^3_b$, defined as the set of points 
$(z,w)\in\mathbb{C}^2$, with the property $\frac{|z|^2}{\tilde{\ell}^2}+\frac{|w|^2}{\ell^2}=1$. The squashing 
parameter $b$ is usually defined as the ratio $b^2={\tilde{\ell}/\ell}$. The parametrization 
$z=\tilde{\ell}\sin\theta\, e^{i\phi_1}$, $w=\ell\cos\theta\, e^{i\phi_2}$, gives the metric
\bea\label{squashedHHH}
ds^2_{\mathbb{S}^3_b}= dzd\bar{z}+ dwd\bar{w}=f(\theta)^2 d\theta^2 + \tilde{\ell}^2\sin^2\theta\, d\phi_1^2+\ell^2 \cos^2\theta\, d\phi_2^2
~,
\eea 
where $f(\theta)^2=\ell^2\sin^2\theta+\tilde{\ell}^2\cos\theta^2$. The coordinates take values in the range 
$\theta\in[0,\pi/2]$ and $\phi_{i}\in [0,2\pi]$ for $i=1,2$. They are {\it toric}, and make manifest the 
$U(1)\times U(1)$ symmetry of the geometry. The north pole at $\theta=0$, and south pole at $\theta=\pi/2$, 
are conventionally defined by the shrinking of the corresponding $\mathbb{S}^1$ cycles. The precise form of 
$f(\theta)$ is not important and all of the following calculations will be valid for a generic regular function 
$g_{\theta\theta}(\theta)$.\footnote{Regularity means any function that asymptotes to 
$\tilde{\ell}$, $\ell$ at $\theta=0$ and $\theta=\pi/2$, respectively.}
The background fields can be taken to be (in a gauge where $V_\mu=0$) as
\beq
\label{ellipsoidprofile}
H=\pm\frac{ i}{g_{\theta\theta}},\qquad 	
A^{(R)}_{\pm} = - \frac{1}{2} 
\left( 1 - \frac{\tilde{\ell}}{g_{\theta\theta}}\right) d\phi_1 \mp \frac{1}{2} \left( 1 - \frac{\ell}{g_{\theta\theta}}\right) d\phi_2
~.
\eeq   
Notice that $A^{(R)}_{\pm}$ is topologically trivial since $A^{(R)}_{\pm\,\phi_1}\rightarrow 0$ at the north pole and 
$A^{(R)}_{\pm\,\phi_2}\rightarrow 0$ at the south pole.\footnote{The background $A^{(R)}$ of \cite{Hama:2011ea} is 
recovered by the substitution $\phi_i\rightarrow-\phi_i$. The difference in the sign is due to 
our choice of $\g$ matrices that differs from the one in \cite{Hama:2011ea}.} 

There are solutions to the Killing spinor equations with both $+$ and $-$ signs. It is then convenient to distinguish 
between {\it positive} and {\it negative} Killing spinors, respectively.

Our immediate task is to obtain the Killing spinors, $\zeta_{\pm}$, $\zetaT_{\pm}$, and calculate the vector fields 
$K^\mu$, $N^\mu$ and $\tilde{K}^\mu$. Notice that $(\theta,\phi_1,\phi_2)$ are not the adapted coordinates 
introduced in the previous section, but since we have coordinate-independent expressions for $K^\mu$, $N^\mu$ 
and $\tilde{K}^\mu$, the choice of coordinates is not an issue. In the frame
\beq
E^1=\ell \cos\theta\, d\phi_2~,\qquad E^2=\tilde{\ell}\sin\theta\, d\phi_1~,\qquad E^3=g_{\theta\theta}\, d\theta~,
\eeq
the explicit expression of the Killing spinors is
\bea\label{U1U1squashing_sp0}
\zeta_{\pm}=\mathfrak{M}_{\left[\, {\pm\,\theta},\,\lp\phi_1\pm\phi_2\rp \right]}\,\eta~,  && \qquad 	
			\eta= \begin{array}{c} \frac{1}{\sqrt{2}}\end{array}  \left(\begin{array}{c} +1 \\ -1\end{array}\right)~,\\  
\zetaT_{\pm}=\mathfrak{M}_{\left[\, {\pm\,\theta},\,\lp\phi_1\pm\phi_2\rp \right]}\,\bar{\eta}~, & & \qquad 
	\bar{\eta}= \begin{array}{c} \frac{1}{\sqrt{2}}\end{array}  \left(\begin{array}{c} +1 \\ +1\end{array}\right)~,
	\rule{0pt}{.8cm}\label{U1U1squashing_sp}
\eea
with the matrix $\mathfrak{M}$ given by 
\beq
\mathfrak{M}_{\left[ \theta,\x\right]}= \exp\left(-i\frac{\theta}{2}\g^3\right) \exp\left(-i \frac{\x}{2} \g^1\right)=
\lp\begin{array}{ccc} e^{-i\theta} \cos\frac{\x}{2} && ie^{-i\theta} \sin\frac{\x}{2}\\   
ie^{+i\theta} \sin\frac{\x}{2} &&e^{+i\theta} \cos\frac{\x}{2}\end{array}\rp~. 
\eeq
In \eqref{U1U1squashing_sp} we have chosen a normalization such that $\zetaT_\pm=\zeta^c_\pm$. In fact, since 
the curved background is real, we are guaranteed that $\zeta^c$ solves the equation of $\zetaT$.  
The Killing vector associated to $\zeta_{\pm}$ is
\beq\label{Killing_spheres}
K^\mu_{\pm}\, \partial_\mu= \zetaT_{\pm}\g^\mu\zeta_{\pm}\,\partial_\mu = \pm \tilde{\ell}^{-1}\partial_{\phi_1}+ \ell^{-1} \partial_{\phi_2} ~,
\eeq
and the novel vectors, $n^\mu\equiv N^\mu/||N||$, and $\tilde{k}^{\mu}=\tilde{K}^{\mu}/||\tilde{K}||$ are
\bea
n_{\pm}^\mu\,\partial_\mu&=&-\frac{1}{g_{\theta\theta}} \partial_\theta~, \\
\tilde{k}_{\pm}^\mu\,\partial_\mu&=&- \tilde{\ell}^{-1}\cot\theta\,\partial_{\phi_1} \pm\,\ell^{-1}\tan\theta\, \partial_{\phi_2} \ \rule{0pt}{.6cm}. \label{KillingT_spheres}
\eea

It is interesting to write the metric in the adapted frame $\{k_\mu,n_\mu,\tilde{k}_\mu\}$. In the case of positive 
Killing spinors, the metric takes the form \eqref{metric_real_param}
\beq\label{metric_Hopf_0}
ds^2_{\mathbb{S}^3_b}=\frac{1}{4} \lp d\psi + \cos\theta_H d\varphi \rp^2 + \frac{1}{4}g_{\theta\theta}^2\, d\theta_H^2 
+\frac{1}{4} \sin^2\theta_H d\varphi^2
~, 
\eeq
where
\beq\label{metric_Hopf}
d\psi= \ell d\phi_2 +\tilde{\ell} d\phi_1~,\qquad d\varphi=\ell d\phi_2 -\tilde{\ell}d\phi_1~,\qquad d\theta_H= 2 d\theta
~.
\eeq

For the round three-sphere $\ell=\tilde{\ell}$ and we recover well known results.
The coordinates $\{\psi,\theta_H,\varphi\}$ coincide with the familiar Hopf coordinates, for which $\mathbb{S}^3$ 
is seen as a $U(1)$ fibration over the two-sphere  $d\theta_H^2+\sin^2\theta_H d\varphi^2$.\footnote{Notice that 
when $\ell\neq\tilde{\ell}$ the periodicities of $\varphi$ and $\psi$ are different from those of the round three-sphere. 
In the coordinates $\psi_H\equiv\phi_1+\phi_2$, $\varphi_H\equiv\phi_2-\phi_1$ the metric of $\mathbb{S}^2_b$ is 
\cite{Benini:2013yva} 
$$
ds^2= \frac{R^2}{4}  \Big[  \lp 1+\mathfrak{b} \cos\theta \rp\,d\theta_H^2 + \frac{1-\mathfrak{b}^2}{1- \mathfrak{b}\cos\theta_H}\, \sin^2\theta\, d\varphi_H^2 \Big] 
					+ \frac{R^2}{4} \lp 1- \mathfrak{b}\cos\theta_H \rp\, 
					\lp d\psi_H + \frac{  \cos\theta_H - \mathfrak{b}}{ 1- \mathfrak{b}\cos\theta_H} d\varphi_H \rp^2\ ,\nn
$$
where $2R^2=\ell^2+\tilde{\ell}^2$ and $\mathfrak{b}=(\tilde{\ell}^2-\ell^2)/(\tilde{\ell}^2+\ell^2)$.}  
The interpretation of the Killing spinors is manifest, $K_+=\partial_\psi$ and sits along the Hopf fiber, whereas 
$K_{-}=\partial_\varphi$ generates the $U(1)$  isometry of $\mathbb{S}^2$. 
Furthermore, in the example of the round three-sphere written in Hopf coordinates, we can write
the $\mathbb{S}^2$ at the base of the fibration as $\mathbb{CP}^1$, and exhibit the THF 
of $\mathbb{S}^3$
\beq\label{THF_three_sphere}
ds^2_{\mathbb{S}^3}= \lp d\tau + \frac{i\,\bar{z}dz}{2(1+|z|^2)}+c.c.\rp+\frac{dz d\bar{z}}{1+|z|^2},
\eeq
where $\tau=(\psi+\varphi)/2$ and $z=\tan(\theta/2)\, e^{i\varphi}$.  It is worth emphasizing that the Killing spinors 
$\zeta_{-}$ and $\zetaT_{-}$, which generates $K_{-}=\partial_\varphi$, become the standard spinor of the 
$\mathbb{S}^2$, after a change of frame. In the next section we will make this statement more precise. 

The Killing spinors $\zeta_{\pm}$ in \eqref{U1U1squashing_sp0} have non-trivial dependence on $\theta$ and $\an$. 
As can be seen from evaluating $\,\mathfrak{M}_{\left[ \theta,\x\right]}\eta\,$, the $\an$ dependence reduces to a phase. 
Following the discussion in the previous section, we can twist away the phase by performing a gauge transformation 
on the background $R$-symmetry connection. To see how this works in practice, let us observe that we can indeed 
decompose $A^{(R)}_{\pm}$ as 
\bea
A^{(R)}_{\pm} =
 +\frac{1}{2} (d\phi_1 \pm  d\phi_2)  + \frac{1}{2g_{\theta\theta}}( \tilde{\ell} d\phi_1\pm \ell d\phi_2)~.
\eea
The background $A^{(R)}_{\pm\, new}$ in which the spinors $\zeta_{\pm}$ are constant along the direction 
of the Killing vectors, can be obtained either by an explicit computation, or by solving the general relation 
$k^\mu A_{\mu\, new}^{(R)} = -iH - k^\mu V_\mu$ from the knowledge of $k^\mu$ and $A_{\mu\, old}^{(R)}$ above. 
The latter strategy implies that
\bea
 A^{(R)}_{\pm\, new} &=& A^{(R)}_{\pm}-\frac{1}{2} (d\phi_1 \pm  d\phi_2)
 = \frac{1}{2g_{\theta\theta}}( \tilde{\ell} d\phi_1\pm \ell d\phi_2)~.
\eea
As we expect $A^{(R)}_{\pm\, new}$ becomes well defined on the ellipsoid $\mathbb{S}^3_b$ with punctures 
at the north and south poles.

\subsubsection{On $U(1)$ fibrations and non A-type geometries} \label{circle_fibration_sec} 

Another class of interesting real curved spaces are the $SU(2)\times U(1)$ squashings of the round three-sphere 
of \cite{Hama:2011ea}. We will consider a slightly more general class of backgrounds, whose metric is given by
\beq\label{General_SU2xU1}
ds^2=\frac{\tilde{\ell}^{2}}{4} \lp d\psi + u(\theta) d\varphi \rp^2 
+ \frac{\ell^2}{4}\left( g_{\theta\theta}^2 d\theta^2 + \sin^2\theta d\varphi^2\right)
~. 
\eeq
When $u(\theta)=\cos\theta$, $g_{\theta\theta}=1$, and $\ell=\tilde{\ell}$ we recover the Hopf fibration of the 
$\mathbb{S}^3$. When $\ell\neq\tilde{\ell}$, the $U(1)$ fiber of the round sphere gets squashed, and the metric 
only preserves the $SU(2)\times U(1)$ subgroup of the original $SO(4)$ isometry group. 
We may also take $u(\theta)=u_0$ constant, and for the particular value $u_0=0$ we recover the metric 
of $\mathbb{S}^2\times \mathbb{S}^1$. 

It will be useful to define the parameter $\beta=\tilde{\ell}/\ell$. It measures the squashing for geometries that are 
deformations of $\mathbb{S}^3$, hereafter $\mathbb{S}^3_\beta$. Also, it measures the inverse temperature for 
geometries of the type $\mathbb{S}^2\times \mathbb{S}^1$.
By a global rescaling we can set $\ell=1$. We will work with the dreibeins
\beq\label{Frame_SU2xU1}
E^1= \frac{1}{2} g_{\theta\theta} d\theta~,\qquad E^2=  \frac{1}{2} \sin\theta d\varphi~,\qquad 
E^3=\frac{\beta}{2} (d\psi +u(\theta) d\varphi)~.
\eeq
The background scalar field $H$ is taken to be purely imaginary, and we turn on
\bea
& V_3&= -i H + \frac{\beta}{g_{\theta\theta}} \frac{u'(\theta)}{\sin\theta}~,\label{generalVSU2xU1}\\
& A_\varphi^{(R)}&=-\frac{1}{2}\frac{\cos\theta}{g_{\theta\theta}} 
-\frac{\beta^2}{2} \frac{u(\theta)}{g_{\theta\theta}}  \frac{u'(\theta)}{\sin\theta}~,\qquad
A_\psi^{(R)}= -\frac{1}{2}-\frac{\beta^2}{2 g_{\theta\theta}} \frac{u'(\theta)}{\sin\theta}~.\label{generalASU2xU1}
\eea

In this setup, the metrics \eqref{General_SU2xU1} admit two Killing spinors of opposite charge
\beq\label{positive_Killing_SU2xU1}
\zeta=e^{-i\psi/2}\left(\begin{array}{c}1\\ 0\end{array}\right), \qquad 
\zeta^c=\zetaT=e^{+i\psi/2}\left(\begin{array}{c}0\\ 1\end{array}\right) 
~.
\eeq
From these Killing spinors we calculate the frame $\{n_\mu,\tilde{k}_\mu,k_\mu\}$, and find
\bea
n^\mu\partial_\mu=\frac{2}{ g_{\theta\theta}} \partial_\theta~,
\qquad k^\mu\partial_\mu=\frac{2}{\beta} \partial_\psi~,\qquad
\tilde{k}^\mu\partial_\mu=\frac{2}{\sin\theta}  \lp u(\theta) \partial_\psi - \partial_\varphi\rp~.
\eea 
We also recognize that the dreibeins $\{E^1, E^2, E^3\}$ correspond to the triple $\{n_\mu,-\tilde{k}_\mu,k_\mu\}$.
The phases $\pm i\psi$ of the spinors $\zeta$ and $\zetaT$ in \eqref{positive_Killing_SU2xU1} can be re-absorbed 
by twisting $A^{(R)}$.  The corresponding gauge transformation leaves $A_\varphi^{(R)}$ invariant and changes 
$A_\psi^{(R)}$ as follows
\beq
A_{\psi}^{(R)}\quad \rightarrow\quad A_{\psi\,new}^{(R)}= -\frac{\beta^2}{2 g_{\theta\theta}} \frac{u'(\theta)}{\sin\theta}
~.
\eeq

Observe that for $\mathbb{S}^2\times\mathbb{S}^1$ geometries, the function $u(\theta)$ is trivial, and therefore
\beq
A_{\psi\, new}^{(R)}=0~,\qquad A_{\varphi}^{(R)}=-\frac{1}{2}\frac{\cos\theta}{g_{\theta\theta}}~.
\eeq
By taking $H=0$ this $\mathbb{S}^2\times\mathbb{S}^1$ background becomes the topologically twisted background 
of \cite{Benini:2015noa}.  

The reasoning that led to the background fields \eqref{generalVSU2xU1} and \eqref{generalASU2xU1} is based 
on simple observations, which we now elucidate. First of all, $A^{(R)}$ and $V$ are real when $H$ is imaginary, 
hence the family of backgrounds is of the $A$-type. For example, considering the round three-sphere, $\beta=1$, 
$u=\cos\theta$, we find $A_\mu^{(R)}=0$, and $V_3=-iH -1$, thus
in the gauge $H=+i$, the spinors $\zeta$ and $\zetaT$ correspond to the positive Killing spinors 
\eqref{U1U1squashing_sp0} and \eqref{U1U1squashing_sp} calculated in the new frame \eqref{Frame_SU2xU1}. 
The more general background fields, \eqref{generalVSU2xU1} and \eqref{generalASU2xU1}, are obtained by solving 
the Killing spinor equation for $A^{(R)}$ and $V$, upon insisting that $\zeta$ in \eqref{positive_Killing_SU2xU1} is a 
solution. By writing the Killing spinor equation in the following form
\beq\label{eq_SU2xU1_special}
( \partial_\mu - i A^{\lp R\rp}_\mu \big) \zeta=-\frac{1}{4}\omega_\mu^{ab}\g_{ab}\zeta 
-\frac{1}{2} \g_{\mu}( H  -i  V_\nu\g^\nu) \zeta
~,
\eeq
we get $V_3$ from the $\theta$ component, $A_\varphi^{(R)}$, and $A_\psi^{(R)}$ from the other two equations. 

Some of the details of this calculation can be seen explicitly in the cases of $\mathbb{S}^3_\beta$ 
and $\mathbb{S}^2\times \mathbb{S}^1$ geometries. The equations \eqref{eq_SU2xU1_special} become   
\bea
\partial_{\theta}\zeta - i A_\theta \zeta &=&+\frac{i}{4}\ (\mathfrak{p}\, \beta+iH - V_3\g_3)\,\g^1 \zeta~,\label{eq_SU2xU1_special_1}\\
\partial_{\psi}\zeta-i A_\psi \zeta &=&-\frac{i}{4}\beta(\mathfrak{p}\, \beta-iH - V_3\g_3)\, \g^3\zeta~,\rule{0pt}{.6cm}\label{eq_SU2xU1_special_2}\\
\partial_\varphi\zeta - i A_\varphi \zeta&=&+\frac{i}{4}\Big[ ((2-\mathfrak{p}\,\beta^2)+iH + V_3\g_3)\g^3\cos\theta\rule{0pt}{.6cm}\nn\\
											 & &\rule{1.2cm}{0pt} + (\mathfrak{p}\,\beta +iH + V_3\g_3)\g^2\sin\theta\ \Big]\zeta~, \rule{0pt}{.5cm} 
											 \label{eq_SU2xU1_special_3}
\eea
where $\mathfrak{p}=1$ for $\mathbb{S}^3_\beta$, and $\mathfrak{p}=0$ for $\mathbb{S}^2\times \mathbb{S}^1$.  
For the round three sphere $\beta=1$, $H=+i$, $V_3=0$ and the r.h.s of \eqref{eq_SU2xU1_special_1} and 
\eqref{eq_SU2xU1_special_3} vanish identically.
The use of the frame fields \eqref{Frame_SU2xU1}, compared to the toric frame of the previous section, makes the 
computation of the positive Killing spinors particularly simple: two out of three
equations can be trivially satisfied, and the remaining one, $\partial_{\psi}\zeta=-\frac{i}{2}\g_3 \zeta$, is solved by 
\eqref{positive_Killing_SU2xU1}. For the $SU(2)\times U(1)$ squashing $\mathbb{S}^3_{\beta}$, 
the background field $V_3$ is tuned in such a way that the r.h.s of \eqref{eq_SU2xU1_special} 
becomes a projector, as one can check from \eqref{eq_SU2xU1_special_1}. Then, the positive Killing spinor of the 
round sphere is promoted to a Killing spinor of the squashed sphere.\footnote{Squashings whose Killing spinors 
reduce to the negative Killing spinors of the round sphere, have been studied in \cite{Imamura:2011wg}. In this case, 
the ansatz for Killing spinors need to be slightly modified.} In this case, the $R$-symmetry background 
is proportional to $A^{(R)}_3$, and it is aligned with $V_3$. 

The analysis of $\mathbb{S}^2\times \mathbb{S}^1$ geometries follows the same logic. 

Before we move on, let us notice that when we consider the negative Killing spinors of the round three-sphere, a different simplification takes place in the equations \eqref{eq_SU2xU1_special_1}-\eqref{eq_SU2xU1_special_3}: the trivial equation becomes $\partial_\psi\zeta=0$, whereas 
the equations along $\theta$, and $\varphi$ become effectively those of the $\mathbb{S}^2$ in its standard 
parametrization \cite{Benini:2012ui}, 
\beq\label{NegativeS2}
\nabla_\theta\zeta=+\frac{i}{2}\g_\theta\zeta~,\qquad \nabla_\varphi\zeta=+\frac{i}{2}\g_\varphi\zeta~.
\eeq
whose solutions are\footnote{These $\mathbb{S}^2$ Killing spinors can be uplifted to $\mathbb{S}^3$, as 
explained in \cite{Closset:2014pda}. In two dimensions, $\g_3$ anti-commutes with $\g^{(2d)}_\mu$. Therefore, 
the positive Killing spinors of the $\mathbb{S}^2$ are proportional to $\g_3\zeta$, with $\zeta$ given 
in \eqref{NegativeS2}.}
\beq
\zeta= C_1\, e^{+ i\frac{\varphi}{2}} \left( \begin{array}{c} \cos\frac{\theta}{2} \\ -i \sin\frac{\theta}{2} \end{array} \right)
		+ C_2\, e^{-i \frac{\varphi}{2}}  \left( \begin{array}{c} \sin\frac{\theta}{2} \\ +i \cos\frac{\theta}{2} \end{array}
		\right)
~.
\eeq

\subsubsection*{$\mathbb{S}^2\times \mathbb{S}^1$ and non A-type geometry}

Metrics of the type $\mathbb{S}^2\times \mathbb{S}^1$ are interesting for a second reason: 
they are perhaps the simplest $3d$ example admitting both a real and a non-real 
structure. The non-real structure is obtained by considering the following background fields
\beq\label{non_real_backg}
H=0~,\qquad V=- \frac{2i}{g_{\theta\theta}}  E^3~, \qquad A^{(R)}= +\frac{i}{g_{\theta\theta}} E^3~,
\eeq
with $E^3=\frac{\beta}{2}(d\psi + u_0 d\varphi)$. The Killing spinor equation \eqref{eq_SU2xU1_special_2} becomes 
trivial: $\partial_\psi\zeta=0$. After $A^{(R)}$ and $V$ have been subtracted, the equations on the $\mathbb{S}^2$ 
base, \eqref{eq_SU2xU1_special_1},  and \eqref{eq_SU2xU1_special_3},  become 
\bea\label{non_real_eq_1}
\nabla_\mu \zeta =+\frac{1}{2g_{\theta\theta}}\g_\mu\g^3\zeta
~.
\eea
The Killing spinor equations for $\zetaT$ are not obtained from \eqref{non_real_eq_1} by charge conjugation. 
Indeed, the background is not real. Instead, from the original Killing spinor equation \eqref{eq_zetaT} we find,  
\beq\label{non_real_eq_2}
\partial_\psi\zetaT=0~\qquad \nabla_\theta\zetaT=-\frac{1}{2g_{\theta\theta}}\g_\theta\g^3\zetaT~,\qquad 
\nabla_\varphi\zetaT=-\frac{1}{2g_{\theta\theta}}\g_\varphi\g^3\zetaT
~.
\eeq
The equations \eqref{non_real_eq_2} appear in the same form in \cite{Doroud:2012xw,Yoshida:2014ssa} 
for $g_{\theta\theta}=1$. The explicit solutions are proportional to the following four spinors
\bea
\label{non_real_sp_example_1}
\zeta_1=e^{-\frac{i}{2}\varphi}\left(\begin{array}{c} \sin\frac{\theta}{2} \\ +\cos\frac{\theta}{2} \end{array}\right)~,&\qquad&
\zeta_2=e^{+\frac{i}{2}\varphi}\left(\begin{array}{c} -\cos\frac{\theta}{2} \\ \sin\frac{\theta}{2} \end{array}\right)~, \\
\zetaT_1=e^{-\frac{i}{2}\varphi}\left(\begin{array}{c} \sin\frac{\theta}{2} \\ -\cos\frac{\theta}{2} \end{array}\right)~,&\qquad&
\zetaT_2=e^{+\frac{i}{2}\varphi}\left(\begin{array}{c} +\cos\frac{\theta}{2} \\ \sin\frac{\theta}{2} \end{array}\right) 
~.
\eea

If the background fields are all purely imaginary (as in \eqref{non_real_backg}), 
it follows from \eqref{eq_zetachargeC} that the charge conjugate spinor $\zeta^c$ is independent of $\zeta$
and solves the same equation. For example, $\zeta_2=\zeta_1^c$ 
in \eqref{non_real_sp_example_1}. The same statement applies to $\zetaT$ and $\zetaT^c$. We conclude that 
if a background admits two Killing spinors of opposite $R$-charge, and all the background fields are purely 
imaginary, by construction it supports $\mathcal{N}=4$ supersymmetry.

\section{Boundary effects in theories with rigid supersymmetry} \label{sec_SUSY_CDFK}

Given a supersymmetric field theory on a compact manifold $\mathcal{M}_3$, defined by an action 
\beq
\mathcal{S}=\int_{\mathcal{M}_3}\mathscr{L}
~,
\eeq 
it is not guaranteed that the action will remain supersymmetric when we insert a boundary
along $\mathcal{M}_2'$, and restrict the fields to the solid torus $\mathcal{T}$. 
In fact, for any symmetry $\delta$ acting on the fields, the Lagrangian is locally invariant up to a total derivative,
$\d \mathscr{L}=\nabla_\mu\TD^\mu$, hence the action restricted on $\mathcal{T}$ will be invariant under the 
symmetry $\d$ iff
\beq\label{variation_boundary}
\d\mathcal{S}=\int_{\mathcal{T}} \nabla_\mu \TD^\mu = \oint_{\mathcal{M}_2'} n_\mu \TD^\mu =0~,
\eeq
where in the last step we used the divergence theorem. Typically, the condition \eqref{variation_boundary} is 
solved by imposing appropriate boundary conditions such that $n_\mu \TD^\mu=0$, or by adding appropriate
degrees of freedom on the boundary. In this paper we consider only the first possibility. In the case of supersymmetry 
$\TD^\mu$ is both a function of the anticommuting Killing spinors, $\ep$ and $\epT$, and the fields of the theory. 
Therefore, in order to solve \eqref{variation_boundary}, one generally synchronizes the boundary conditions on 
the fields with certain conditions on the spinors. For example, if we assume that a certain projection on the spinors 
realizes a specific sub-algebra of the bulk supersymmetry, we can insert this knowledge into $n_\mu \TD^\mu$ to 
simplify the problem and deduce definite boundary conditions for the fields of the theory.

For example, in the case of boundary conditions in two-dimensional $\NN=(2,2)$ 
theories on a strip \cite{Ooguri:1996ck, Hori:2000ck}, one can consider two different types of $\frac{1}{2}$-BPS 
boundary conditions, called $A$- and $B$-type. They are characterized by the spinor projections 
\begin{itemize}
\item[$\bullet$]  $\ \epb_+=+e^{i\a} \ep_-\ $ for A-type~,   
\item[$\bullet$]  $\ \ep_+=-e^{i\a} \ep_-\ $ for B-type~.  
\end{itemize}
$\ep_{\pm}$ and $\epb_{\pm}$ are the complex components of the $2d$ Weyl spinors $\ep$ and $\epb$, 
and $\epb$ is the complex conjugate of $\ep$. The phase $\a$ is an arbitrary constant and the minus sign 
is a convention. An $\NN=(2,2)$ theory has $4$ real supercharges and the $1/2$-BPS projections preserve $(1,1)$ or 
$(2,0)$ supersymmetry, for A-type or B-type, respectively. Such conditions play an important role in D-brane physics
described by setups with $\NN=(2,2)$ worldsheet supersymmetry.
In $3d$ theories with $\NN=2$ supersymmetry similar projections (and corresponding boundary conditions) 
have been formulated in flat space in \cite{Okazaki:2013kaa}.

When one attempts to apply this standard logic to a theory on a curved background, as in this paper, 
one encounters inevitably some obvious difficulties. Most notably, on curved backgrounds many of the simplifications of 
constant flat space spinors are absent. The Killing spinors $\ep$, $\epT$ 
are, in general, non-trivial functions of the coordinates and an $A$- or a $B$-type projection cannot be imposed in 
the simple standard flat space form written above.

In what follows we will describe how to impose a direct generalization of the $A$-type condition on the anticommuting 
spinors $\ep$ and $\epT$ in a generic three-dimensional $A$-type background. We will do so by introducing a 
``canonical" formalism that builds on the observations of the previous two sections.
We anticipate that a similar generic formulation exists also for $B$-type projections. However, in this paper we will 
focus exclusively on $A$-type boundary conditions leaving $B$-type projections and $B$-type boundary conditions 
to a separate treatment in future work.

\subsection{Generalized $A$-type projections on supersymmetry}\label{Canonical_Formalism}

Out of the commuting spinors $\zeta$ and $\zetaT$ we construct two natural projectors, $\Pr$ and $\Prb$
\beq\label{Projector_Spinors}
\Pr\psi= \frac{1}{\Omega} (\zetaT\psi)\zeta,\qquad \Prb\psi=\frac{1}{\Omega} (\psi\zeta)\zetaT,\qquad \forall\,\psi.
\eeq
It is simple to check that $\Pr^2=\Pr$, $\Prb^2=\Prb$ and $\Pr+\Prb=\mathbb{I}$. Since the Killing spinors 
$\zeta, \zetaT$ are nowhere vanishing these projectors are everywhere well defined. Moreover, both 
$\Pr$ and $\Prb$ are invariant under the symmetry $\zeta\rightarrow \lambda\zeta$, 
$\zetaT\rightarrow \lambda^{-1}\zetaT$, with $\lambda\in\mathbb{C}$. 

By acting with $\Pr$ and $\Prb$ on both $\ep$ and $\epT$ we formulate the generalized $A$-type conditions
\beq
\label{projectaa}
\Prb \ep = 0~, ~~ \Pr \epT =0
~,
\eeq
\beq
\label{projectab}
(\Pr \ep \, \zetaT) = (\zeta \, \Prb \epT)~,\qquad \zetaT=\zeta^c
~.
\eeq
Defining the parameters
\beq
 \vartheta\equiv\frac{1}{\Omega}(\zetaT\ep)~, \qquad \widetilde{\vartheta}\equiv\frac{1}{\Omega}\lp\epT\,\zeta\rp~ .
\eeq
the above relations become
\beq
\label{projectac}
\ep = \vartheta \zeta~,~~
\epT = \widetilde{\vartheta} \zetaT
~,
\eeq
\beq\label{general_Atype}
\vartheta = \widetilde{\vartheta}
~.
\eeq

The restriction $\zetaT=\zeta^c$ (which is possible in $A$-type backgrounds) is imposed here because 
the scalar product $(\epT\,\zeta)=(\zetaT\ep)$ alone does 
not enforce a relation between $\ep$ and $\epT$. Indeed, by rescaling $\epT\rightarrow\a\epT$ and 
$\ep\rightarrow\b\ep$, with arbitrary $\a,\b\in\mathbb{C}$, it is always possible to find  
two representatives of the commuting spinors, $\lm\zeta$ and $\lm^{-1}\zetaT$, for which the relation $(\epT\,\zeta)=(\zetaT\ep)$ is satisfied. The condition $\zetaT=\zeta^c$ is needed to break the invariance under the 
rescalings by $\lm\in\mathbb{C}$ to a residual $U(1)$. 

As a simple check that exhibits why this is the natural curved space generalization of the $A$-type projection 
we notice that for constant spinors in flat space the relation \eqref{general_Atype} reduces to the familiar 
$A$-type condition $\epT_+=+e^{i\a} \ep_-$. Indeed, in flat space, we may set $\zeta=(1,0)$, $\zetaT=(0,1)$, 
$\Omega=1$, and then the relation $\lp\epT\,\zeta\rp=(\epT\,\zeta)$ becomes the expected $\epT_+=\ep_-$. 
The residual $U(1)$ transformation gives the most general boundary condition, which is precisely 
$\epT_+=e^{i\a} \ep_-$.
 
We emphasize that the curved space version of the above $A$-type condition is, by construction, 
compatible only with $A$-type curved manifolds, for which $\zetaT=\zeta^c$. The projections \eqref{projectac}, 
\eqref{general_Atype} reduce the amount of supersymmetry by one half.

\vspace{0.3cm}

In sections \ref{preview}-\ref{SUSYbcsII}, we will demostrate how the input of the projections
\eqref{projectac}, \eqref{general_Atype} affects the (in)variance of a generic $\mathcal{N}=2$ field theory
under supersymmetry, and we will study corresponding general $A$-type boundary conditions on 
$\NN=2$ supersymmetric gauge theories that preserve half of the bulk supersymmetry at the boundary.

\subsection{Bulk ${A}$-type supersymmetries and BPS equations} 

Having understood how to project the anticommuting Killing spinors of a generic $A$-type background, we now 
go back to the supersymmetry transformations of chiral and vector superfields, and reformulate them accordingly. 
First we spell out the supersymmetry transformations with generic $\vartheta$ and $\widetilde{\vartheta}$, and then 
we study what happens upon enforcing the projection $\vartheta=\widetilde{\vartheta}$.

Before entering the details we point out that we can decompose any spinor $\psi$ as
\beq
\psi= \frac{1}{\Omega} (\zetaT\psi)\zeta + \frac{1}{\Omega} (\psi\zeta)\zetaT~.
\eeq
Moreover, we notice the useful identities
\beq
\begin{array}{l}
(\g^\mu\zeta)_\a = \frac{\zetaT\g^\mu\zeta}{\Omega}\, \zeta_\a - \frac{\zeta\g^\mu\zeta}{\Omega}\,\zetaT_\a= k^\mu\,\zeta_\a - U^\mu\,\zetaT_\a~, \\
(\g^\mu\zetaT)_\a=\frac{\zetaT\g^\mu\zetaT}{\Omega}\, \zeta_\a - \frac{\zetaT\g^\mu\zeta}{\Omega}\, \zetaT_\a= \tilde{U}^\mu\, \zeta_\a - k^\mu \,\zetaT_\a~. \rule{0pt}{.7cm}
\end{array}
\eeq

\vspace{0.3cm}

\subsubsection*{Chiral and anti-chiral multiplets}

In \eqref{susyrules_chiral} and \eqref{susyrules_antichiral} we wrote down the supersymmetric transformation rules for
chiral and anti-chiral multiplets for generic Killing spinors. When we further specialize the supersymmetry to an 
$A$-type background we obtain the following expressions.
\begin{itemize}
\item For a chiral multiplet: 
\end{itemize}
\beq
\begin{array}{cccl}
\frac{1}{\sqrt{2}}&\d\phi&=&  +\,\vartheta\, \zeta\psi~,\\
\frac{1}{\sqrt{2}}&\d\psi_\a&=&+\,\vartheta~F\, \zeta_\a   + i \, \widetilde{\vartheta} ~\Big[\Big( k^\mu\D_\mu\phi - ir (iH)\phi - (z- q\s)\phi\Big)\zetaT_\a - (\widetilde{U}^\mu\D_\mu\phi)\zeta_\a \Big]~,\rule{0pt}{.5cm}\\
\frac{1}{\sqrt{2}}&\d F&=& +i \, \widetilde{\vartheta}~\Big[\, \Big( k^\mu (\D_\mu -\frac{i}{2} V_\mu)\psi -i (r-\frac{1}{2})(iH)\psi - (z-q\s)\psi \Big)\zetaT \rule{0pt}{.5cm}\\
&& & 						\rule{6cm}{0pt}  + \widetilde{U}^\mu\, \zeta(\D_\mu-\frac{i}{2}V_\mu)\psi + \sqrt{2} q\, \zetaT\lmb\,\phi \Big]\rule{0pt}{.5cm}
~.
\end{array}
\eeq
\begin{itemize}
\item For an anti-chiral multiplet: 
\end{itemize}
\beq
\begin{array}{cccl}
\frac{1}{\sqrt{2}}&\d\phiT&=& -\, \widetilde{\vartheta}\, \zetaT\psi~,\\
\frac{1}{\sqrt{2}}&\d\psiT_\a&=& +\,\widetilde{\vartheta}~\FT\, \zetaT_\a   + i \, {\vartheta} ~\Big[\Big( k^\mu\D_\mu\phiT + ir (iH)\phiT + (z- q\s)\phiT\Big)\zeta_\a - ({U}^\mu\D_\mu\phiT)\zetaT_\a \Big]~,\rule{0pt}{.5cm}\\
\frac{1}{\sqrt{2}}&\d \FT&=& -i\,\vartheta~\Big[\, \Big( k^\mu (\D_\mu +\frac{i}{2} V_\mu)\psiT +i (r-\frac{1}{2})(iH)\psiT + (z-q\s)\psiT \Big)\zeta + \rule{0pt}{.5cm}\\
					&& & \rule{6cm}{0pt} + {U}^\mu\, \zetaT(\D_\mu+\frac{i}{2}V_\mu)\psiT - \sqrt{2} q\, \zeta\lm\, \phiT\Big]  \rule{0pt}{.5cm}~.
\end{array}
\eeq

It is clear, in particular, that the fixed point (BPS) equations, in which the fermions are set to zero and the bosons 
satisfy $\d f=0$ for any fermion $f$ of the multiplet, depend on the assumption we make about $\vartheta$ and 
$\widetilde{\vartheta}$. For the $A$-type projection, $\vartheta=\widetilde{\vartheta}$, we obtain
\bea
k^\mu\D_\mu\phi - ir (iH)\phi - (z- q\s)\phi =0~, & \rule{1cm}{0pt} & i\widetilde{U}^\mu\D_\mu\phi- F =0 ~,\label{first_line_BPS_chiral} \\
k^\mu\D_\mu\phiT + ir (iH)\phiT + (z- q\s)\phiT =0~, & & i{U}^\mu\D_\mu\phiT - \FT =0 
~.
\eea
Further assuming the reality conditions 
$\phiT=\phi^\star$ and $\FT=F^\star$, these equations reduce to
\beq\label{BPS_chiral_equation}
k^\mu\D_\mu\phi - ir (iH)\phi=0\qquad \& \qquad (z- q\s)\phi =0 \qquad \& \qquad  i{U}^\mu\D_\mu\phi- F =0~.
\eeq
We obtained the last equation using the property $\widetilde{U}=-U^\star$. In the case of arbitrary $\vartheta$ and 
$\widetilde{\vartheta}$ we would have instead $F=\FT=0$ and ${U}^\mu\D_\mu\phi=0$ independently. In the presence 
of a superpotential, we should integrate out $F^a$ in favor of $g^{a\bar{c}}\partial_{\bar{c}}\widetilde{W}$. Recalling that 
$U^\mu=e^{i\x}(n^\mu-i\tilde{k}^\mu)$, we see that the equation $i{U}^\mu\D_\mu\phi^a- F^a=0$ becomes the natural 
$3d$ generalization  of the domain wall equations in two dimensional $(2,2)$ theories.

\subsubsection*{Real and gauge multiplets}

The supersymmetric transformation rules for the gauge field were discussed in subsection \eqref{realmultiplets}. 
There we made a connection between the real multiplet and the gauge multiplet:
\beq\label{rec_def_Smult}
j_\mu=-\frac{i}{2}\varepsilon_{\mu\nu\rho}\F^{\nu\rho},~\qquad
a_\mu=-j_\mu-\s V_\mu~,\qquad \psi_\Sigma=i \lmb~,\qquad \psiT_\Sigma=-i\lm~.
\eeq
Here we use the real multiplet parametrization for the fermions, and write the field strength $\mathcal{F}$ in 
terms of the vector $a_\mu$. The supersymmetry transformations on an $A$-type background then takes the 
following form.
\begin{itemize}
\item For the $\ep$ variation of the bosons
\end{itemize}
\beq
\begin{array}{ccl}
\d_\ep\s&=&+i \vartheta\, (\zeta\psi_\Sigma)~,\\
\d_\ep \A_\mu&=& - \vartheta\, \big[ +e^{i\x} n_\mu~(\zetaT\psi_\Sigma) -i e^{i\x} \tilde{k}_\mu (\zetaT\psi_\Sigma) - k_\mu~(\zeta\psi) \big]~, \rule{0pt}{.7cm}\\
\d_\ep D&=& + i \vartheta~\Big[ \zeta\big( k^\mu(\D_\mu-\frac{i}{2} V_\mu) -\frac{1}{2}H \big)\psi_\Sigma - \zetaT\, U^\mu (\D_\mu-\frac{i}{2} V_\mu)\psi_\Sigma \Big]
\rule{0pt}{.7cm}~.
\end{array}
\eeq
\begin{itemize}
\item For the $\epT$ variation of the bosons
\end{itemize}
\beq
\begin{array}{ccl}
\d_{\epT}\s&=&+i\widetilde{\vartheta}~(\zetaT\psiT_\Sigma)~,\\
\d_{\epT} \A_\mu&=& - \widetilde{\vartheta} \big[ -e^{-i\x} n_\mu ~(\zeta\psiT_\Sigma) - ie^{-i\x} \tilde{k}_\mu ~(\zeta\psiT_\Sigma) - k_\mu~(\zetaT\psiT_\Sigma) \big]~, \rule{0pt}{.7cm}\\
\d_{\epT} D&=&- i \widetilde{\vartheta}~\Big[ \zetaT \big( k^\mu (\D_\mu + \frac{i}{2} V_\mu) +\frac{1}{2} H \big)\psiT_\Sigma- \zeta\, \widetilde{U}^\mu (\D_\mu+\frac{i}{2} V_\mu)\psiT_\Sigma \Big]
\rule{0pt}{.7cm}~.
\end{array}
\eeq
\begin{itemize}
\item For the fermionic fields
\end{itemize}
\beq
\begin{array}{ccl}
\d\psi_{\Sigma}&=& \widetilde{\vartheta}\,\Big[ \big[ D-i \s (iH + k^\mu V_\mu)- i k^\mu (j_\mu +i \partial_\mu\s)\big]\zetaT- i \widetilde{U}^\mu (a_\mu -i \partial_\mu\s) \zeta  \Big]~, \\
\d\psiT_{\Sigma}&=& \vartheta\, \Big[ \big[ D-i \s (iH + k^\mu V_\mu)- i k^\mu (j_\mu -i \partial_\mu\s)\big]\zetaT- i {U}^\mu (a_\mu +i \partial_\mu\s) \zeta  \Big]\rule{0pt}{.7cm}~.
\end{array}
\eeq
The fixed point equations $\psi_{\Sigma}=\psiT_{\Sigma}=0$ and $\d\psi_{\Sigma}=\d\psiT_{\Sigma}=0$ are 
\bea
D-i \s (iH + k^\mu V_\mu)- i k^\mu (j_\mu +i \partial_\mu\s)= (n^\mu+i \tilde{k}^\mu) (a_\mu -i \partial_\mu\s)=0 ~,\label{fixptS_1}\\
D-i \s (iH + k^\mu V_\mu)- i k^\mu (j_\mu -i \partial_\mu\s)=(n^\mu-i \tilde{k}^\mu) (a_\mu +i \partial_\mu\s) =0 ~.\label{fixptS_2}
\eea
Let us notice that, from its definition \eqref{rec_def_Smult}, $j_\mu$ (and thus $a_\mu$) is imaginary if the 
gauge field $\A$ is real, and real if $\A$ is imaginary. 
The solutions to \eqref{fixptS_1} and \eqref{fixptS_2} include the `Coulomb branch' solution
\beq
D-i \s (iH + k^\mu V_\mu)=0~,\qquad \partial_\mu\s=a_\mu=0~,
\eeq
and the solution
\bea
n^\mu a_\mu + \tilde{k}^\mu\partial_\mu \s =0~, &\qquad & k^\mu\partial_\mu\s=0~, \label{fixptS_3}\\
\tilde{k}^\mu a_\mu - n^\mu \partial_\mu\s=0~,&\qquad& D-i \s (iH + k^\mu V_\mu)=i k^\mu j_\mu~. \label{fixptS_4} 
\eea 
The equations \eqref{fixptS_3} and \eqref{fixptS_4} generalize to arbitrary $A$-type backgrounds those of 
\cite{Benini:2013yva}\cite{Benini:2015noa}.

\section{$\NN=2$ Lagrangians} \label{sec_Lagrangians_N=2}

With all the geometric prerequisites in place we need one more element before we can start discussing concretely 
how to treat $\NN=2$ supersymmetric field theories on $A$-type curved backgrounds with boundaries. We need
to collect all the surface terms that arise in the supersymmetric variation of explicit Lagrangians. This is
the main purpose of this section.

\subsection{$\NN=2$ non-linear sigma models} 

In this subsection we study first the most general (classical) $\NN=2$ theory of chiral superfields 
on $A$-type curved manifolds. In flat space such theories are characterized in standard fashion 
by an action governed by a K\"ahler potential $K$ and a superpotential $W$. The curved space generalization
of this action is straightforward. We spell out the details for a non-linear sigma model (NL$\s$) of $\numb$ {\it elementary} chiral superfields $\{\phi^a,\psi^a_\a,F^a\}$, and their conjugate 
$\{\phiT^{\bar{c}},\psiT^{\bar{c}}_\a,\tilde{F}^{\bar{c}} \}$, with a generic superpotential. 
As far as we know, some of the following calculations are not listed in the literature.

\subsubsection{General K\"ahler interactions}

In flat space, supersymmetry turns a generic target space into a K\"ahler manifold. This continues to be true 
in curved space. In addition, the Lagrangian contains a set of new couplings between the dynamical fields and the 
background fields $H$ and $V^\mu$. By following the strategy outlined in the review section \ref{sec:CDFK}, 
the Lagrangian of the curved non-linear sigma model is obtained from the curved D-term combination \eqref{Def_SUSY_DLagrangian} 
evaluated on the composite multiplet
\beq
\mathcal{K}= 
\{ K, \chi^{(K)}, \tilde{\chi}^{(K)}, M^{(K)}, \tilde{M}^{(K)}, a_\mu^{(K)}, \s^{(K)}, \lm^{(K)}, \lmb^{(K)}, D^{(K)} \}~,
\eeq
whose bottom component is the generic real function $K=K(\phi^{a},\phiT^{\bar{c}})$. Derivatives of $K$ w.r.t.\ the 
fields will be indicated by $K_{I_1 I_2,\ldots I_n}$, where $I$ can be either an unbarred or a barred index. For 
$n>1$ the tensor $K_{I_1 I_2,\ldots I_n}$ is totally symmetric. The assignment of R- and central charges is 
\beq
\label{Rzcharges}
R[\phi^a]=r^a,\quad R[\phiT^{\bar{c}}]=-r^{\bar{c}},\quad Z[\phi^a]=z^a,\quad Z[\phiT^{\bar{c}}]=-z^{\bar{c}},
\eeq
and the Lagrangian takes the form
\beq\label{Lagra_NLSM}
\begin{array}{ccl}
\mathscr{L}_{NL\s} & =& -\frac{1}{2} \lp D^{(K)} - a^{(K)}_\mu V^\mu - \s^{(K)} H \rp \\
 &=& \mathscr{L}^{flat}- \frac{\mathfrak{R}}{8} \lp r^a K_a \phi^a + r^{\bar{c}} K_{\bar{c}}\phiT^{\bar{c}}\rp +\mathscr{L}^{bos}_H + \mathscr{L}^{ferm}_H+ \mathscr{L}^{bos}_V+ \mathscr{L}^{ferm}_V,\rule{0pt}{0.6cm}
\end{array}
\eeq
where $\mathfrak{R}$ is the curvature of the background manifold and we have defined:
\begin{subequations}
\bea
\mathscr{L}^{flat}&=&  +g^{\mu\nu} D_\mu \phi^{a} K_{a\bar{c}}  D_\nu\phiT^{\bar{c}}-\frac{i}{2}\,  K_{a\bar{c}}\, \psiT^{\bar{c}}\g^{\mu}(\mathbb{D}_\mu\psi^a)
   	 					 +\frac{i}{2} ( \mathbb{D}_\mu \psiT^{\bar{c}}) \g^{\mu}\psi^a K_{a\bar{c}} \\
			   &   & - F^a \FT^{\bar{c}} K_{a\bar{c}}  -\frac{1}{2} F^a  K_{\bar{c}\bar{n}a} ( \psiT^{\bar{c}}\psiT^{\bar{n}} ) + \frac{1}{2} \FT^{\bar{c}} K_{\bar{c} am}(\psi^m\psi^a )
			  + \frac{1}{4}  K_{\bar{c}\bar{n}{a}m} \psiT^{\bar{c}}\psiT^{\bar{n}} \psi^a\psi^m~,
	\nn\\
\mathscr{L}^{bos}_H &=& +\lp H r^{\bar{c}}- z^{\bar{c}}\rp \lp H r^{a}- z^{a}\rp \phiT^{\bar{c}} K_{\bar{c}a}\phi^a 
\nn\\
 				&& 
				-\frac{H}{4} \Big[ \lp H r^a-z^a\rp K_a\phi^a + \lp H r^{\bar{c}} - z^{\bar{c}}\rp K_{\bar{c}}\phiT^{\bar{c}}\Big]
								+ \frac{3H}{4} \lp z^a K_a \phi^a + z^{\bar{c}} K_{\bar{c}}\phiT^{\bar{c}} \rp ,  
\\
\mathscr{L}^{ferm}_H &=& -\frac{i}{2} \Big[  \lp H  \lp r^{\bar{c}} -\frac{1}{2}\rp - z^{\bar{c}} \rp K_{a\bar{c}}   + \lp H r^m - z^m \rp \phi^m K_{\bar{c}ma}    \Big] \psi^a\psiT^{\bar{c}} 
\nn\\
			& &	
			-\frac{i}{2} \Big[  \lp H \lp r^a -\frac{1}{2}\rp- z^{\bar{c}} \rp  K_{a\bar{c}} +  \lp  H r^{\bar{n}} - z^{\bar{n}}  \rp  \phiT^{\bar{n}} K_{a\bar{n}\bar{c}} \Big]  \psi^a\psiT^{\bar{c}}~ , 
			\\
\mathscr{L}^{bos}_V&=& + iV_\mu \Big[   \lp r^{\bar{c}}\phiT^{\bar{c}} K_{\bar{c}a} -\frac{1}{2} K_a \rp D^\mu\phi^a - \lp r^a\phi^a K_{a\bar{c}} - \frac{1}{2}K_{\bar{c}}\rp D^\mu \phiT^{\bar{c}}  
\nn\\
				  & & \rule{3cm}{0pt}  +\frac{i}{4} \ V^\mu \lp  r^a K_a\phi^a +r^{\bar{c}}K_{\bar{c}}\phiT^{\bar{c}} - 4 r^a\phi^a r^{\bar{c}}\phiT^{\bar{c}} \rp \Big] ~,
		\\
\mathscr{L}^{ferm}_V&=& +\frac{ i}{2} V_\mu \Big[ i\psiT^{\bar{c}} \g^\mu \lp \lp r^a-\frac{1}{2} \rp \psi^a + \G^a_{mn} r^m\phi^m\psi^n\rp 
\nn\\
         && \rule{3cm}{0pt} +i \lp \lp r^{\bar{c}}-\frac{1}{2}\rp \psi^{\bar{c}} +  \G^{\bar{c}}_{\bar{m}\bar{n}} r^{\bar{m}}\phiT^{\bar{m}}\psi^{\bar{n}} \rp\g^\mu\psi^a \Big] K_{a\bar{c}}~ .\rule{0pt}{.6cm} 
\eea
\end{subequations}
In \eqref{Lagra_NLSM} we are using the covariant derivatives
\beq
\begin{array}{ccl}
D_\mu \varphi_{(r,z)}&=& \D_\mu \varphi_{(r,z)}+ i r V_\mu \varphi_{(r,z)} \\
							 &=&\nabla_\mu\varphi_{(r,z)} -ir A_\mu^{(R)} \varphi_{(r,z)},  \rule{0pt}{.5cm}\\
\mathbb{D}_\mu \psi^a  &=& D_\mu\psi^a +  K^{a\bar{c}}K_{\bar{c}mn} D_\mu\phi^m\psi^n, \rule{0pt}{.5cm}\\
\mathbb{D}_\mu \psiT^{\bar{c}}  &=& D_\mu\psiT^{\bar{c}} +  K^{\bar{c}a}K_{a\bar{m}\bar{n}} D_\mu\phib^{\bar{m}}\psi^{\bar{n}}. \rule{0pt}{.5cm}
\end{array}
\eeq
The background connection appearing in $D_\mu$ is $A^{(R)}=A-\frac{3}{2} V$. 
Let us also mention that the $R$-charges of the derivatives of the composite fields are
\beq\label{RchargesKahler}
R[K_a]=-r_a,\qquad R[K_{\bar{c}}]=r^{\bar{c}},\qquad R[K_{a\bar{c}}]=-r^a+r^{\bar{c}}
~.
\eeq
As in flat space, the function $K$ defines a K\"ahler potential for the metric $G_{a\bar{c}}\equiv K_{a\bar{c}}$. 
Consistency of the supersymmetric transformation rules requires $K$ to be a quasi-homogeneous function of 
vanishing $R$- and central charge.\footnote{This restriction can be understood from the computation of $\s^{(K)}$. 
Extracting $\s^{(K)}$ from $\d\chi^{(K)}$ and $\d\tilde{\chi}^{(K)}$ leads to two different expressions: 
\beq
\s^{(K)}= -2 \lp r^a H -z^a\rp K_a \phi^a + i K_{a\bar{c}}\,\psi^a\psiT^{\bar{c}} =  -2\lp r^{\bar{c}} H -z^{\bar{c}}\rp K_{\bar{c}} \phiT^{\bar{c}} + i K_{a\bar{c}}\, \psi^a\psiT^{\bar{c}} ~.
\eeq
In order for $\s^{(K)}$ to be well defined, $K$ has to be quasi-homogeneous of the type
\eqref{extra_homogeneity_kahler}.}
Collecting the fields $\phi^a$ and $\phiT^{\bar{c}}$ under the variable $C^I=(\phi^a,\phiT^{\bar{c}})$, the two conditions on $K$ are
\beq\label{extra_homogeneity_kahler}
\begin{array}{lcl}
\sum_I r^I C^I K_I =0,& \rule{.5cm}{0pt} & r^I= (r^a,-r^{\bar{c}}), \\
\sum_I z^I C^I K_I=0, &\rule{.5cm}{0pt} & z^I= (z^a,-z^{\bar{c}})~. \rule{0pt}{.5cm} 
\end{array}
\eeq
These extra conditions on the K\"ahler potential arise from coupling the theory to the background field $H$.

\subsubsection{Superpotential interactions}

Superpotential interactions are introduced as F-terms for a chiral multiplet $\Omega_W=(W,\psi^{(W)},F^{(W)})$, 
where $W$ is a holomorphic function of the chiral fields $\phi^a$. The resulting Lagrangian in components is
\beq\label{Lagra_W}
\mathscr{L}_{W}  = F^m\partial_m\W -\frac{1}{2} \psi^i\,\psi^j\partial_i\partial_j\W + \FT^{\nb}\partial_{\nb}\Wb +\frac{1}{2}\psiT^{\nb}\psiT^{\bar{m}}\partial_{\nb}\partial_{\bar{m}}\Wb
~. 
\eeq
Invariance under supersymmetry requires $W$ to be a quasi-homogeneus function of the $\phi^a$ of degree $2$
\beq
-2 W + \sum_i r^a\phi^a \partial_aW=0
~.
\eeq
In a similar way $\Wb$ is quasi-homogeneous of degree $-2$. The $R$-charges of $\partial_a W$ and 
$\partial_c \Wb$ are 
\beq\label{RchargesW}
R[\partial_a W]=2-r^a,\qquad R[\partial_{\bar{c}}W]=r^{\bar{c}}-2
~.
\eeq

The most general Lagrangian for a set of chiral superfields is then specified by the two functions $K$ and $W$, and by 
the assignment of charges. Schematically, from \eqref{Lagra_NLSM} and \eqref{Lagra_W} we find
\beq
\mathscr{L}_{\NLs}=\mathscr{L}_{K}+ \mathscr{L}_W
~.
\eeq

\subsubsection{Variation under supersymmetry}

Given $\mathscr{L}_{\NLs}$, the object of interest for us is the total derivative that arises in a supersymmetric variation
\beq
\d \mathscr{L}_{\NLs}+ \tilde{\d}\mathscr{L}_{\NLs} =\nabla_\mu(\TD^\mu_{\NLs}) .
\eeq
The supervariation can be obtained either by varying the action explicitly or by evaluating 
\eqref{Def_SUSY_DLagrangian} and \eqref{Def_SUSY_FLagrangian} for the multiplets 
$\mathcal{K}$, $\Omega_W$ and $\widetilde{\Omega}_{\widetilde{W}}$. The result in both cases is
\bea\nn
\sqrt{2}\,\TD^\mu_{\NLs}& = & 
						+\ep\Big[ \g^{\mu} \g^\nu \psi^a\, K_{a\bar{c}} \D_\nu\phiT^{\bar{c}} 
									-  (r^{\bar{c}} H-z^{\bar{c}})\, \g^\mu \psi^a\, K_{a \bar{c}}\phiT^{\bar{c}}- i V^\mu \psi^a K_a
									- 2i \g^\mu\psiT^{\bar{c}} \partial_{\bar{c}} \Wb \Big] \nn\\
&&						-\epT\Big[ \g^\mu\g^\nu \psiT^{\bar{c}}\, K_{\bar{c}a} \D_\nu\phi^a	
								-  (r^{a}H- z^{a} )\, \g^\mu\psiT^{\bar{c}}\, K_{\bar{c}a} \phi^a + i V^\mu \psiT^{\bar{c}} K_{\bar{c}} 
								+ 2i \g^\mu\psi^{a} \partial_{a} \W \Big]\rule{0pt}{.5cm}\nn \\
&& 				+ i  \ep\,\g^\mu \psiT^{\bar{c}}\lp F^a K_{\bar{c}a}  - \frac{1}{2} K_{am\bar{c}}(\psi^a\psi^m)  \rp
			+ i\, \epT\,\g^\mu\psi^a \lp  \FT^{\bar{c}} K_{\bar{c}a} +\frac{1}{2} K_{\bar{c}\bar{n}a}(\psiT^{\bar{c}}\psiT^{\bar{n}})\rp \nn \rule{0pt}{.5cm}~.\\ \label{WZgen1}		 	
\eea
The equations of motion of the auxiliary fields $F^a$ and $\bar{F}^{\bar{c}}$  are
\beq
\begin{array}{ccl}
K_{a\bar{c}} \FT^{\bar{c}} + \frac{1}{2} K_{a\bar{c}\bar{n}} (\psiT^{\bar{c}}\psiT^{\bar{n}}) &=& \partial_a W~, \\
F^a K_{a\bar{c}} - \frac{1}{2} K _{\bar{c} a m} (\psi^a\psi^m) &=&  \partial_{\bar{c}} \Wb~. \rule{0pt}{.5cm}
\end{array}
\eeq
Integrating out $F^a$ and $\bar{F}^{\bar{c}}$ we obtain the final expression
\bea\nn
 \sqrt{2}\,\TD^\mu_{\NLs}&=&	
 					+\ep\Big[ \g^{\mu} \g^\nu \psi^a\,  \D_\nu\phiT^{\bar{c}} 
					-  (r^{\bar{c}} H-z^{\bar{c}})\, \g^\mu \psi^a\, \phiT^{\bar{c}}- i V^\mu \psi^a K^{\bar{c}}
					- i \g^\mu\psiT^{\bar{c}} W^a \Big] K_{a\bar{c}} \nn\\
&& 				-\epT\Big[ \g^\mu\g^\nu \psiT^{\bar{c}}\,  \D_\nu\phi^a	
					-  (r^{a}H- z^{a} )\, \g^\mu\psiT^{\bar{c}}\,  \phi^a + i V^\mu \psiT^{\bar{c}} K^a 
					+ i \g^\mu\psi^a \Wb^{\bar{c}}  \Big] K_{a\bar{c}}~, \rule{0pt}{.5cm} \rule{.8cm}{0pt}
					 \label{LG_SuperV}
\eea
where we have defined the vectors 
\beq
W^a\equiv K^{a\bar{c}}\partial_{\bar{c}}\Wb, \qquad \widetilde{W}^{\bar{c}}\equiv K^{\bar{c}a}\partial_aW, \qquad {K}^{\bar{c}}\equiv K^{\bar{c}a}K_{a},\qquad K^a \equiv K^{a\bar{c}}K_{\bar{c}}~. 
\label{Vectors_Invers_LG}
\eeq
The R-charges of these vectors can be deduced from \eqref{RchargesKahler} and \eqref{RchargesW}: 
$R[W^a]=r^a-2$, $R[\Wb^{\bar{c}}]=2-r^{\bar{c}}$, and so on. 
Observe that the bilinears appearing in $\TD_{\NLs}$, are the  
most general bilinears of vanishing $R$-charge with the correct index structure built out of $\ep$ and $\epT$, 
$\psi$ and $\psiT$, and the corresponding bosonic fields. For example, it is obvious that derivatives of the 
superpotential $W^a$ only couple to $\ep\g^\mu\psiT$, not to $\ep\g^\mu\psi$.

\subsubsection{Digression on target space geometry}\label{sec_Target_Space}

In differential geometry, a K\"ahler manifold is defined as a symplectic (real) manifold $(\mathcal{N},\omega)$,  
equipped with a complex structure $J$ such that $G(\cdot,\cdot)\equiv\omega(\cdot, J\,\cdot )$ is a Riemaniann 
metric on $\mathcal{TN}$. The last condition is called {\it $\omega$-compatibility} \cite{daSilva}. 
In a local description with coordinates $(\phi^a,\phib^{\bar{c}})$ the metric is represented as
\beq\label{Kmetric_geo}
ds^2_{\mathcal{N}}=  G_{a\bar{c}} d\phi^a d\phib^{\bar{c}} + G_{\bar{c}a} d\phi^a d\phib^{\bar{c}} 
= 2 G_{a\bar{c}} d\phi^a d\phib^{\bar{c}}~,\qquad 
G_{a\bar{c}}= G^\star_{\bar{a}c}~,
\eeq
and the two-form $\omega$ is represented as $\omega_{\a\bar{c}}\propto G_{a\bar{c}}d\phi^a\wedge d\phi^{\bar{c}}$.
The target space of the non-linear sigma model, listed above, is such a K\"ahler manifold.

For many of the explicit computations in the following sections, a different parametrization will turn out to be 
especially useful. This involves the change of variables 
\beq\label{change_to_real}
\phi^a= \Phi^a+ i \Phi^{a+\numb},\qquad \phib^{\bar{c}}= \Phi^{\bar c} - i \Phi^{\bar c+\numb},
\eeq
where $a,\bar c=1,\dots, \numb$. When the reality condition 
$\phib^{\bar{c}}=\phi^{c\, \star}$ holds, the fields $\Phi^I$ are real. However, in general, we may 
consider $\phi$ and $\phib$ as two independent complex variables. Then the fields $\Phi^I$ are also complex and 
\eqref{change_to_real} is a standard change of variables in $GL(2\mathfrak{s},\mathbb{C})$.

Collecting the labels of the type $(a,a+\numb)$ into one index $I=1,\ldots, 2\numb$, the matrix that represents the 
change of variable is
\beq
\left( \begin{array}{c} \phi^a \\ \phib^{\bar{c}}\end{array}\right)
=\mathfrak{M} \left( \begin{array}{l} \Phi^i \\ \Phi^{i+\numb} \end{array}\right),\qquad 
\mathfrak{M}=\lp\begin{array}{ccc} \d_i^a & &+i \d^a_{i+\numb}\\ \d^{\bar{c}}_{i} & &-i\d^{\bar{c}}_{i+\numb} \end{array} \rp,\qquad 
\mathfrak{M}^{-1}=\begin{array}{c}\frac{1}{2}\end{array} \mathfrak{M}^\dagger,
\eeq
where the symbol $\d^a_{i+\numb}$ stands for a diagonal matrix in the off-diagonal blocks of $\mathfrak{M}$, and is 
defined to be $\d^a_{i+\numb}=1$ (or $0$) if $a=i$ (or $a\neq i$), as is clear from \eqref{change_to_real}. 
The metric changes accordingly
\beq
 ds^2_{\mathcal{N}}= G_{IJ}\,d\Phi^I d\Phi^J,\qquad  
G_{IJ}=\lp \begin{array}{ccc} \d^{\{a}_i \d^{\bar{c}\}}_j\, G_{a\bar{c}} & & -i \d^{[a}_i \d^{\bar{c}]}_{j+\numb} \, G_{a\bar{c}} \\
+i \d^{[a}_{i+\numb} \d^{\bar{c}]}_{j} \, G_{a\bar{c}}	\rule{0pt}{.5cm} &  & \d^{\{a}_{i+\numb} \d^{\bar{c}\}}_{j+\numb}\, G_{a\bar{c}} \end{array}\rp
~.
\eeq
The matrix $G_{IJ}$ is real and symmetric, $G=G^T$. On the other hand, the complex structure and the two-form are 
given by
\beq
J^{M}_{\ N}= \lp \begin{array}{ccc} 0 & \d^m_{n+\numb} \\ -\d^{m}_{n+\numb} & 0 \end{array} \rp,\qquad \omega_{MN}=  -G_{MI} J^I_{\ N}\ . \label{C_structure_Omega}
\eeq 
Important relations are $J=-J^T$, $J^2=-\mathbb{I}$, and $G=J G J^T$.\footnote{When we write matrix products 
we always understand row by column multiplication, from right to left.} 
The second one in \eqref{C_structure_Omega} is precisely the condition of $\omega$-compatibility, which is part 
of the definition of $\mathcal{N}$. 

By construction, two types of ``products" exist on a K\"ahler manifold, one is the symplectic product defined from the 
tensor $\omega_{IJ}$ and the other one is the metric. In components, we find 
\bea
\lp v^a \bar{w}^{\bar{c}} + w^a \bar{v}^{\bar{c}} \rp G_{a\bar{c}}&=&  V^I G_{IJ} W^J ~,\label{scalar_metric_Phi}\\
\lp v^a \bar{w}^{\bar{c}} - w^a \bar{v}^{\bar{c}}\rp G_{a\bar{c}}&=& i V^I \omega_{IJ} W^J~, \rule{0pt}{.5cm}\label{scalar_symplectic_Phi}
\eea
for any pair of vectors $V^I$, $W^J$.
The formulae \eqref{scalar_metric_Phi} and \eqref{scalar_symplectic_Phi} will be useful in several occasions. Here 
we mention one simple application regarding the kinetic energy, which in the new variables $\Phi^I$ is the sum of 
both the metric and the symplectic product. Because of the following identity
\bea
&& G_{a\bar{c}} (\partial_\mu \phi^a -i r^a a_\mu \phi^a) (\partial^\mu \phib^{\bar{c}} + i r^{\bar{c}} a^\mu \phib^{\bar{c}}) =\nn \\
&& \rule{1cm}{0pt} = \frac{1}{2} G_{MN} (\partial_\mu \Phi^M + a_\mu \sum_I J^M_{\ I} r^I\Phi^I)(\partial^\mu \Phi^N + a^\mu \sum_K J^{N}_{\ K} r^K\Phi^K)
\eea
valid for any connection $a_{\mu}$, 
it is possible to introduce the analog of the covariant derivatives $(\D_\mu\phi^a, \D_\mu\phib^{\bar{c}})$ acting on 
$\Phi^I$. In particular, we define
\beq
(\D_\mu\phi^a,\D_\mu\phi^{\bar{c}})\  \rightarrow\  \partial_\mu \bPhi +\lp A_\mu-\frac{1}{2} V_\mu \rp J\,  \mathcal{R}\,\bPhi + J\, \mathcal{Z}\, \bPhi\ 
~,
\eeq
where  
the bold symbol $\bPhi$ represents the vector $\Phi^I$ and the matrices $\mathcal{R}$ and $\mathcal{Z}$ are given by 
\beq
\mathcal{R}_{IJ} = \lp\begin{array}{cc} r^a\,\d^a_i\d^{a}_j  & 0 \\ 0 & r^{\bar{c}}\,\d^{\bar{c}}_i\d^{\bar{c}}_j  \end{array}\rp,\qquad
\mathcal{Z}_{IJ}=\lp\begin{array}{cc} z^a\,\d^a_i\d^{a}_j  & 0 \\ 0 & z^{\bar{c}}\,\d^{\bar{c}}_i\d^{\bar{c}}_j  \end{array}\rp\ 
~.
\eeq
Notice the absence of negative signs in the right bottom corner of $\mathcal{R}$ ($\mathcal{Z}$), corresponding to
$r^{\bar{c}}$ ($z^{\bar{c}}$). The bold symbols $\bPsi$, $\bW$, and $\bK$, will  be used to describe the vectors 
corresponding to $\Psi^I$, $W^I$ and $K^I$, that indeed appear in the supervariation \eqref{LG_SuperV}.

\subsection{$\NN=2$ gauge theories coupled to matter} \label{Gauge_Actions}

\subsubsection{YM and CS theories}

Next, consider a vector multiplet $\mathcal{V}=\{ \A_\mu, \lambda,\lmb, \s, D \}$ valued in the Lie algebra 
$\mathfrak{g}$ of a gauge group $\mathfrak{S}$, possibly non-abelian. The field strength $\F_{\mu\nu}$ of the 
gauge field and the covariant derivatives of the various fields in the vector multiplet are
\beq
\label{gaugecov}
\begin{array}{ccl}
\F_{\mu\nu}&=&\partial_\mu\A_\nu -\partial_\nu\A_\mu - i  [\A_\mu,\A_\nu] ~,\\ 
D_\mu\lm&=&\D_\mu\lm + i V_\mu\lm- i  [\A_\mu,\lm]~,\rule{0pt}{.6cm}\\
D_\mu\lmb&=&\D_\mu\lmb - i V_\mu\lm +i  [\lmb, \A_\mu]~,\rule{0pt}{.6cm}\\
D_\mu\s & =& \partial_\mu \s - i [\A_\mu,\s] \rule{0pt}{.6cm}
~.
\end{array}
\eeq 

In three dimensions a gauge field admits both Yang-Mills (YM) kinetic terms and Chern-Simons (CS) kinetic terms. 
For abelian theories the supersymmetric Lagrangian is obtained as the curved D-term of the composite multiplet 
$-\frac{1}{\gYM^2}\Sigma^2$, where $\Sigma$ is the real multiplet associated to $\mathcal{V}$, and $\gYM$ is the 
coupling constant. The non-abelian Lagrangian is the standard generalization of this construction, and the result in 
components is
\bea
\gYM^2\,\mathscr{L}_{YM}&=& \mathrm{Tr}\Big\{\, \frac{1}{4} \F_{\mu\nu}\F^{\mu\nu}+\frac{1}{2}D^\mu\s D_\mu\s 
									-\frac{i}{2} \lmb (\g^\mu D_\mu\lm) + \frac{i}{2} (D_\mu\lmb\,\g^\mu)\lm \nn  \\
			& &\rule{1cm}{0pt}+i\lmb[\s,\lm]-\frac{1}{2}\lp D+\s H\rp^2 + \frac{i}{2} H \lmb\lm + \mathscr{L}_{YM}^V \,\Big\}~,  \label{L_Yang_Mills} \\
\gYM^2\,\mathscr{L}_{YM}^V&=&	 +\frac{i}{2}V_\mu \Big\{ \s\vep^{\mu\nu\rho}\F_{\nu\rho} -\frac{1}{2} V^\mu \s^2 + \frac{i}{2} \lmb\g^\mu\lm \Big\}	~.			
\eea
For CS theories the supersymmetric Lagrangian is
\beq\label{L_ChernSimon}
\mathscr{L}_{CS}=\frac{k}{4\pi} \mathrm{Tr} \left\{ i\vep^{\mu\nu\rho} \lp \A_\mu\partial_\nu\A_\rho + \frac{2}{3} \A_\mu \A_\nu \A_\rho\rp - 2 D\s + 2 i \lmb\lm \right\}~.
\eeq
Finally, if the gauge group contains a (product of) $U(1)$ factors 
we can add for each abelian factor the corresponding FI term
\beq\label{L_FI}
\mathscr{L}_{FI}=+\frac{1}{2} \xi (D -\A_\mu V^\mu - \s H)~.
\eeq

\subsubsection{Matter couplings}

Matter can be added both to CS and YM theories by coupling the vector multiplet to chiral and anti-chiral
superfields in arbitrary representations of the gauge group $\mathfrak{S}$. We consider 
matter superfields $\Phi^{\bf a}$ and $\widetilde{\Phi}^{\bf \bar{c}}$ labelled by a bold index which collectively 
indicates both the color index $a$ and flavor index $m$, i.e.\ ${\bf a}=(a,m)$. 
The color indices are contracted in scalar products defined in the appropriate representation of the chiral 
and anti-chiral fields. Similarly, the components of the gauge multiplets act on the matter fields 
according to their representation, and the covariant derivatives contain both the background and the gauge fields, 
$D_\mu \varphi_{(r,z)}= \D_\mu \varphi_{(r,z)}+ i r V_\mu \varphi_{(r,z)}- i \A_\mu \varphi_{(r,z)}$ for any field 
$\varphi_{(r,z)}$. 

The gauge invariant interactions among different flavors are fixed by a choice of K\"ahler potential and superpotential. 
For the simplicity of the presentation, we will consider a canonical K\"ahler potential. Each flavor may also have 
different background $R$-charge $r^m$ and central charge $z^m$. Assuming that chiral and anti-chiral superfields have 
opposite charges, it is convenient to define the diagonal matrices of $\mathcal{R}$- and $\mathcal{Z}$-charges. 
The Lagrangian is
\beq
\label{matterab}
\mathscr{L}_{matter} = \mathscr{L}_{K} + \mathscr{L}_W 
~,
\eeq
where $\mathscr{L}_W$ contains a gauge invariant superpotential, \eqref{Lagra_W}, and $\mathscr{L}_{K} $ is given by
\beq
\label{matterac}
\mathscr{L}_K = 
\mathscr{L}^{flat}
- \begin{array}{l} \frac{\mathfrak{R}}{4} \end{array}\phiT \mathcal{R} \phi  
+\mathscr{L}^{bos}_H + \mathscr{L}^{ferm}_H+ \mathscr{L}^{bos}_V+ \mathscr{L}^{ferm}_V
~.
\eeq
In this formula $\mathfrak{R}$ is the curvature of the background manifold and
\begin{subequations}
\bea
\label{matterad}
\mathscr{L}^{flat} &=& g^{\mu\nu} D_\mu \phi D_\nu\phiT 
-\frac{i}{2}\, \psiT\g^{\mu} (D_\mu\psi)
+\frac{i}{2} (D_\mu \psiT) \g^{\mu} \psi 
- F \FT 
+ \sqrt{2} i \lp \phiT \lambda \psi + \psiT \tilde \lambda \phi \rp
+\phiT  D \phi
~, \nonumber\\
\\
\label{matterae}
\mathscr{L}^{bos}_H &=& \phiT
\lp H^2  \mathcal{R}\lp\mathcal{R}-\frac{1}{2}\rp + (\mathcal{Z} + \sigma) ^2 - 2 H (\mathcal{R}-1) (\mathcal{Z}-\s) \rp \phi
 +H  \phiT \sigma \phi
~,\nonumber\\
\\
\label{matteraf}
\mathscr{L}^{ferm}_H &=& 
- i  \psi
\lp H  \lp \mathcal{R} -\frac{1}{2}\rp - ( \mathcal{Z}- \sigma) \rp \psiT  
~,\nonumber\\
\\
\label{matterag}
\mathscr{L}^{bos}_V&=& iV_\mu
 \lp    \phiT \lp \mathcal{R} -\frac{1}{2} \rp  D^\mu\phi
 -  \phi  \lp \mathcal{R} -\frac{1}{2} \rp D^\mu \phiT
-\frac{i}{2} \ V^\mu \phiT\,
 \mathcal{R} \lp \mathcal{R} - \frac{1}{2} \rp 
\phi \rp 
~,\nonumber\\
\\
\label{matterae}
\mathscr{L}^{ferm}_V&=& - V_\mu
  \psiT \lp \mathcal{R} - \frac{1}{2} \rp \g^\mu \psi
~.
\eea
\end{subequations}

Equipped with the precise form of $\mathscr{L}_{YM}$, $\mathscr{L}_{CS}$ and $\mathscr{L}_{matter}$ it is 
possible to write down the most generic quiver gauge theory. In this case, the gauge fields will be also labelled by 
a bold index of the type, $\bold{m}=(a,m)$, where $m$ labels the nodes of the quiver theory, and $a$ labels the 
generators of the gauge group $\mathfrak{S}_m$ at the node $m$. Considering normalized generators for 
the gauge groups, the CS coupling $\kappa$ is promoted to a matrix of the form 
$\kappa_{\bold{mn}}=\d_{ac}\otimes \kappa_{mn}$, with $\kappa_{mn}$ a symmetric tensor.

\subsubsection{Variation under supersymmetry}

The supersymmetric variation of the actions $\mathscr{L}_{YM}$, $\mathscr{L}_{CS}$ and $\mathscr{L}_{matter}$
has the following properties. Let us begin with the non-abelian YM theory. The change in the action under a 
supersymmetric transformation is given by the total derivative of
\bea\label{YM_SuperV}
\gYM^2\, \TD^\mu_{YM}&=&  \Tr\Big[+\frac{1}{4} \ep\,\g^\mu\g^\rho\lmb\, ( \Fh_\rho +2i \s V_\rho) 
									-\frac{1}{2}\ep\,\g^\mu\g^\rho\lmb\,\partial_\rho\s 
											\nn\\
						& & \rule{1cm}{0pt} +\frac{1}{4} \epT\,\g^\mu\g^\rho\lm\,( \Fh_\rho +2i \s V_\rho) 
								+ \frac{1}{2}\epT\,\g^\mu\g^\rho\lm\,\partial_\rho\s 	
											\nn\\
						&&		 \rule{1.5cm}{0pt}	+\frac{1}{2}  \ep\,\g^\mu\psi_{\Sigma}\, (iD+\s ( iH))
						+\frac{1}{2}\epT\,\g^\mu\psiT_{\Sigma}\, (iD+\s ( iH)) \Big]~, 
\eea
where $\Fh_\rho=\varepsilon_{\mu\nu\rho}\F^{\mu\nu}$. 
In the real multiplet parametrization,
\beq
\label{vectornotation}
j_\rho= -\frac{i}{2} \Fh_\rho~, \qquad a_\rho = -j_\rho - \s V_\rho~, \qquad\lm=i\psiT_{\Sigma}~, \qquad \lmb=-i\psi_{\Sigma}
~,
\eeq
we can rewrite $\TD^\mu_{YM}$ in a more compact form as follows
\bea
\label{YM_SuperV_real}
\gYM^2\, \TD^\mu_{YM}&=&  \Tr\Big[ 
-\frac{1}{2}\Big(\ep \g^\mu\g^\rho\psi_\Sigma\, \lp a_\rho -i\partial_\rho\s\rp  
-\ep \g^\mu\psi_\Sigma\, \big (i D + (iH)\s  \big) \Big) 
\nn\\
& &  \rule{1cm}{0pt} + \frac{1}{2} \Big(\epT \g^\mu\g^\rho\psiT_\Sigma\, \lp  a_\rho+i\partial_\rho\s\rp
+ \epT \g^\mu\psiT_\Sigma\, \big(i D + (iH)\s \big)\Big)\ \Big]~.
\eea
For the CS action \eqref{L_ChernSimon} the variation under supersymmetry gives
\bea
\label{CSUaa}
\TD^\mu_{CS}
&=&+\frac{i}{4\pi}\kappa_{\bf ac}\, \Big[ \varepsilon^{\mu\nu\rho}\,( \ep\g_\rho\psi_{\Sigma}^{\bf a}  - \epT\g_\rho\psiT_{\Sigma}^{\bf a})\A_\nu^{\bf c}
+ 2 (\ep\g^\mu\psi_{\Sigma}^{\bf a}  + \epT\g^\mu\psiT_{\Sigma}^{\bf a})\s^{\bf c} \Big]~.
\eea
The case of the FI Lagrangian \eqref{L_FI} is straightforward, and we obtain
\bea
\TD^\mu_{FI}&=& +\frac{1}{2}\,\xi\,  (\ep\g^\mu\lmb-\epT\g^\mu\lm)=
-\frac{i}{2} \xi \big[ \ep\g^\mu\psiT_{\Sigma} + \epT\g^\mu\psi_{\Sigma}\big]
~.
\eea

Finally, the variation of the matter action generates 
\bea 
 \sqrt{2}\,\TD^\mu_{matter}&=&	
 					+\ep\Big[ \g^{\mu} \g^\nu \psi^{\bf a}\,  \D_\nu\phiT^{\bf \bar{c}} 
					-  (r^{\bf \bar{c}} H-z^{\bf \bar{c}})\, \g^\mu \psi^{\bf a}\, \phiT^{\bf \bar{c}}\,- i V^\mu \psi^{\bf a} \phiT^{\bf \bar{c}} + i\,\g^\mu \psiT^{\bf \bar{c}} F^{\bf a}
					\Big] G_{\bf a\bar{c}} \nn\\
&& 				-\epT\Big[ \g^\mu\g^\nu \psiT^{\bf \bar{c}}\,  \D_\nu\phi^{\bf a}
					-  (r^{\bf a}H- z^{\bf a} )\, \g^\mu\psiT^{\bar{c}}\,  \phi^{\bf a} 
					+ i V^\mu \psiT^{\bar{c}} \phi^{\bf a} - i\, \g^\mu\psi^{\bf a}   \FT^{\bf \bar{c}}\, 
					\Big] G_{\bf a\bar{c}} \rule{0pt}{.5cm} \rule{.8cm}{0pt} \\
&& -\ep\, \g^\mu\psi^{\bf a} (\s\phiT)^{\bar{c}} G_{\bf a\bar{c}} + \epT\,\g^\mu\psiT^{\bf \bar{c}} (\s\phi)^{\bf a} G_{\bf a\bar{c}} - i\sqrt{2} \Big[ \ep\,\g^\mu \phiT^{\bf \bar{c}}\,(\psi_{\Sigma}\, \phi)^{\bf a} + 	\epT\,\g^\mu \phiT^{\bf \bar{c}}\,(\psiT_{\Sigma}\,\phi)^{\bf a} \Big]	G_{\bf a\bar{c}}	~. \nn 
\eea
The contraction of the color and flavor indices is packaged into $G_{\bf a\bar{c}}$. Notice that in the last line $\s$, 
$\psi_{\Sigma}$ and $\psiT_{\Sigma}$ act appropriately on color indices.

\section{Boundary conditions: a preview}\label{preview}

In the previous sections we made precise two key elements of our initial discussion: we decomposed any compact 
$A$-type background $\mathcal{M}_3$ into the union of submanifolds with boundary, called $\mathcal{T}$, and 
we wrote down supersymmetric field theories for $\NN=2$ chiral and vector superfields on $\mathcal{M}_3$, explicitly 
calculating the expressions for the supersymmetric variation $\TD^\mu$. When these field theories are restricted 
on $\mathcal{T}$, the action can only be invariant under a subset of the bulk supersymmetries if there are 
boundary conditions solving the corresponding constraints $\TD^\perp=0$. 

In addition, a well-defined classical problem requires appropriate boundary conditions that annihilate all the 
surface contributions in the Euler-Lagrange variation of the system.
Schematically, given a field $\Phi$, and a bulk action $\mathcal{S}=\int_{\mathcal{M}_3}\mathscr{L}[\Phi]$ the 
equations of motion of the theory require $\d\mathcal{S}=0$, where
\beq
\d \mathcal{S} = \int_{\mathcal{M}_3} \, \d\Phi \left[ \frac{\partial\mathscr{L}}{\partial\Phi} -\partial_\mu \lp \frac{\partial\mathscr{L}}{\partial\Phi_\mu} \rp\right] + \int_{\mathcal{M}_3} \partial_\mu \lp  \frac{\partial\mathscr{L}}{\partial\Phi_\mu} \d\Phi \rp~,\qquad \Phi_\mu\equiv\partial_\mu\Phi~.
\eeq 
On a space with boundary, one demands simultaneously
\beq
\mathbb{E}[\Phi]= \frac{\partial\mathscr{L}}{\partial\Phi} -\partial_\mu \lp \frac{\partial\mathscr{L}}{\partial\Phi_\mu} \rp=0~,\qquad  
\mathbb{B}[\Phi,\d\Phi]=n_\mu\frac{\partial\mathscr{L}}{\partial\Phi_\mu} \d\Phi\Big|_{\mathrm{bdy}}=0~.
\eeq

A priori, the boundary equations $\mathbb{B}=0$ are a set of on-shell equations. In what follows,
some of these boundary equations will be required to hold also off-shell and will be used to find solutions of 
$\VV^\perp=0$, which is our main goal.

\subsection{The boundary value problem}

%
\subsubsection{Fermions}

Let us focus first on the boundary value problem for the fermions in $\mathscr{L}_{YM}$ and $\mathscr{L}_{matter}$, 
respectively. In $\mathscr{L}_{CS}$ the fermions do not have a kinetic term and do not contribute boundary terms.
The corresponding boundary contributions are
\beq
\label{ELaa}
-\frac{i}{2} \int_{\mathcal{M}_2'} \Tr \lp \tilde \lambda \g^\perp \delta \lambda
-\delta \tilde \lambda \g^\perp \lambda \rp\, \subset \,
\delta \mathcal{S}_{YM} 
~,
\eeq
\beq
\label{ELab}
-\frac{i}{2} \int_{\mathcal{M}_2'} K_{a\tilde c} \lp \psiT^{\bar c} \g^\perp \delta \psi^a 
-\delta \psiT^{\bar c} \g^\perp \psi^a \rp\,  \subset \,
\delta \mathcal{S}_{matter}
~.
\eeq
It is convenient to rewrite both terms in a uniform way. Defining the doublet $\d\bPsi=(\d\psi^a,\d\psiT^{\bar{c}})$ 
and $\bPsi=(\psi^a,\psiT^{\bar{c}})$, we obtain the expression
\beq\label{expr_form_chiral_ferm}
\mathbb{B}^f[\bPsi,\d\bPsi]=-\frac{i}{2} \bPsi^T \lp\begin{array}{cc} 0 & K_{a\bar{c}} \\ K_{\bar{c}a}&0\end{array}\rp\otimes\g^\perp\, \d\bPsi\ .
\eeq
The form \eqref{expr_form_chiral_ferm} also covers the case of \eqref{ELaa}. It is convenient to use $\psi_\Sigma$ 
instead of $\lm$. If the generators of the Lie algrebra $\{{\bf t}^a\}$ are normalized so that $\Tr[{\bf t}^a{\bf t}^c]=\d^{ac}$, 
the corresponding metric $K$ is the identity. In the real notation of subsection \ref{sec_Target_Space} 
both \eqref{ELaa} and \eqref{ELab} can be written in the compact form
\beq
\label{ELac}
\mathbb{B}^f[\bPsi,\d\bPsi]=-\frac{i}{2} G_{IJ}\bPsi^I \g^\perp\, \d\bPsi^J ~,
\eeq
where $G$ is the appropriate metric. Notice that because of the anti-symmetry of $ \bPsi^I\g^\perp\, \d\bPsi^J$, the 
boundary term $\mathbb{B}[\bPsi,\d\bPsi]$ is a $2$-form on the space of fermions, i.e.\ 
$\mathbb{B}[\bPsi,\d\bPsi]=-\mathbb{B}[\d\bPsi,\bPsi]$. 

As we did before, we decompose
\beq
\bPsi= \frac{\zetaT\bPsi}{\Omega}\zeta + \frac{\bPsi\zeta}{\Omega}\zetaT~,\qquad  
\d\bPsi= \frac{\zetaT\d\bPsi}{\Omega}\zeta + \frac{\d\bPsi\zeta}{\Omega}\zetaT~.
\eeq
Then, the equation $\mathbb{B}[\bPsi,\d\bPsi]=0$ becomes
\beq
G_{IJ}\, \frac{\zeta\g^\perp\zeta}{\,\Omega} \, (\zetaT\bPsi)^I\, (\zetaT\d\bPsi)^J 
+  G_{IJ}\,\frac{\zetaT\g^\perp\zetaT}{\,\Omega} \, (\bPsi\zeta)^I\, (\d\bPsi\zeta)^J=0~.
\eeq
Recalling \eqref{Bilinears_2}, we solve this equation by requiring the boundary conditions
\bea
\label{ELad}
\frac{\zeta\g^\perp\zeta}{\,\Omega}\, (\zetaT\bPsi)^I = M^I_{\ K}\, (\bPsi\zeta)^K
\eea
with a general (possibly field-dependent) matrix $M$ that has the property 
\beq
\label{ELae}
M^T G M = G
~.
\eeq
The boundary condition \eqref{ELad} respects the $R$-symmetry whatever $r$-charge is assigned to $\bPsi$.

\subsubsection{Vectors}\label{sec_vectors_form}

There are two possible actions for a vector field $\A_\mu$ in $3d$: $\mathscr{L}_{CS}$, and $\mathscr{L}_{YM}$. 
The Euler-Lagrange variation with respect to $\A_\mu$, yields the boundary terms
\beq
\label{ELag}
 -\frac{i}{4\pi }\int_{\mathcal{M}_2'}  {\kappa_{mn}}\, \Tr\, \Big[ \vep^{\perp\nu\rho} \A^m_\nu\d\A^n_\rho\Big]\, \subset\, \d \mathcal{S}_{CS}
~,
\eeq
\beq
\label{ELaf}
\int_{\mathcal{M}_2'} \Tr \Big[  
\Big( \F^{\perp\nu} + i\,  \varepsilon^{\perp\nu\rho} V_\rho \, \sigma \Big) \delta \A_\nu 
\Big]=\int_{\mathcal{M}_2'} \Tr \Big[  
+ i \varepsilon^{\perp\rho\nu} a_\rho \delta \A_\nu\Big] \,\subset\, \d\mathcal{S}_{YM}
~.
\eeq
$\F$ is the full non-abelian field strength and $a_\rho$ is defined in eq.\ \eqref{vectornotation}. 
For a given set of generators $\{{\bf t}^a\}$ of the gauge group, we can write $\A=\A^c {\bf t}^c$ and $a=a^c {\bf t}^c$.
Then, both \eqref{ELag} and \eqref{ELaf} can be expressed in terms of the tensor 
\beq
\label{ELafa}
\mathbb{B}^v[ \mathcal{V}, \d\A] = G_{\bold{mn}}\, \varepsilon^{\perp\rho\nu}\, \mathcal{V}^{\, \bold{m}}_\rho\, \delta \A^{\, \bold{n}}_\nu
\eeq
with $\mathcal{V}_\rho=\A_\rho $ for CS, and $\mathcal{V}_\rho=a_\rho $ for YM. We introduced bold indices 
$\bold{m}$ and $\bold{n}$ to describe general quiver gauge theories. Specifically, $\bold{m}=(a,m)$ is a double 
index where $m$ labels the nodes of the quiver and $a$ labels the generators of the gauge group $\mathfrak{S}_m$, 
that refers to the node $m$ of the quiver. Considering orthonormal generators, the matrix $G$ is 
$G_{\bold{mn}}=\d_{ac}\otimes \kappa_{mn}$.

In the orthogonal frame $\{k_\mu,\tilde{k}_\mu\}$ on  $T\mathcal{M}_2'$, we can further decompose $\mathcal{V}$ 
and $\d\A$ along $k$ and $\tilde{k}$ to obtain
\beq\label{vector_form_CS}
\mathbb{B}^v[ \mathcal{V}, \d\A] = -G_{\bold{mn}} \left( \begin{array}{c} \mathcal{V}^{\, \bold{m}}_{\tilde{k}} \\ \mathcal{V}^{\, \bold{m}}_k \end{array}\right)^T 
\left(\begin{array}{cc} 0& +1 \\ -1& 0 \end{array}\right) \left( \begin{array}{c} \d\A_{\tilde{k}}^{\,\bold{n}} \\ \d\A^{\,\bold{n}} _k\end{array} \right)
~.
\eeq 
We used $\varepsilon^{\perp\rho\nu}\tilde{k}_\rho k_\nu =-1$. 

After tracing over the bold indices, $\mathbb{B}^v[ \mathcal{V}, \d\A]$ becomes a $2$-form on the cotangent space 
of $\mathcal{M}_2'$. Equation $\mathbb{B}^v[ \mathcal{V}, \d\A]=0$ is solved by finding appropriate Lagrangian 
submanifolds associated to this $2$-form. Concretely, we may pick any $Sp(2,\mathbb{C})$ matrix with unit 
determinant, call it $M$, and impose the boundary conditions
\beq
\label{CSEL}
(1-M)\d\A=(1-M)\mathcal{V}=0\qquad \forall\, p\in\mathcal{M}_2'~.
\eeq
When $\mathcal{M}_2'$ is endowed with a complex structure, the action of $Sp(2,\mathbb{C})$ has a natural 
interpretation. By construction, these solutions are valid both for CS and YM gauge theories.

We point out that an additional interesting solution of $\mathbb{B}^v[ \mathcal{V}, \d\A]=0$ is available in the case of 
CS theories. In general, the tensor $\kappa_{mn}$ is symmetric, but need not be positive definite. In that case, it 
may have isotropic subspaces. On this subspaces $\mathbb{B}^v[ \mathcal{V}, \d\A]$ vanishes automatically, 
independently of the coordinate dependence of $\mathcal{V}$ and $\d\A$. For example, given an isotropic vector 
$v^m$ such that $v^m \kappa_{mn} v^n=0$, we may consider boundary conditions $\d\A=v^m\d\A^m_\mu dx^\mu$
and $\mathcal{V}=v^m \mathcal{V}^m_\mu dx^\mu$ with arbitrary components $\d\A_\mu$ and $\mathcal{V}_\mu$ 
on $\mathcal{M}_2'$. For a general treatment of such boundary conditions in CS theory we refer the
reader to \cite{Kapustin:2010hk}.

\subsubsection{Scalars}\label{scalar_form_boundary}

In the non-linear sigma model, the variation of $\mathscr{L}_{scalar}$ with respect to $\phi^a$, $\phiT^{\bar c}$, yields the 
result\footnote{We remind the reader that in this, and the next two sections, we are referring to a flat target 
space for which the coefficients $K_{a\bar c}$ are constants independent of the field profiles 
$\phi^a$, $\phiT^{\bar c}$.}
\bea
\label{ELai}
\delta \mathcal{S}_{NL\s} &\supset& -\int_{\mathcal{M}_2'}\, 
\Big( \D^\perp \phiT^{\bar c}\,K_{a\bar c}\,  \delta \phi^a  +  \delta\phiT^{\bar c}\, K_{a\bar c}\,  \D^\perp \phi^a \Big)-
i V^\perp\Big(\begin{array}{l}\frac{1}{2}\end{array} K_a \d\phi^a - \begin{array}{l}\frac{1}{2}\end{array}K_{\bar{c}} \d\phiT^{\bar{c}} \Big)
 \, .
\eea
The term proportional to $V^\perp$ does not contribute, because $V^\perp=0$ at the boundary. 
The first term can be written in compact notation as
\beq\label{scalar_BC_from_action}
\d\bPhi^T G\, \D^\perp\bPhi~,\qquad G=\lp\begin{array}{cc} 0 & K_{a\bar{c}} \\ K_{\bar{c}a}&0\end{array}\rp
\eeq
where $G$ is the target space metric and $\bPhi$ the vector of scalars, introduced in section~\ref{sec_Target_Space}. 
We can set \eqref{scalar_BC_from_action} to zero by assuming that the two vectors $\d\bPhi$ and $\D^\perp\bPhi=0$ 
are orthogonal. The standard way to do this, is to consider Dirichlet, $\d\phi^a=0$, or Neumann, $\D^\perp\phi^a=0$,
boundary conditions (and similarly for the scalars $\phiT$). Notice that in general $\D^\perp$ contains 
non-vanishing normal components of a gauge connection. 

In supersymmetric YM theories, the gauge multiplet contains a real scalar $\s$ in the adjoint representation of the 
gauge group. The variation $\d\s$ of the action yields the boundary term $\Tr\,(\d\s D^\perp\s)$. This term is similar to 
\eqref{scalar_BC_from_action}, and can be set to zero in the same way.

\subsection{Path integral and closure under supersymmetry}\label{closure}

We conclude this section with an additional remark. 
In the ensuing sections \ref{Bound_Sec_1} and \ref{SUSYbcsII} we solve the equations
$\TD^\perp=0$ to obtain half-BPS boundary conditions for general supersymmetric gauge theories. 
This is sufficient for the purposes of the classical problem.

In the quantum problem we are integrating over generic field configurations in a path integral. In the presence of a 
boundary the integration is further restricted to configurations with specific boundary conditions. Consequently, in 
this context the invariance of the path integral with respect to a given symmetry requires that the boundary conditions 
are also invariant under the symmetry in question. In general, this is not automatic and it may lead to further restrictions 
on the boundary conditions. 

Although we are mainly interested in the classical problem in this paper, we will partially address the issue of the
closure of boundary conditions under supersymmetry in the following sections.

\section{Boundary conditions I}\label{Bound_Sec_1}

In this section we address the precise form of $A$-type boundary conditions in general three-dimensional 
non-linear sigma models. A good prototype for this exercise are $A$-type boundary conditions in $2d$ 
$\NN=(2,2)$ non-linear sigma models on the strip that define D-branes in a K\"ahler target space $\mathcal{X}$.
In that case we know, \cite{Hori:2000ck}, that the solution of the $A$-type boundary conditions is describing D-branes
wrapping Lagrangian submanifolds in $\mathcal{X}$. We will describe how similar solutions arise in 
three-dimensional theories. We work out first the case of a flat space background, and then explain how things are
modified when the $3d$ theory is placed on a general curved $A$-type background.

\subsection{Non-linear sigma models}\label{sec_WZ_model}

\subsubsection{General equations}

The action  $\mathcal{S}$ of a supersymmetric non-linear sigma model is specified by a K\"ahler potential $K$, 
a superpotential $W$, and finally the $R$-charges and central charges of the chiral superfields. 
In this subsection $K$ is generic (a flat K\"ahler potential for chiral superfields will be considered in the
ensuing section \ref{SUSYbcsII}). We continue to call the target space $\XX$.

In section~\ref{sec_Lagrangians_N=2} we calculated the variation 
of $\mathcal{S}$ under supersymmetry, and found a generic expression for $\TDWZ^\mu$. Here we 
are interested in solutions of the equations $\TDWZ^\perp=0$ at $\mathcal{M}_2'$. We have 
\bea\nn
 \sqrt{2}\,\TDWZ^\perp&=&	
 					+\ep\Big[ \g^{\perp} \g^\nu \psi^a\,  \D_\nu\phiT^{\bar{c}} 
					-  (r^{\bar{c}} H-z^{\bar{c}})\, \g^\perp \psi^a\, \phiT^{\bar{c}}- i V^\perp \psi^a K^{\bar{c}}
					- i \g^\perp\psiT^{\bar{c}} W^a \Big] K_{a\bar{c}} \nn\\
&& 				-\epT\Big[ \g^\perp\g^\nu \psiT^{\bar{c}}\,  \D_\nu\phi^a	
					-  (r^{a}H- z^{a} )\, \g^\perp\psiT^{\bar{c}}\,  \phi^a + i V^\perp \psiT^{\bar{c}} K^a 
					+ i \g^\perp\psi^a \Wb^{\bar{c}}  \Big] G_{a\bar{c}}~. \rule{0pt}{.5cm} \rule{.8cm}{0pt}
					 \label{ula1}
\eea
The indices $a,\bar{c}$ run from $1$ to $\mathfrak{s}$, where $2\mathfrak{s}$ is the real dimension of the target space $\mathcal{X}$. 
It is convenient to use the identity $\g^\mu\g^\nu=g^{\mu\nu}+\g^{\mu\nu}$ and rewrite
\bea
+\ep\g^{\perp} \g^\nu \psi^a\,  \D_\nu\phiT^{\bar{c}} &=& +\ep\psi^a \D^\perp\phiT^{\bar{c}} + \ep\g^{\perp\nu} \psi^a\,  \D_\nu\phiT^{\bar{c}}~, \label{ula2}\\
-\epT\g^\perp\g^\nu \psiT^{\bar{c}}\,  \D_\nu\phi^a &=& - \epT\psiT^{\bar{c}} \D^\perp\phi^a -\epT\g^{\perp\nu} \psiT^{\bar{c}}\,  \D_\nu\phi^a ~. \label{ula3}
\eea
In equations \eqref{ula1}-\eqref{ula3} we recognize the combinations 
\beq
V^\perp\left[  K_{\bar{c}}(\d\phiT^{\bar{c}})_{susy} - K_a(\d\phi^a)_{susy} \right]\qquad \& \qquad
(\d\phi)_{susy}  \D^\perp\phiT + (\d\phiT)_{susy} \D^\perp\phi~,
\eeq
which appeared in the analysis of $\mathbb{B}^s[\Phi,\d\Phi]$ \eqref{ELai}. 
This is expected because on-shell we can always use the Noether current to rewrite $\TD^\perp$. 

Following the discussion in section~\ref{preview}, we require $\TDWZ^\perp=0$. 
The analysis of this equation reduces naturally to the study of four types of terms: 
\bea
\label{uia}
\TD_1&=& +\left[ \ep\psi^a \D^\perp\phiT^{\bar{c}} - \epT\psiT^{\bar{c}} \D^\perp\phi^a\right] K_{a\bar{c}}~,\\
\label{uib}
\TD_2&=& + \left[ \ep\g^{\perp\nu} \psi^a\,  \D_\nu\phiT^{\bar{c}}  
-\epT\g^{\perp\nu} \psiT^{\bar{c}}\,  \D_\nu\phi^a\right] K_{a\bar{c}}~, \\
\label{uic}
\TD_3&=& -\left[\ep\,\g^\perp \psi^{a}\, \phiT^{\bar{c}} 
-\epT\,\g^\perp \psiT^{\bar{c}}\, \phi^a \right]  K_{a\bar{c}}~,\\
\label{uid}
\TD_4&=&-\left[\ep \g^\perp\psiT^{\bar{c}} W^a + \epT\g^\perp\psi^a \Wb^{\bar{c}} \right] K_{a\bar{c}}
~.
\eea
In order to obtain explicit boundary conditions for the fields that appear in these equations we have to 
disentangle the spinorial and target space structures. The reader can find the details of this computation in 
appendix \ref{bilinearFactor}. Here we outline the main steps. 

Firstly, the anticommuting spinors are decomposed in components using the projectors $\Pr$ and $\Prb$.
As a result, all the geometric information can be packaged into the bilinears \eqref{Bilinears_2}
\beq
\begin{array}{lcl}
\zeta\g^\perp\zeta&\equiv&\Omega\, e^{i\x}~,\\
\zeta\g^\perp\g^{\nu_\parallel}\zeta&=&\Omega\, e^{i\x}\, k^{\nu_\parallel}~,   \rule{0pt}{.6cm} \\
\zetaT\g^\perp\g^{\nu_\parallel}\zeta&=& -i\Omega\, \tilde{k}^{\nu_\parallel}~,  \rule{0pt}{.6cm}
\end{array}
\qquad
\begin{array}{lcl}
\zetaT\g^\perp\zetaT&=&-\Omega\, e^{-i\x}~,\\
\zetaT\g^\perp\g^{\nu_\parallel}\zetaT&=&\Omega\, e^{-i\x}\, k^{\nu_\parallel}~.    \rule{0pt}{.6cm} \\
\zeta\g^\perp\g^{\nu_\parallel}\zetaT&=& -i\Omega\, \tilde{k}^{\nu_\parallel}~.  \rule{0pt}{.6cm}
\end{array}
\eeq
Secondly, we impose the $A$-type projection on the spinors $\ep$ and $\epT$,
\beq
\zetaT\ep
=\epT\zeta
~.
\eeq
Finally, we impose the boundary condition \eqref{ELad} on the spinors, i.e. 
$e^{i\x} (\zetaT\bPsi)^I = M^{I}_{\,K}(\bPsi\zeta)^K$. 
These manipulations introduce the orthogonal matrix $M$ and the phase $\x$ in $\TD_i$.
At the end, the $\TD_i$ depend only on $\zetaT\ep$ and $\bPsi\zeta$. Hence, a bilinear $\ep\bPsi$ 
common in all terms can be factorized out, and the result for $\TDWZ^\perp$ can be understood as a condition 
on the bosons. This is nicely expressed in the matrix notation $\bPhi$ and $\bPsi$ of section~\ref{sec_Target_Space}. 
As a simple example of these manipulations, we obtain 
\beq\nn
\TD_1=+\ep\psi^a K_{a\bar{c}} \D^\perp\phiT^{\bar{c}} - \epT\psiT^{\bar{c}} K_{a\bar{c}} \D^\perp\phi^a
=(\ep\bPsi)^T\lp \frac{1-iJ}{2} + e^{-i\x} M^T\frac{1-iJ}{2}\rp G D^\perp\bPhi ~. 
\eeq
The complete result is
\begin{subequations}
\bea
\TDWZ^\perp=& + (\ep\bPsi)^T & 
						\left[ (\matrUnity -iJ)\right]\, G\, P_M^{(\x,+)}\, \Big[  n^\mu\D_\mu\bPhi  + J\, \tilde{k}^\mu \D_\mu \bPhi \Big] \label{final_result_NLSM1}\\
					&  + (\ep\bPsi)^T  &
									\left[ e^{-i\x} (\matrUnity +iJ) \right]\, G\, P^{(\x,-)}_M\, \Big[\, k^\mu \D_\mu \bPhi +  J\, ( iH)\, \mathcal{R}\, \bPhi - iJ\, \mathcal{Z}\, \bPhi \Big]	\rule{0pt}{.6cm}	\label{final_result_NLSM2}\\
					& - (\ep\bPsi)^T  &\left[ e^{-i\x}(\matrUnity -iJ)\right]\, P_{M^T}^{(-\x,+)}\, J\Big[G\, \bW\Big] ~,	\rule{0pt}{.6cm}  					 \label{final_result_NLSM3}								 
\eea
\label{final_result_NLSM}
\end{subequations}
where the matrix $P_M^{(\x,\pm)}$  is a {\it target space projector} defined as
\bea
\label{curraf}
 \label{projT}
P_{M}^{\lp \x ,\pm \rp } &\equiv& \frac{1}{2} \Big( \matrUnity \pm M[\an] \Big)~,\\
M[\an]&\equiv& M\matrRot[\an]=\matrRot[-\an/2] M \matrRot[\an/2]~, \rule{0pt}{.6cm} \label{rewrite_M_x}\\
\matrRot[\an]&\equiv&  \cos\an\, \matrUnity + \sin\an\, J~. \rule{0pt}{.6cm} 
\eea 
In deriving \eqref{final_result_NLSM} we imposed $\{M,J\}=0$ from which \eqref{rewrite_M_x} follows. 
With this condition, $P_M$ is a projector if $M^2=1$. Collecting all requirements, the matrix $M$ is an 
orthogonal matrix with the properties 
\beq
\label{Mproperties}
M^2=1~,~~ 
\{M,J\}=0
~.
\eeq 
The matrix $\matrRot[\an]$ is the matrix of local $R$-symmetry.

\subsubsection{Solutions in flat space}

Having obtained the general formula \eqref{final_result_NLSM} we are now in position to study solutions to 
equation $\TDWZ^{\perp}=0$. Flat space is of course a special case of our discussion. It is instructive to exhibit
first how Lagrangian `$D$-branes' come out of \eqref{final_result_NLSM} for a theory defined on a euclidean 
$3d$ half-plane. In this case the boundary leaf $\mathcal{M}_2'$ is a 2-plane.

In flat space the profile of the background fields is trivial, and  
the covariant derivative $\D_\mu$ reduces to the standard partial derivative $\partial_\mu$. 
In what follows we will also set, for convenience, $\mathcal{Z}=0$ for the central charges. The role of 
$\mathcal{Z}$ in \eqref{final_result_NLSM2} is the same as that of a real mass obtained by giving a vev to the 
bottom component of a real multiplet coupled to $\bPhi$. We will consider such masses in relation to YM and CS
theories in section \ref{SUSYbcsII}.

Before going into the details of the solution, it is worth emphasizing two simplifying special properties of flat space:
\begin{itemize} 
\item[1)] There is always a choice of coordinates, say $\{\theta,x,\tilde x\}$, such that the frame 
$\{n_\mu,k_\mu,\tilde{k}_\mu\}$ is precisely $\{\partial_\theta,\partial_x,\partial_{\tilde x}\}$. 
The boundary is placed at a fixed value of $\theta$.
\item[2)] The phase $\x$ appearing in $M[\x]$ is a constant. 
\end{itemize}
Both of these features are generically absent in curved space because of the background curvature. 

Focusing on the vanishing of the components \eqref{final_result_NLSM1}-\eqref{final_result_NLSM2}, 
we obtain the conditions
\bea
&& \,\partial_\theta\bPhi\,\in\mathrm{Ker}(1+M[\x])~ \label{condition_flat_Dbrane1} \\
&& \partial_x \bPhi\, \in \mathrm{Ker}(1-M[\x]) \qquad\&\qquad 
\partial_{\tilde x}\bPhi\,\in \mathrm{Ker}(1-M[\x])~. \label{condition_flat_Dbrane2}
\eea
Since $M^2=1$, the eigenvalues of the matrix $M$ are $\pm1$. Moreover, since $\{ M,J\}=0$, the complex structure 
of the target space is a bijection between $\mathrm{Ker}(1-M)$ and $ \mathrm{Ker}(1+M)$. As a result, 
$\mathrm{Ker}(1\pm M)$ is middle dimensional in the target space, and the direct sum 
$\mathrm{Ker}(1-M)\oplus \mathrm{Ker}(1+M)$ is a basis for $\mathcal{TX}$. The submanifold corresponding to 
the distribution $\mathrm{Ker}(1-M)$ is a {\it Lagrangian} submanifold $\Lagr$.\footnote{For the convenience of
the reader we remind that a {\it Lagrangian} submanifold $\Lagr$ (defined on a symplectic manifold 
$(\mathcal{N},\omega)$, where $\omega$ is the symplectic form) is characterized by the two conditions: 
\beq
\omega\big|_{\mathcal{TL}}=0,\qquad \mathrm{dim}\,{\Lagr}=\frac{1}{2}\mathrm{dim}\,\mathcal{N}
~. 
\eeq
When the symplectic manifold $\mathcal{N}$ is K\"ahler, the Riemaniann metric $G_{IJ}$ can be used to 
characterize $\Lagr$, and the definition just given is equivalent to the condition 
\beq
\mathcal{TL}^\perp=J\,\mathcal{TL},\qquad \mathcal{TL}^\perp
=\{ \vec{v} \in \mathcal{TN}\, |\, v^I G_{IJ}w^J=0\,\,\forall\, \vec{w}\in\mathcal{TL}\}
~.
\eeq
} 
The effect of the matrix $\matrRot[\an]$ is to change the orientation of the Lagrangian submanifold by a 
constant angle $\x$. 

The Lagrangian submanifold just described contains $\bPhi(\mathcal{M}_2')$, the image of $\mathcal{M}_2'$ 
under the maps $\bPhi$. Both $M$ and the derivatives of $\bPhi$ are objects in $\mathcal{TX}$. The solutions \eqref{condition_flat_Dbrane1}-\eqref{condition_flat_Dbrane2} 
transform correctly under a change of coordinates in the target space. 
Locally, we may take a chart such that the Lagrangian submanifold is described by mixed Dirichlet and Neumann 
boundary conditions. We impose Neumann boundary conditions along the directions parallel to the submanifold, and 
Dirichlet conditions along the directions transverse to the submanifold. 

In the simplest situation, in which $\mathcal{X}$ is an affine vector space and the K\"ahler potential is canonical, 
the Neumann and Dirichlet boundary conditions can be seen explicitly by solving 
\eqref{condition_flat_Dbrane1}-\eqref{condition_flat_Dbrane2}. This is done by considering a basis 
$\{v_i \}_{i=1}^{\mathfrak{s}}$ of ~$\mathrm{Ker}(1+M)$, and writing 
$\bPhi= \sum_{i=1}^{\mathfrak{s}} \big[ f_i v^i + g_i Jv^i\big]$ with $f_i$ and $g_i$ functions of the coordinates. 
The solution to \eqref{condition_flat_Dbrane1} is $\partial_\theta g_i=0$ at the boundary, i.e.\ Neumann boundary 
conditions along the direction of the submanifold. The generic solution to \eqref{condition_flat_Dbrane2} is 
$f_i=f_i(\theta)$, and therefore $f=const$ at the boundary, i.e.\ Dirichlet boundary conditions in the direction 
transverse to the submanifold. The worldvolume of $\mathcal{L}$ is along the span of $\{Jv_i\}_{i=1}^{\mathfrak{s}}$. 
The case with $\x\neq0$, is solved by rotating the fields accordingly with the projector. The latter can be written as
\beq\label{bcond_on_W}
P_M^{(\x,\pm)}= \matrRot[-\an/2] \frac{(1\pm M)}{2} \matrRot[\an/2]~,
\eeq
and the solution is $\bPhi\equiv\matrRot[-\an/2]\bPhi'$, where $\bPhi'$ satisfies the Neumann/Dirichlet boundary 
conditions that depend on $M$. 

Along similar lines consider the boundary conditions derived from the superpotential term, namely the equation 
that arises by requiring the last term \eqref{final_result_NLSM3} to vanish,
\beq\label{SuperP_BC_1}
P_{M^T}^{(-\x,+)}\, J\,G\, \bW=\matrRot[\an/2] (1+M^T) \matrRot[-\an/2]\, J\,G\, \bW = 0
~.
\eeq
In this case the projector depends on $M^T\matrRot[-\an]$, in agreement with $R$-symmetry considerations. 
The vector $\bW$ was defined in section \ref{sec_Target_Space}, and in the complex basis it has components 
$W^a= K^{a\bar{c}}\partial_{\bar{c}}\Wb$, $\widetilde{W}^{\bar{c}}= K^{\bar{c}a}\partial_aW$.
Since $W=\mathrm{Re}\,W+ i \mathrm{Im}\,W$ is a holomorphic function of the fields, 
the Cauchy-Riemann equations imply the relations
\bea
\left[\begin{array}{c} \partial_m\, \mathrm{Re}\, W \\ \partial_{m+\numb}\, \mathrm{Re}\, W\end{array}\right]  =
									J\, \left[\begin{array}{c} \partial_m\, \mathrm{Im}\, W \\ \partial_{m+\numb}\, \mathrm{Im}\, W\end{array}\right],\qquad 
\frac{\partial}{\partial\phi^m} W=\frac{\partial}{\partial\Phi^m} \mathrm{Re}\, W - i \frac{\partial}{\partial\Phi^{m+\numb}} \mathrm{Re}\, W~.					
\eea
The quantity $G\, \bW$ is 
\beq
G\,\bW=\left[\begin{array}{c} \partial_m\,\mathrm{Re}\,W(\bPhi) \\ 
\partial_{m+\numb}\, \mathrm{Re}\,W(\bPhi) \end{array}\right]
~,
\eeq 
where $\partial_i$ is shorthand notation for $\partial_i=\partial/\partial\Phi^i$. Implementing the rotation 
$\bPhi=\matrRot[-\an/2]\bPhi'$, we obtain from \eqref{SuperP_BC_1} the projection equations
\beq\label{bc_ImW}
(1+M^T)\left[\begin{array}{c} \partial'_m\, \mathrm{Im}\, W(\bPhi') \\ \partial'_{m+\numb}\, \mathrm{Im}\, W(\bPhi')\end{array}\right]=0,
\eeq
where $\partial'={\partial}/{\partial\Phi'}$. Because $\mathrm{Ker}(1\pm M)$ span the tangent space 
$T\mathcal{M}$, and $J$ is a bijection between these two kernels, we can understand the boundary condition 
\eqref{bc_ImW} by considering the action of $v^T(1+M^T)$ and $(Jv)^T(1+M^T)$ on $\partial'\mathrm{Im}\, W(\bPhi')$, 
for any $v\in\mathrm{Ker}(1+M)$. By definition $v^T(1+M^T)=0$, thus only $(Jv)^T(1+M^T)$ is non-trivial. The latter 
can be calculated explicitly $(Jv)^T(1+M^T)= 2 (Jv)^T$, and from \eqref{bc_ImW} we obtain the boundary condition
\beq
(Jv)^I \partial'_I \mathrm{Im}\, W(\bPhi')=0~, 
\eeq
which translates into the statement that $\partial'_I \mathrm{Im}W(\bPhi')$ has no component along the span of 
$\{Jv_i\}_{i=1}^{\mathfrak{s}}$ and therefore $\mathrm{Im}W(\bPhi')$ is constant along the wordvolume of the 
submanifold $\Lagr$.

\subsubsection{Solutions in curved space}\label{sol_curv_NLSM}

In the previous section, we solved the equations $\TDWZ^\perp=0$ relying on two special features of flat space: 
the fact that the phase $\x$ is constant, and the fact that there is a coordinate-adapted orthogonal basis 
in $T\mathcal{M}_3$. In curved space we do not expect in general these two features to hold. 

For example, in the case of the ellipsoid in toric coordinates with background fields \eqref{ellipsoidprofile}
\beq
\label{ellipsoidprofile2}
H=\pm\frac{ i}{g_{\theta\theta}},\qquad 	
A^{(R)}_{\pm} = - \frac{1}{2} \left( 1 - \frac{\tilde{\ell}}{g_{\theta\theta}}\right) d\phi_1 
\mp \frac{1}{2} \left( 1 - \frac{\ell}{g_{\theta\theta}}\right) d\phi_2
~,
\eeq 
we find $\x_\pm =\psi$ with frame vectors
\bea\label{useful_about_ktilde_squashS3}
n_{\pm}^\mu\,\partial_\mu&=&-\frac{1}{g_{\theta\theta}} \partial_\theta~, \\
k^\mu_{\pm}\, \partial_\mu&=& \pm  \tilde{\ell}^{-1}\partial_{\phi_1}+ {\ell}^{-1}\partial_{\phi_2}~,\rule{0pt}{.5cm}\\
\tilde{k}^\mu_{\pm}\partial_\mu&=& 
\mp\, \cot(2\theta)\, k^\mu_{\pm}\partial_\mu \, \pm \,\frac{1}{\sin(2\theta)} \left( \frac{1}{\ell} \partial_{\phi_2} \mp \frac{1}{\tilde{\ell}} \partial_{\phi_1}\right)~.
\rule{0pt}{.5cm}
\label{ktilde_sec_NLSM}
\eea

Consider now a more general manifold $\mathcal{M}_3$ in toric coordinates $(\theta,\phi_1,\phi_2)$, in similar notation
to the one above for the ellipsoid. By definition, the Killing vector $k=\frac{1}{\Omega}\partial_\psi$ is expressed
as a combination of $\partial_{\phi_1}$ and $\partial_{\phi_2}$, and $\x$ is only a function of $\psi$. 
The triple of vectors $(k^\mu, n^\mu, \tilde{k}^\mu)$ takes the form
\beq\label{torictriple}
k= \frac{1}{\Omega} \partial_\psi~, ~~
n=f_{n}\, \partial_\theta~,~~ 
\tilde{k}=\tilde{f}\, \partial_\psi + v^\mu\partial_\mu
~.
\eeq 
The functions $\tilde{f}$, $f_n$ and  $v^\mu$ depend on the details of the background, however, the integrability 
condition \eqref{Frobenius_constr} implies $[v^\mu\partial_\mu,\partial_\psi]=0$.
$\mathcal{M}_3$ is decomposed in solid tori as before, $\mathcal{M}_3\cong\mathcal{T}_1\#\mathcal{T}_2$, and 
the fields are restricted on one of the solid tori, call it $\mathcal{T}$ for simplicity.

In this case, the general solution of $\TD^\perp=0$ has (see \eqref{final_result_NLSM1}, \eqref{final_result_NLSM2})
\bea
&& \,\partial_\theta\bPhi\,\in\mathrm{Ker}(1+M[\x])~ \label{condition_flat_Dbrane11} \\
&& \D_\psi \bPhi +  J\, ( iH)\, \mathcal{R}\, \bPhi\, \in \mathrm{Ker}(1-M[\x])~,\qquad \label{condition_flat_Dbrane21}
 \tilde{k}^\mu \D_\mu \bPhi\,\in \mathrm{Ker}(1-M[\x])~. 
\eea
In the first line we used, for illustration purposes, the simplifying assumption $A_\perp^{(R)}=0$, which clearly
holds for the example of the ellipsoid \eqref{ellipsoidprofile2}. The covariant derivatives in 
\eqref{condition_flat_Dbrane21} are
\bea
\tilde{k}^\mu \D_\mu\bPhi&=& \tilde{f}\, \partial_\psi\bPhi + v^\mu \partial_\mu \bPhi + \tilde{k}^\mu A_\mu^{(R)}\, J \mathcal{R}\, \bPhi ~,\\
\label{use_twisting_relation_backg}
k^\mu \D_\mu\bPhi + J (iH) \mathcal{R} \bPhi &=& 
k^\mu\partial_\mu\bPhi + \Big[ k^\mu A_\mu^{(R)} + ( iH + k^\mu V_\mu) \Big] J \mathcal{R}\, \bPhi ~.\rule{0pt}{.6cm}
\eea
Regardless of whether the phase $\x$ is constant or coordinate dependent $\partial_\theta\, \matrRot[\an]=0$. 
Thus, we can always solve \eqref{condition_flat_Dbrane11} with Neumann boundary conditions along the directions 
of the submanifold $\mathrm{Ker}(1-M[\x])$. The solution of the other two equations instead depends on $\x$. 

Consider first the case of background fields where $\x$ is constant. As we explained in section 
\ref{sec_twisting_phases}, this can be achieved from a general background with a gauge transformation of 
the original $A_\mu^{(R)}$ to a new $R$-symmetry background $A^{(R)}_{new}$. 
In that case, the term that appears inside the parenthesis on the r.h.s.\ of equation \eqref{use_twisting_relation_backg}, 
with the substitution $A^{(R)}\rightarrow A^{(R)}_{new}$, vanishes because of the condition we found in \eqref{twisting_relation_back}. Consequently, we obtain as in flat space
\beq
k^\mu \D_\mu\bPhi + J (iH) \mathcal{R} \bPhi\,\Big|_{twisted} = k^\mu\partial_\mu\bPhi \in \mathrm{Ker}(1-M[\x])~.
\eeq
The analysis of $\tilde{k}^\mu \D_\mu\bPhi$ requires more detailed knowledge of $\tilde{f}$ and 
$\tilde{k}^\mu A^{(R)}_{\mu\,new}$. To be concrete, in the case of the geometries introduced in section 
\ref{SUSY_EXAMPLES} we obtain the following expressions:
\begin{itemize}
\item For the ellipsoid,  $A_{\mu\, new}^{(R)}$ and its scalar product with $\tilde{k}^\mu$, given in \eqref{ktilde_sec_NLSM}, are
\bea
A^{(R)}_{\pm\, new} &=& A^{(R)}_{\pm}-\frac{1}{2} (d\phi_1 \pm  d\phi_2)= \frac{1}{2g_{\theta\theta}}( \tilde{\ell} d\phi_1\pm \ell d\phi_2),\\
\tilde{k}^\mu A_{\mu\, new}^{(R)} \Big|_{\pm}&=& \pm \cot(2\theta)\, iH_{\pm}~. \rule{0pt}{.6cm}
\eea
The function $\tilde{f}$ is also proportional to $\cot(2\theta)$. 
\item For the circle bundles of section \ref{circle_fibration_sec}, we find
\bea
A_{\mu\, new}^{(R)} dx^\mu&=& - \lp \frac{\cos\theta}{2 g_{\theta\theta}} + \frac{\beta^2}{2} \frac{u(\theta)}{g_{\theta\theta}}  \frac{u'(\theta)}{\sin\theta}\rp d\varphi
-\frac{\beta^2}{2 g_{\theta\theta}} \frac{u'(\theta)}{\sin\theta}\, d\psi~, \\
\tilde{k}&=&\frac{2}{\sin\theta}  \lp u(\theta) \partial_\psi - \partial_\varphi\rp~, \rule{0pt}{.6cm}\\
\tilde{k}^\mu A_{\mu\, new}^{(R)}&=& \frac{\cot\theta}{g_{\theta\theta}}~. \rule{0pt}{.6cm}
\eea
The function $\tilde{f}$ is proportional to $u(\theta)/\sin\theta$.
\end{itemize}

We notice that the boundary condition from $\tilde{k}^\mu\D_\mu\bPhi$ simplifies when the boundary is placed 
at the {\it equator} of the corresponding geometries, because at that point $\tilde{k}^\mu A_{\mu\, new}^{(R)}=0$. 
When this happens, the covariant derivative $\tilde{k}^\mu\D_\mu\bPhi$ becomes a combination of partial derivatives, 
and again we can solve the boundary conditions as in flat space. Namely, we impose Neumann boundary conditions 
along the directions parallel to the submanifold, and Dirichlet for the directions transverse to the submanifold. The value 
of $\tilde{f}$ at the equator is not important in this statement. When the boundary is placed away from the 
equator a more complicated boundary condition \eqref{condition_flat_Dbrane21} has to be imposed.

In more general setups, a background $\mathcal{M}_3$ exhibits a coordinate-dependent phase $\x$. In that case 
the boundary equations \eqref{condition_flat_Dbrane11}-\eqref{condition_flat_Dbrane21} 
\bea
&& \,\partial_\theta\bPhi\,\in\mathrm{Ker}(1+M[\x])~,\\
&& \D_\psi \bPhi +  J\, ( iH)\, \mathcal{R}\, \bPhi\, \in \mathrm{Ker}(1-M[\x])~,\qquad 
 \tilde{k}^\mu \D_\mu \bPhi\,\in \mathrm{Ker}(1-M[\x]) 
\eea
do not exhibit any simplification in the covariant derivatives, and the boundary conditions are functionals of both 
the derivatives and the values of the fields. As a result, the direct geometric meaning of the boundary conditions
in target space, that was present in backgrounds with constant $\x$, is now lost. Nevertheless, one can still 
solve the boundary conditions by diagonalizing $M[\x]$, for a given choice of $M$, and arranging the combinations  
\eqref{condition_flat_Dbrane11}-\eqref{condition_flat_Dbrane21} to belong to $\mathrm{Ker}(1\pm M[\x])$. Since 
$[M,J]\neq0$, the eigenvectors of $M[\x]$ are not the ones of $M=M[\x=0]$. Consequently, as we move along the 
orbit of the Killing vector, or more generically, along the boundary $\mathcal{M}_2'$, these eigenvectors change 
according to their $\x$ dependence.

\vspace{0.3cm}

\subsection{Real multiplets}\label{Real_Mult_sec}

Before we tackle general gauge theories, there is another comparatively simple example we would like to discuss.
It is well-known that in $3d$ flat space there is a simple duality between a chiral superfield and an abelian gauge 
field.\footnote{In the simplest case, the duality is obtained by considering 
$\int d^4\theta( \Sigma^2 - \Sigma(\Phi+\bPhi))$, where $\Phi$ is a chiral superfield and $\Sigma$ a generic superfield. 
Integrating out $\Phi$ constrains $\Sigma$ to be a real superfield and produces a $U(1)$ gauge theory. Alternatively,
integrating out $\Sigma$ gives the action of a chiral superfield. It is interesting to reconsider this exercise in curved 
spaces.} 
We expect the corresponding boundary conditions to be mapped trivially under this duality. With this in mind,  
in this subsection we present $A$-type boundary conditions for $\NN=2$ theories of $\mathfrak{s}$ abelian vector 
superfields interacting via a constant target space metric $G$.

The supersymmetric variation $\TD^\mu$ is expressed most conveniently in terms of the real parametrization of 
the abelian vector superfields in \eqref{YM_SuperV_real}:
\bea\label{real_multi_var}
\TD^\mu_{real}&=&  
-\frac{1}{2}\Big(\ep \g^\mu\g^\rho\psi^a_\Sigma\, \lp a_\rho -i\partial_\rho\s\rp^c  
-\ep \g^\mu\psi_\Sigma^a\, \big (i D + (iH)\s \big)^c  \Big) G_{ac}
\nn\\
& &  \rule{1cm}{0pt} + \frac{1}{2} \Big(\epT \g^\mu\g^\rho\psiT^a_\Sigma\, \lp  a_\rho+i\partial_\rho\s\rp^c
+ \epT \g^\mu\psiT^a_\Sigma\, \big(i D + (iH)\s \big)^c \Big)G_{ac}\ ~.
\eea
We can further rearrange $\TD^\perp_{real}$ by borrowing results from the study of $\TD^\perp_{\NLs}$ 
in the previous section. In particular, let us define the two complex combinations: 
$\partial_\rho\phiT_{\Sigma}\equiv a_\rho-i\partial_\rho\s$ 
and $\mathrm{Im}\,\varphi_\Sigma\equiv (D+(H)\s)$, $\mathrm{Re}\,\varphi_\Sigma\equiv 0$. 
Then, we can rewrite $\TD^\perp_{real}$ as
\bea
\TD^\mu_{real}&=&  
-\frac{1}{2}\Big(\ep \g^\mu\g^\rho\psi^a_\Sigma\, \partial_\rho\phiT_{\Sigma}^c  
+\ep \g^\mu\psi_\Sigma^a\, \widetilde{\varphi}_{\Sigma}^c  \Big) G_{ac}
+ \frac{1}{2} \Big(\epT \g^\mu\g^\rho\psiT^a_\Sigma\,\partial_\rho\phi_{\Sigma}^c  
+ \epT \g^\mu\psiT^a_\Sigma\, \varphi_{\Sigma}^c \Big)G_{ac}\ ~,\nn
\eea
and with an obvious change of variables, it is clear that we have obtained an expression that is essentially the sum 
of $\TD_1$, $\TD_2$ and $\TD_3$, given in \eqref{uia}, \eqref{uib}, \eqref{uic}, respectively. 

Consequently, the surface term $\TD^\perp_{real}$ takes the suggestive form
\bea\label{Supervariation_Perp_YM}
\begin{array}{ccl}
\TD_{real}^\perp=& - (\ep\bPsi_\Sigma)^T & 
				\left[ (\matrUnity -iJ) G\right]P_M^{(\x,+)}\,  \lp n^\mu  \left[\begin{array}{c} a_\mu \\ \partial_\mu\s \end{array}\right] 
				+ J\, \tilde{k}^\mu  \left[\begin{array}{c} a_\mu \\ \partial_\mu\s \end{array}\right] \rp \\
	&  - (\ep\bPsi_{\Sigma})^T  &
	\left[ e^{-i\x} (\matrUnity +iJ) G\right] P^{(\x,-)}_M\, \lp 
	k^\mu  \left[\begin{array}{c} a_\mu \\ \partial_\mu\s \end{array}\right] -  \left[\begin{array}{c} iD + (iH) \s \\ 0 \end{array}\right] \rp
	\rule{0pt}{.8cm} \\
\end{array}				
\eea
where the matrix $M$ fixes the spinor boundary conditions $e^{i\x} \zetaT\bPsi_{\Sigma}= M \bPsi_\Sigma \zeta$. 
From the definition of $a_\mu=-j_\mu-\s V_\mu$, and the fact that $V^\perp=0$, we finally obtain the boundary 
conditions
\bea
&& n^\mu  \left[\begin{array}{c} j_\mu \\ \partial_\mu\s \end{array}\right]\ \in \mathrm{Ker} (1 + M[\an]) ~,\quad
\left\{ \tilde{k}^\mu \left[\begin{array}{c} a_\mu \\ \partial_\mu\s \end{array}\right]~,~ k^\mu \left[\begin{array}{c} j_\mu 
\\ \partial_\mu\s \end{array}\right]\right\} \ \in \mathrm{Ker}(1-M[\an])~,\nn\\
&& D-i\s( (iH)+ k^\mu V_\mu)=0~.
\eea
The last condition is correctly invariant under the shift symmetry \eqref{shift_inv}. 
Assuming $\tilde{k}^\mu V_\mu=0$, the boundary conditions for $j_\mu$ and $\partial_\mu\s$ 
are arranged as those of a neutral chiral multiplet.

\subsection{Closure under supersymmetry}

As we noted in subsection \ref{closure} the boundary conditions may transform non-trivially under supersymmetry. 
We would like to know if the boundary conditions that were formulated above are invariant under the $A$-type
supersymmetries, and if not, whether invariance can be restored by imposing further constraints.
Since the boundary conditions on the fermions are algebraic, it is immediately possible to examine how things
work in some generality. In particular, when the matrix $M$ is field independent we find that supersymmetry 
invariance of the fermion boundary conditions does not impose any new constraints.

In contrast, the analysis of the transformation of the boson boundary conditions is
more involved and case-dependent. Since the boson boundary conditions involve derivatives of the bosons, 
their transformation leads to expressions that involve derivatives of the corresponding fermions. The details
of the resulting expressions depend on the specifics of the differential operators and, in general, have
to be analyzed case by case. For that reason, and in order to keep the discussion as generic as possible, 
in what follows we will concentrate mostly on the transformation properties of the fermion boundary conditions.

\subsubsection*{Chiral and anti-chiral multiplets}

The supersymmetry transformations of the fermions $(\psi, \psiT)$ in a chiral multiplet are
\beq
\begin{array}{ccl}
\d\psi_\a&=&+\,\vartheta~F\, \zeta_\a   + i \, \widetilde{\vartheta} ~\Big[\Big( k^\mu\D_\mu\phi - ir (iH)\phi \Big)\zetaT_\a - (\widetilde{U}^\mu\D_\mu\phi)\zeta_\a \Big]~,\\
\d\psiT_\a&=& +\,\widetilde{\vartheta}~\FT\, \zetaT_\a   + i \, {\vartheta} ~\Big[\Big( k^\mu\D_\mu\phiT + ir (iH)\phiT \Big)\zeta_\a - ({U}^\mu\D_\mu\phiT)\zetaT_\a \Big]~. \rule{0pt}{.6cm}
\end{array}
\eeq
For $A$-type supersymmetries, we set $\theta=\widetilde{\theta}$. We want to examine how $A$-type supersymmetry
transforms the boundary conditions $e^{i\x} \zetaT \bPsi=M\, \bPsi\zeta$. Assuming for simplicity that the matrix $M$
is invariant we only need to consider the bilinears $\delta \bPsi \zeta$, $\zetaT \delta \bPsi$. Straightforward 
manipulations yield the scalar products
\bea
\d\bPsi\zeta\ &=&
+i \frac{1+i J}{2} \Big( k^\mu\D_\mu + r (iH)J\Big) \bPhi + \frac{1-iJ}{2} \bold{F} - i\frac{1-iJ}{2} e^{+i\x}(n^\mu-ik^\mu) \D_\mu\bPhi ~, \\
\zetaT\d\bPsi\ &=&     
+ i \frac{1-iJ}{2} \Big( k^\mu\D_\mu + r (iH) J\Big)\bPhi   +\frac{1+iJ}{2} \bold{F}  +i\frac{1+iJ}{2}e^{-i\x}(n^\mu+ik^\mu)\D_\mu\bPhi ~.
\eea
Consequently, the condition $e^{i\x} \zetaT\d\bPsi=M\, \d\bPsi\zeta$ holds if the following equations are satisfied
\beq\label{check_BC_Lagra_ferm}
\begin{array}{l}
(1+iJ)(1-M \matrRot[\an]) \Big( k^\mu\D_\mu + r (iH)J\Big) \bPhi=0~,\\
(1-iJ)(1+M\matrRot[\an]) (n^\mu \D_\mu\bPhi - J \tilde{k}^\mu\D_\mu\bPhi )=0~,\rule{0pt}{.6cm}\\
(1-iJ)(1-M\matrRot[-\an])  \bold{F} =0~.\rule{0pt}{.6cm}
\end{array}
\eeq

In these formulae we recognize the boundary conditions that we derived previously for the bosons. 
As a minor difference comparing \eqref{check_BC_Lagra_ferm} to the original boundary condition 
\eqref{final_result_NLSM1}, we notice a sign change in front of the term $J\tilde{k}^\mu\D_\mu\bPhi$. 
This sign difference, however, is irrelevant in the final boundary conditions, since the two terms in the second
equation in \eqref{check_BC_Lagra_ferm} have to vanish independently. We note that both 
$n^\mu \D_\mu\bPhi $ and $J \tilde{k}^\mu\D_\mu\bPhi$ belong in $\mathrm{Ker}(1+M[\x])$ and their relative 
normalization is not fixed by the boundary conditions.

We conclude that the supersymmetry invariance of the fermion boundary conditions does not impose any new
constraints when $M$ is separately invariant. In the more general case of a field dependent $M$ one needs
to consider extra contributions from the supersymmetric variation of $M$.

Finally, regarding the variation of the bosons at the boundary, it is possible to prove in complete generality the 
orthogonality condition $\d\bPhi\, G\, \D^\perp\bPhi=0$. From the $A$-type supersymmetry, the definition of 
$\d\phi$ and $\d\phiT$, and the boundary condition on the spinors $e^{i\an} \zetaT\bPsi=\bPsi\zeta$, we deduce 
the boundary variation
\beq
\d\bPhi = \frac{1}{2} \Big( (1+iJ) + (1-iJ) e^{-i\an}M\Big)\ep\bPsi = P_M^{(\an,+)} \frac{1+iJ}{2} \ep\bPsi~.
\eeq 
Therefore, $P_M^{(\an,-)}\d\bPhi=0$ and $\d\bPhi$ belongs to $\mathrm{Ker}(1-M[\an])$. From the orthogonality of the 
two kernels  $\mathrm{Ker}(1\pm M[\an])$, the condition $\d\bPhi\, G\, \D^\perp\bPhi=0$ follows. Let us notice that 
on-shell this orthogonality condition corresponds to  $\mathbb{B}^s[\d\bPhi,\bPhi]=0$ (see \eqref{ELai}).

\subsubsection*{Real multiplets}

The supersymmetry transformations $\d\psi_{\Sigma}$ and $\d\psiT_{\Sigma}$ in a real multiplet are very 
similar to the ones of the chiral fermions $\d\psi$ and $\d\psiT$. The only difference is the contribution of the $D$-term 
\beq\nn
\begin{array}{ccl}
\d\psi_{\Sigma}&=& \vartheta\,\Big[ \big[ D-i \s (iH + k^\mu V_\mu)- i k^\mu (j_\mu +i \partial_\mu\s)\big]\zetaT + i e^{-i\x} (n^\mu+i\tilde{k}^\mu) (a_\mu -i \partial_\mu\s) \zeta  \Big]~, \\
\d\psiT_{\Sigma}&=& \vartheta\, \Big[ \big[ D-i \s (iH + k^\mu V_\mu)- i k^\mu (j_\mu -i \partial_\mu\s)\big]\zeta- i e^{+i\x} (n^\mu-i\tilde{k}^\mu) (a_\mu +i \partial_\mu\s) \zetaT  \Big]~.\rule{0pt}{.7cm}\
\end{array}
\eeq
Repeating the evaluation of $e^{i\x} \zetaT\d\bPsi_{\Sigma}=M\, \d\bPsi_\Sigma \zeta$ we obtain results similar to the
chiral multiplet case. Also in this case supersymmetry invariance of the fermion boundary conditions does not impose 
any further constraints.

\section{Boundary conditions II}\label{SUSYbcsII}

In this section we study $A$-type boundary conditions in general (non-abelian) $\NN=2$ supersymmetric 
CS/YM-matter theories. The corresponding actions on curved backgrounds and their supersymmetric variations 
$\TD^\mu$ were obtained in section \ref{Gauge_Actions}. Our analysis recovers previously known results in 
special cases, e.g.\ flat space, and extends them to general $A$-type backgrounds $\TT$ with a solid torus topology.

We discuss first the conditions arising from the supersymmetric variation in the 
gauge sector. The corresponding conditions in the matter sector are presented separately.

\subsection{Gauge sector}\label{sc_gauge_sector}\label{sec_gaugesec_I}

\subsubsection{Description and summary of results}

From the supersymmetric variation of the Yang-Mills and Chern-Simons actions, respectively, we obtain the 
boundary terms
\bea\label{gsec1}
\gYM^2\, \TD^\perp_{YM}&=&  \Tr\Big[ -\frac{i}{4} \ep\,\g^\perp\g^\rho\psi_{\Sigma}\, ( \Fh_\rho +2i \s V_\rho) 
									+\frac{i}{2}\ep\,\g^\perp\g^\rho\psi_{\Sigma}\,\partial_\rho\s 
											\nn\\
						& & \rule{1cm}{0pt} +\frac{i}{4} \epT\,\g^\perp\g^\rho\psiT_{\Sigma}\,( \Fh_\rho +2i \s V_\rho) 
								+ \frac{i}{2}\epT\,\g^\perp\g^\rho\psiT_{\Sigma}\,\partial_\rho\s 
											\nn\\
						&&		 \rule{1.5cm}{0pt}			+\frac{1}{2}  \ep\,\g^\perp\psi_{\Sigma}\, (iD+\s ( iH))
						+\frac{1}{2}\epT\,\g^\perp\psiT_{\Sigma}\, (iD+\s ( iH)) \Big]~, \\
						\nn\\
\TD^\perp_{CS}
&=&+\frac{i}{4\pi}\kappa_{\bf ac}\, \Big[ \varepsilon^{\perp\nu\rho}\,( \ep\g_\rho\psi_{\Sigma}^{\bf a}  - \epT\g_\rho\psiT_{\Sigma}^{\bf a})\A_\nu^{\bf c}
+ 2 (\ep\g^\perp\psi_{\Sigma}^{\bf a}  + \epT\g^\perp\psiT_{\Sigma}^{\bf a})\s^{\bf c} \Big]~.	\label{gsec2}							\eea
In the gauge sector analysis we will also include a term coming from the vector-matter couplings 
\beq\label{gsec3}
\TD^\perp_{matter} 
\supset  -i \langle\phiT\, {\bf t}^a \phi\rangle \big[ \ep\g^\perp\psi_{\Sigma}^a+ \epT\g^\perp\psiT_{\Sigma}^a\big]~.
\eeq
$\{{\bf t}^a\}$ is a basis for the generators of the gauge groups in play, and $\langle\phi\, {\bf t}^a \phiT\rangle$ denotes 
the action of the adjoint fermions $\lm=\lm^a {\bf t}^a$ and $\lmb=\lmb^a {\bf t}^a$ in the representation of each 
of the matter fields $\phi$ and $\phiT$. Recall that we use bold indices $\bold{a}$ to describe general quiver gauge
theories. In the multi-index $\bold{a}=(a,m)$ $m$ labels the nodes of the quiver theory, and $a$ the generators of the 
gauge group $\mathfrak{S}_m$ at the node $m$. For any set $\{{\bf t}^a\}$ of generators we set 
$\Tr[ {\bf t}^a {\bf t}^c]=G_{ac}$, and $\kappa_{\bold{ac}}=G_{ac}\otimes \kappa_{mn}$. For canonically  
normalized generators  $G_{ac}=\d_{ac}$. 

The expressions in \eqref{gsec1}-\eqref{gsec3} are a collection of all the terms in 
$\TD^\perp_{YM+matter}$ and $\TD^\perp_{CS+matter}$ that are functions of the spinors $\psi_{\Sigma}$ and 
$\psiT_{\Sigma}$ of the $\NN=2$ vector multiplets. In what follows, we refer to the sum of \eqref{gsec1} and
\eqref{gsec3} as $\TD^\perp_{gs\, YM}$, and the sum of \eqref{gsec2} and \eqref{gsec3} as $\TD^\perp_{gs\, CS}$
($gs$ stands for gauge sector).\\   

To analyze the supersymmetric variations $\TD^\mu_{gs\, YM}$ and  $\TD^\mu_{gs\, CS}$ we need to disentangle 
the geometric and spinorial structures. This can be achieved, as before, by using the projectors $\Pr$, $\Prb$, and the 
A-type projection on $\ep$ and $\epT$. On the anti-commuting spinors $\bPsi_{\Sigma}=(\psi_{\Sigma},\psiT_{\Sigma})$ 
we impose the general boundary condition 
\beq\label{psiSigma_BC}
e^{i\an} \zetaT\bPsi_{\Sigma}=M\,\bPsi_{\Sigma}\zeta
~.
\eeq 
Supersymmetry will soon fix some of the properties of the matrix $M$, as we found for the non-linear sigma 
model in sec.\ \ref{Bound_Sec_1}. Nevertheless, the case of the non-linear sigma model and the case of the general 
gauge theory discussed here exhibit conceptually different properties. Let us highlight the origin of these differences. 

In the boundary condition \eqref{psiSigma_BC}, the spinors of the vector multiplets $\psi_{\Sigma}$, $\psiT_{\Sigma}$
have been arranged conveniently as a doublet $\bPsi_{\Sigma}$. The same doublet can be formed in non-linear 
sigma models out of the fermions in the chiral superfields. In that case we can also form naturally a 
corresponding doublet of bosons $\bPhi=(\phi,\phiT)$. This is also possible 
for abelian real superfields, where the role of $\bPhi$ is played by the complex combination of the dual photon and 
the real scalar $\s$. In the case of a non-abelian gauge theory, however, there is no obvious natural bosonic $\bPhi$ 
that we can associate to $\bPsi_{\Sigma}$. 
As a result, we cannot proceed identically to the non-linear sigma model case thinking in terms of a 
generalized target space structure on the gauge indices. 

An alternative approach is suggested by the 2-form
\beq\label{recall_B}
\BB^v[ \mathcal{V}, \d\A] = G_{\bold{mn}}\, \varepsilon^{\perp\rho\nu}\, \mathcal{V}^{\, \bold{m}}_\rho\, \delta \A^{\, \bold{n}}_\nu~,
\eeq
that appears in the on-shell boundary value problem for vectors. In \eqref{recall_B} both $\mathcal{V}$ and $\d\A$ 
are $1$-forms on the boundary. For example, $\BB^v$ appears in the Euler-Lagrange variation of 
CS theories, \eqref{ELafa}, as well as in the supersymmetric variation $\TD^\perp_{CS}$, \eqref{CSUaa},
\beq\label{cs1_gs}
\BB^v[\d\A,\A]\ \propto\ \kappa_{\bf a c}\,\varepsilon^{\perp\nu\rho}\,\d\A^{\bf a}_\rho\A^{\bf c}_\nu \ =\ \kappa_{\bf a c}\,\varepsilon^{\perp\nu\rho}\,(  \epT\g_\rho\psiT_{\Sigma}^{\bf a}- \ep\g_\rho\psi_{\Sigma}^{\bf a}  )\A_\nu^{\bf c}~. 
\eeq
In this equation $\BB^v$ couples the two components of $\A_\nu$ in the boundary directions to a combination of 
the spinors. It is therefore natural to think in terms of doublets distinguished
by the spacetime indices of vectors parallel to the boundary. 

Similar manipulations can be employed in $\TD^\perp_{YM}$ using the identity $\g^\mu\g^\nu=g^{\mu\nu}+\g^{\mu\nu}$ 
to rewrite the kinetic terms as follows
\begin{subequations}
\bea
 \TD^\perp_{YM}\supset&   + \frac{1}{4\gYM^2}\, G_{ac}\, \varepsilon^{\perp\nu\rho}\, ( \ep\g_\rho\psi^a_{\Sigma} -\epT\g_\rho\psiT^a_{\Sigma}) \Fh^c_\nu 
      			&-\frac{i}{4\gYM^2}\, G_{ac}\, ( \ep\psi^a_{\Sigma} - \epT\psiT^a_{\Sigma} )\, g^{\perp\nu} \Fh^c_\nu \label{ym1_gs}\\
			&		   - \frac{1}{2\gYM^2}\, G_{ac}\, \varepsilon^{\perp\nu\rho}( \ep\g_\rho\psi^a_{\Sigma} +\epT\g_\rho\psiT^a_{\Sigma}) \D_\nu\s^c 
						&+\frac{i}{2\gYM^2}\, G_{ac}\, ( \ep\psi^a_{\Sigma} + \epT\psiT^a_{\Sigma} )\, g^{\perp\nu} \D_\nu\s^c~. \rule{0pt}{.7cm}
\label{ym2_gs}
\eea
\end{subequations}
The first two terms in \eqref{ym1_gs} and \eqref{ym2_gs} have the same structure as $\BB^v$ in \eqref{recall_B}
(up to a difference in $\pm$ signs).

Introducing the notation
\beq\label{source_Ipm}
\mathfrak{I}_{\pm}(\bPsi,\mathcal{V})\equiv \frac{1}{2} \varepsilon^{\perp\nu\rho}\big[\ep\g_\rho\psi^{\bf a} \pm \epT\g_\rho\psiT^{\bf a} \big] G_{\bf ac} \mathcal{V}_\nu^{\bf c}
\eeq
we show in the next subsection (see eqs.\ \eqref{jminus}, \eqref{jplus}) that 
$\mathfrak{I}_{\pm}(\bPsi_{\Sigma},\mathcal{V})$ is closely related to 
$\BB^v[ \ep\bPsi_{\Sigma}, P_U \mathcal{V}]$,
where $P_U$ is a certain projector depending on a matrix $U$ that has only gauge indices and satisfies 
$U^TGU=G$ and $U^2=1$. 
The interplay between A-type supersymmetry and the geometry of the form $\BB^v$ fixes the relation between $U$ and $M$ by
setting
\beq\label{recap_BC_lam}
M=\Id_{2\times 2}\otimes U=\lp\begin{array}{cc} U&0\\0&U\end{array}\rp~,\qquad
e^{i\an} \zetaT\psi_{\Sigma}^a = U^a_c \psi_{\Sigma}^c\zeta~,\qquad 
e^{i\an} \zetaT\psiT_{\Sigma}^a = U^a_c \psiT_{\Sigma}^c \zeta~.
\eeq
Note that unlike the boundary conditions in the non-linear sigma model case, \eqref{ELad}, \eqref{Mproperties}, in 
\eqref{recap_BC_lam} the gauge indices of $\psi_\Sigma$ and $\psiT_\Sigma$ do not mix.

With these boundary conditions and the standard, by now, manipulations on spinor bilinears we arrive at compact
expressions for $\TD^\perp_{gs\, CS}$ and $\TD^\perp_{gs\, YM}$. In order to keep the notation simple and most
transparent, let us quote the pertinent results in the case of a single gauge group. In Chern-Simons-matter theories we 
obtain 
\beq
\TD^\perp_{gs\, CS}= \frac{\kappa}{2\pi} \Big[ 
\left( P_U \AA \right)^a G_{ac}
- i\Big(  \s^a G_{ac}- \, \frac{2\pi}{\kappa}\langle\phi\, {\bf t}^a \phiT\rangle \Big)\, U^a_{\, c}\, \Big] \big[  \ep\psi_{\Sigma}^c +  \epT\psiT_{\Sigma}^c\big]~,
			\label{final_perp_cs}
\eeq
where $P_U\A=\tilde{k}^\mu \A_\mu +i U k^{\mu} \A_\mu$.
In the case of Yang-Mills theories
\bea
\TD^\perp_{gs\, YM}& = & 
+\frac{i}{2\gYM^2}\, \Big[ 
-\frac{i}{2} \left( P_U \hat \FF \right)^a
G_{ac}\, +  \D_\perp\s^a\, G_{ac}\,  \nn\\
&&			\rule{1.5cm}{0pt}		-  U^a_{\ b}\big( D^b- i\s^b (\, iH + k^\mu V_\mu\,)\big)  G_{ac} + 2{\bf e}^2\, U^a_{\ c} \langle\phi\, {\bf t}^a \phiT\rangle\Big] \big[ \ep\psi_{\Sigma}^c + \epT\psiT_{\Sigma}^c\big] \nn \\			
& &		  +\frac{1}{2\gYM^2}\, G_{ac}\,\Big[  j^a_\perp  - (P_U \DD\sigma)^a
\Big] 	  \big[ \ep\psi^c_{\Sigma} - \epT\psiT^c_{\Sigma} \big]~, \label{final_perp_ym}
\eea
where 
\bea
P_U\Fh= \tilde{k}^\mu (\Fh_\mu+2i\sigma V_\mu) +i U k^{\mu} \Fh_\mu~,&\qquad &
P_U\D\s=\tilde{k}^\mu\D_\mu \s + i U k^{\mu} \D_\mu\s~,\\
j_{\perp}=-\frac{i}{2} \varepsilon_{\perp\nu\rho}\F^{\nu\rho}= -\frac{i}{2} \Fh_\perp~. & &
\eea
When the gauge group has an abelian component, a FI term can also be added to the Lagrangian. Since the variation 
of this term is of the type
\beq
\TD^\perp_{FI}=
+\frac{1}{2}  \xi ( \ep\g^\perp\lmb- \epT\g^\perp\lm)=
-\frac{i}{2} \xi \big[ \ep\g^\mu\psiT_{\Sigma} + \epT\g^\mu\psi_{\Sigma}\big]
~, 
\eeq
we can easily include the FI parameters in \eqref{final_perp_ym} by considering the shift $D\rightarrow D-\xi$. 

In summary, without assuming any further constraints on the spinors $\bPsi_{\Sigma}$ other than 
\eqref{recap_BC_lam}, the most generic boundary conditions on the bosonic fields of the gauge multiplet are
\bea
\mathrm{CS-theories}:&\rule{.2cm}{0pt} & P_U\A  -  i U\big( \s -  \, \frac{2\pi}{\kappa}\,  \langle\phi\, {\bf t}\,\phiT\rangle \big) =0\label{cs_cond_sg1}~,\\
\mathrm{YM-theories}:&\rule{.2cm}{0pt} \rule{0pt}{.8cm} & D_\perp\s  \begin{array}{c}-\frac{i}{2} \end{array} P_U\Fh -
										U\big( D- i\s (\, iH + k^\mu V_\mu\,)\big)  + 2{\bf e}^2\, U \langle\phi\, {\bf t}\, \phiT\rangle=0\label{cs_cond_sg_21}\\
						&\rule{0cm}{0.7cm} & ~j_\perp - P_U\D\s=0~. \label{cs_cond_sg_22}
\eea

As a special solution, one can further impose $P_U\mathcal{V}=0$ both in CS and YM theories. 
In the next subsection we show that this is equivalent to requiring $\mathfrak{I}_\pm=0$.
This projection, which is natural from the point of view of the Euler-Lagrange variations
in Chern-Simons theory \eqref{CSEL}, selects a Lagrangian submanifold of $\BB^v$, as we explain in the next 
section. The remaining conditions yield:
\begin{itemize}
\item
In the case of CS theory, \eqref{cs_cond_sg1} reduces to the algebraic equation of motion of the auxiliary field $D$,
\beq
\d\mathscr{L}_{CS-matter} \supset \lp -\frac{\kappa}{2\pi}  \s^a + \langle\phi\, {\bf t}^a \phiT\rangle\rp \d D^a=0~.
\eeq
\end{itemize}
\begin{itemize}
\item In the case of YM, the condition $\Fh_\perp=0$ translates into  $\varepsilon_{\perp\mu\nu}\F^{\mu\nu}=0$, 
where the free indices are constrained to run over the boundary indices by anti-symmetry. Then, $\Fh_\perp=0$ 
is satisfied if the non-abelian connection is flat at the boundary, namely $\F=0$ at the boundary. In components, 
the boundary condition on $\s$ becomes
\beq
\partial_\perp\s^a-i[\A_\perp,\s]^a=U^a_{\ b}\big( D^b- i\s^b (\, iH + k^\mu V_\mu\,)\big) - 2{\bf e}^2 \langle\phi\, U^a_{\ b}{\bf t}^b \phiT\rangle~.
\eeq
\end{itemize}

\subsubsection{Technical details}\label{sec_gaugesec_II}

Let us elaborate further on the details that led to the above boundary conditions. The key quantity is 
$\mathfrak{I}_{\pm}(\bPsi,\mathcal{V})$ defined in \eqref{source_Ipm}. We re-express this quantity using the 
$A$-type projection on $\ep$ and $\epT$. Leaving the label $\Sigma$ of the spinors 
implicit, the resulting expression is 
\bea\nn
\begin{array}{l}					
 \mathfrak{I}_{\pm}(\bPsi,\mathcal{V})=\\
\qquad +\frac{1}{2} \frac{\zetaT\ep}{\Omega}\  \varepsilon^{\perp\nu\rho} \tilde{k}_\rho k_{\nu} \Big[ \big[ -e^{i\x} \zetaT\psi^{\bf m} \mp e^{-i\x} \psiT^{\bf m}\zeta\,\big]\, (i \mathcal{V}_k)  	
	- \big[ +\psi^{\bf m}\zeta \pm \zetaT\psiT^{\bf m}\, \big] \, \mathcal{V}^{\bf n}_{\tilde{k}}\ \Big] G_{\bf mn} ~,    \rule{0pt}{.7cm}								
\end{array}\\
\label{manip_Ipm}
\eea
or equivalently in matrix notation (with $\varepsilon^{\perp\nu\rho}\tilde{k}_\rho k_\nu =+1$)
\beq
\mathfrak{I}_{\pm}=
+ \frac{1}{2} \frac{\zetaT\ep}{\Omega}\  
G_{\bf mn}\ \left( \begin{array}{c}  - e^{i\x} \zetaT\psi^{\bf m} \mp e^{-i\x} \psiT^{\bf m}\zeta \\  + \psi^{\bf m}\zeta \pm \zetaT\psiT^{\bf m}  \rule{0pt}{.5cm} \end{array}\right)^T
 \lp \begin{array}{cc} 0 & +\Id \\ -\Id & 0 \end{array}\rp\, 
\left( \begin{array}{c}  \mathcal{V}^{\bf n}_{\tilde{k}} \\  (i \mathcal{V}^{\bf n}_k)  \rule{0pt}{.5cm} \end{array} \right)~.\label{combo_Ipm}
\eeq 
It is clear that $\mathfrak{I}_{\pm}$ is similar in form to $\BB^v$ evaluated on specific complex combinations of 
the components of $\mathcal{V}$ and the spinors. We mentioned in sec.~\ref{sec_vectors_form} that the most 
general solution to the equations $\BB^v=0$, are the Lagrangian submanifolds of the two-form $\BB^v$. 
In special cases the general $A$-type boundary conditions \eqref{cs_cond_sg1}-\eqref{cs_cond_sg_22} are 
solved by these Lagrangian submanifolds. We proceed to examine this aspect more closely.

Starting with $\mathfrak{I}_-$, which appears in the CS case, we notice that we can rewrite the fermions in \eqref{combo_Ipm} as follows
\bea
\nn
&&\left( \begin{array}{c}  - e^{i\x} \zetaT\psi^{\bf m} + e^{-i\x} \psiT^{\bf m}\zeta \\   \rule{0pt}{.5cm} 
+ \psi^{\bf m}\zeta - \zetaT\psiT^{\bf m}  \end{array}\right)= \\
&& -  \left(\begin{array}{cc} e^{+i\x/2} & 0 \\ 0 & e^{-i\x/2} \end{array}\right)
				      \, (\zetaT\bPsi)^{\bf m}\, e^{i\x/2} 
 +   \left(\begin{array}{cc} e^{-i\x/2} & 0 \\ 0 & e^{+i\x/2} \end{array}\right)\,  \s_1\, 
 					(\bPsi\zeta)^{\bf m}\, e^{-i\x/2}
					 ~.\\ \nn
					 \label{vector_ferm_CS}
\eea
Then, imposing the boundary condition 
$
e^{i\an} \zetaT\bPsi= M\, \bPsi\zeta,
$
on the spinors $\psi$ and $\psiT$ we obtain 
\bea
\frac{1}{2} \frac{\zetaT\ep}{\Omega}
 \left( \begin{array}{c}  - e^{i\x} \zetaT\psi^{\bf m} + e^{-i\x} \psiT^{\bf m}\zeta \\   \rule{0pt}{.5cm} + \psi^{\bf m}\zeta - \zetaT\psiT^{\bf m}  \end{array}\right)
&=& 
-  \frac{1}{2} \lp \Id -\s_1  \,e^{+i \s_3 \frac{\x}{2} } \, M^{-1} e^{-i \s_3 \frac{\x}{2}}\rp\,  e^{+i\s_3\frac{\x}{2}} \, ({\ep\bPsi})\, e^{i\x/2} ~.\nn 
\eea
The last expression can be written as a projector $P_M^{-}$ acting on $\ep\bPsi$ with  
\bea
P_M^{\pm} &\equiv& +\frac{1}{2} \lp \Id \pm \s_1  \,e^{+i \s_3 \frac{\x}{2} } \, M^{-1} e^{-i \s_3 \frac{\x}{2}}\rp~.
\eea
The matrix $M$, which acts on the doublet $\bPsi=(\psi^{\bf m} ,\psiT^{\bf m} )$, is of the general form 
$M=R_{(2\times 2)}\otimes U$, where $U$ acts on the gauge indices and $R$ is a $2$-by-$2$ matrix. 
$P^{\pm}_M$ is a projector only if $R=\pm\Id$. Choosing $R=+\Id$ for concreteness, (the $R=-\Id$ choice is very 
similar), $P_M$ becomes
\beq
P_U^{\pm} = +\frac{1}{2} \lp \Id \pm \s_1\otimes U \rp\,  ~,
\eeq
and the matrix $U$ is required to be orthogonal with respect to $G$, and to satisfy $U^2=1$. 
The quantity $\mathfrak{I}_-$ takes the final form
\bea
\label{jminus}
\mathfrak{I}_- &=&  + e^{i\x/2}\,e^{+i\s_3\frac{\x}{2}}\, \left( \begin{array}{c}  \ep\psi^{\bf m'}  \\   \rule{0pt}{.5cm} \ep\psiT^{\bf m'}  \end{array}\right)^T
					(i\s_2)\, 
					\lp\begin{array}{cc} \Id & U_{\bf m'}^{\bf{m}} \\ U_{\bf m'}^{\bf{m}} & \Id \end{array}\rp\, G_{\bf mn}
 				\, \left( \begin{array}{c} \mathcal{V}^{\bf n}_{\tilde{k}} \\ i \mathcal{V}^{\bf n}_{k}  \rule{0pt}{.5cm}  \end{array} \right)\, 	~,\\
(i\s_2)&=&\lp \begin{array}{cc} 0 & +\Id \\ -\Id & 0 \end{array}\rp~.
\eea
Since $U$ is orthogonal with respect to $G$, the condition $\mathfrak{I}_-=0$ can be achieved by setting
\beq\label{BC_Lagra_tangent}
\mathcal{V}^{\bf n}_{\tilde{k}}+ U^{\bf n}_{\ \bf c}\  i\mathcal{V}^{\bf c}_k =0~.   
\eeq
Notice that the dependence on the phase $\x$ has disappeared in the above manipulations. 

The calculation of $\mathfrak{I}_+$ proceeds along similar lines. In this case, the relevant combination of spinors in
\eqref{combo_Ipm} can be recast as
\bea
\frac{1}{2} \frac{\zetaT\ep}{\Omega}
\left( \begin{array}{c}  - e^{i\x} \zetaT\psi^{\bf m} - e^{-i\x} \psiT^{\bf m} \zeta \\  + \psi^{\bf m}\zeta + \zetaT\psiT^{\bf m} \rule{0pt}{.5cm} \end{array}\right)&=&
 -\s_3  ( \Id + \s_1\, e^{+i \s_3 \frac{\x}{2} }\,  M^{-1} e^{-i \s_3 \frac{\x}{2}})\, e^{+i\s_3\frac{\x}{2}}\, (\ep\bPsi)\, e^{i\x/2}\, ~,  \nn
\eea
yielding the final expression
\bea
\label{jplus}
\mathfrak{I}_+ = +e^{i\x/2} \,e^{+i\s_3\frac{\x}{2}}\,  \left( \begin{array}{c}  \ep\psi^{\bf m'}  \\   
\rule{0pt}{.5cm} \ep\psiT^{\bf m'}  \end{array}\right)^T\, 
					\s_3\,(i\s_2)\,\lp\begin{array}{cc} \Id & U_{\bf m'}^{\bf{m}} \\ U_{\bf m'}^{\bf{m}} & \Id \end{array}\rp\, G_{\bf mn}
			\left( \begin{array}{c} \VV^{\bf n}_{\tilde{k}} \\ i \VV^{\bf n}_{k}  \rule{0pt}{.5cm}\end{array} \right)~.
\eea

In conclusion, with $R=+\Id$, both conditions $\mathfrak{I}_{+}=0$ and $\mathfrak{I}_{-}=0$ lead to
\eqref{BC_Lagra_tangent}. Considering instead the choice $R=-\Id$ would lead to 
$\mathcal{V}^{\bf n}_{\tilde{k}}- U^{\bf n}_{\ \bf c}\  i\mathcal{V}^{\bf c}_k =0$. Clearly, this choice  
is equivalent to the substitution $U\rightarrow - U$.\\

We noted in the previous subsection that by setting $\mathfrak{I}_{\pm}=0$ in $\TD^\perp_{CS}$ or $\TD^\perp_{YM}$
we are led to a special solution of the boundary conditions \eqref{cs_cond_sg1}-\eqref{cs_cond_sg_22} where 
$P_U \VV=0$:
\begin{itemize}
\item In CS theories, where $\VV = \AA$, this is equivalent to a single boundary condition on the gauge field 
\beq\label{BC_Lagra_tangent_CS}
\mathcal{A}^{\bf n}_{\tilde{k}}+ U^{\bf n}_{\ \bf c}\  i\mathcal{A}^{\bf c}_k  =0~.
\eeq
\item In YM theories, where $\mathcal{V}=\Fh, \D\s$, one obtains two separate boundary conditions: one on 
the non-abelian field strength and another one on $\D_\mu\s$
\beq\label{BC_Lagra_tangent_YM}
\Fh^{\bf n}_{\tilde{k}}+ U^{\bf n}_{\ \bf c}\  i\Fh^{\bf c}_k= \tilde{k}^\mu \D_\mu \s^{\bf n} + U^{\bf n}_{\ \bf c}\  i k^\mu\D_\mu\s^{\bf c}=0~.
\eeq
These equations are natural covariant generalizations of corresponding boundary conditions in flat space 
that set components parallel to the boundary of the dual field strength $\hat \FF_\mu$ and $\DD_\mu \sigma$ to zero.
\end{itemize}

\subsection{Matter Sector}

Next we focus on terms that arise from the supersymmetric variation of the matter sector of the gauge theory.
These terms are functions of the spinors $\psi$ and $\psiT$ of the chiral and anti-chiral multiplet. The
relevant boundary contributions can be summarized in the expression
\bea 
&& \sqrt{2}\,\TD^\perp_{matter}=	
 					+\ep\Big[ \g^{\perp} \g^\nu \psi^{\bf a}\,  \D_\nu\phiT^{\bf \bar{c}} 
					-  (r^{\bf \bar{c}} H-(z^{\bf \bar{c}}-q^{\bf \bar{c}} \s))\, \g^\perp \psi^{\bf a}\, \phiT^{\bf \bar{c}}\,- i V^\perp \psi^{\bf a} \phiT^{\bf \bar{c}} + i\,\g^\perp \psiT^{\bf \bar{c}} F^{\bf a}
					\Big] G_{\bf a\bar{c}} \nn\\					
&&~~~~ 				-\epT\Big[ \g^\perp\g^\nu \psiT^{\bf \bar{c}}\,  \D_\nu\phi^{\bf a}
					-  (r^{\bf a}H- (z^{\bf a}-q^{\bf a}\s) )\, \g^\perp\psiT^{\bar{c}}\,  \phi^{\bf a} + i V^\perp \psiT^{\bar{c}} \phi^{\bf a} - i\, \g^\perp\psi^{\bf a}   \FT^{\bf \bar{c}}\,
					\Big] G_{\bf a\bar{c}} ~.\rule{0pt}{.5cm} \rule{.8cm}{0pt}					 \label{matter_V}				
\eea
The effects of a gauge invariant superpotential $W$ can be incorporated, as already done in \eqref{WZgen1}, 
by considering the on-shell relations 
\beq
G_{a\bar{c}} \FT^{\bar{c}} = \partial_a W~, \qquad
F^a G_{a\bar{c}} =  \partial_{\bar{c}} \Wb~. 
\eeq
In \eqref{matter_V} the chiral and anti-chiral superfields transform in arbitrary representations of the gauge group. In 
the bold multi-indices ${\bf a}=(a,m)$, $a$ is a color index and $m$ a flavor index. The metric $G$ is the scalar 
product in the combined flavor/color index space. In non-abelian theories $\s$ acts on $\phi$ $(\phiT)$ and $\psi$ 
$(\psiT)$ according to their representations. We will make a slight abuse of notation where the specifics of this action 
are suppressed. 

By making use of the standard identity $\g^{\perp}\g^\nu=g^{\perp\nu}+i\, \varepsilon^{\perp\nu\rho}\g_\rho$ and the 
fact that $V^\perp=0$ at the boundary, $\TD^\perp_{matter}$ can be rewritten in the form:
\bea 
 \sqrt{2}\,\TD^\perp_{matter}&=&	
 +i \Big[ \varepsilon^{\perp\nu\rho} \ep\g_\rho\psi^{\bf a} \D_\nu\phiT^{\bf \bar{c}} 
						+ (iH) \ep\g^\perp \psi^{\bf a}  \,  r^{\bf \bar{c}} \phiT^{\bf \bar{c}} \Big] G_{\bf a\bar{c}}  \nn\\
&&		  \rule{.5cm}{0pt}  +\Big[\ep\psi^{\bf a}\,  \D^\perp\phiT^{\bf \bar{c}} 
					+ (z^{\bf \bar{c}}-q^{\bf \bar{c}} \s)\, \ep\g^\perp \psi^{\bf a}\, \phiT^{\bf \bar{c}}\Big] G_{\bf a\bar{c}} 
					 + i\,\ep\g^\perp \psiT^{\bf \bar{c}}\, G_{\bf \bar{c}a}\, F^{\bf a} \nn\\
&& 				-  i \Big[ \varepsilon^{\perp\nu\rho}\epT \g_\rho \psiT^{\bf \bar{c}}\, \D_\nu\phi^{\bf a}
					+  (iH) \epT \g^\perp \psiT^{\bf \bar{c}}\, r^{\bf a } \phi^{\bf a} \Big]G_{\bf \bar{c} a} \nn  \rule{0pt}{.7cm} \\
&&		  \rule{.5cm}{0pt} - \Big[  \epT \psiT^{\bf \bar{c}}\,  \D^\perp\phi^{\bf a}
					+ (z^{\bf a}-q^{\bf a}\s )\, \epT\g^\perp\psiT^{\bar{c}}\,  \phi^{\bf a} \, \Big]G_{\bf \bar{c} a}
					+ i\, \epT\g^\perp\psi^{\bf a} \,G_{\bf a\bar{c}}   \FT^{\bf \bar{c}}~.\qquad
					 \label{Vmatter_gauge}		
\eea

The analysis of $\TD^\perp_{CS}$ and $\TD^\perp_{YM}$ selected boundary conditions in the gauge sector on the 
basis of the two form $\BB^v$. Even though $\TD^\perp_{matter}$ can still be thought of as $\TD_{NLs}^\perp$, 
on the basis of the flavor indices, in this subsection we will not follow the approach of section \ref{Bound_Sec_1}. 
Instead, we will explore the extension of the manipulations of the previous subsection \ref{sec_gaugesec_I} to 
the matter sector.
Accordingly, we assume from the start the following boundary conditions on the matter fermions
\beq\label{matter_CS_YM_bc_sp}
\frac{\zeta\g^\perp\zeta}{\Omega} (\zetaT\bPsi) =  M (\bPsi\zeta)~,\qquad  M= \lp\begin{array}{cc}  \S& 0 \\ 0 & \widetilde{\S}  \end{array}\rp~.
\eeq
The matrices $\S$ and $\widetilde{\S}$ act on the representation space of the matter. They are required to have 
the properties $\S^2=\widetilde{\S}^2=1$, and $\S G \widetilde{\S}=G$. $M$ acts diagonally on the doublet 
$\bPsi=(\psi^{\bf a},\psiT^{\bf \bar{c}})$. This is the same type of ansatz that emerged in the gauge sector. Here 
two possibly different $\S$ and $\widetilde{\S}$ are allowed because of the two chiralities. 

With standard manipulations of the spinor bilinears, we recast $\TD^\perp_{matter}$ in terms of two independent 
spinor components $\ep\psi$ and $\epT\psiT$,
\bea
\label{V_matter_general}
 \sqrt{2}\,\TD^\perp_{matter}&=&
		+\ep\psi^{\bf a}\, G_{\bf a\bar{c}}\ \Big[ 
		 i P_{\widetilde{\S}}\, \D\phiT^{\bf \bar{c}} 
		+ \big( \D^\perp \phiT^{\bf \bar{c}} -\widetilde{\S} ^{\bf \bar{c}}_{\, \bf \bar{n}} (z^{\bf \bar{n}}-q^{\bf \bar{n}}\s) \phiT^{\bf \bar{n}}\big)
											+ i e^{-i\x} \FT^{\bf\bar{c}} \Big] \nn\\
&&		+\epT\psiT^{\bf \bar{c}}\, G_{\bf \bar{c}a}\ \Big[  
		  i P_{\S}\, \D\phi^{\bf a}
		 	- \big(   \D^\perp\phi^{\bf a} - \S^{\bf a}_{\, \bf m}(z^{\bf m}-q^{\bf m} \s)\phi^{\bf m} \big) + i e^{i\x} F^{\bf a} \Big]~.													
\eea
We defined 
\bea
P_{\S}\, \D\phi^{\bf a} &\equiv& \D_{\tilde{k}}\phi^{\bf a}  +i \S^{\bf a}_{\, \bf m} (\D_k \phi^{\bf m}- i r^{\bf m} (iH)\phi^{\bf m})~, \\
P_{\widetilde{\S}}\, \D\phiT^{\bf \bar{c}}&\equiv&
\D_{\tilde{k}}\phiT^{\bf \bar{c}}  +i \widetilde{\S} ^{\bf \bar{c}}_{\, \bf \bar{n}}  (\D_k \phiT^{\bf \bar{n}}+ i r^{\bf \bar{n}} (iH)\phiT^{\bf \bar{n}})~.
\eea
The projectors $P_S$ and $P_{\widetilde{S}}$ are the analog of $P_U$ in the gauge sector. For matter charged under the $R$-symmetry, we see that the terms in $\TD^\perp_{matter}$ proportional to the $R$-charges, $\pm r (iH)$, correctly combine with the covariant derivatives along the Killing vector.   

The expression \eqref{V_matter_general} allows us to read off the following general boundary conditions on 
the matter sector
\bea
\mathrm{chiral}:&\rule{.5cm}{0pt}&  i P_{\widetilde{\S}}\, \D\phiT^{\bf \bar{c}} 
		+ \big( \D^\perp \phiT^{\bf \bar{c}} -\widetilde{\S} ^{\bf \bar{c}}_{\, \bf \bar{n}} (z^{\bf \bar{n}}-q^{\bf \bar{n}}\s) \phiT^{\bf \bar{n}}\big)
											+ i e^{-i\x} \FT^{\bf\bar{c}}=0~,\qquad \label{bc_matter_1}\\
\mathrm{anti-chiral}:&\rule{.5cm}{0pt} \rule{0pt}{.8cm}  &i P_{\S}\, \D\phi^{\bf a}
		 	- \big(   \D^\perp\phi^{\bf a} - \S^{\bf a}_{\, \bf m}(z^{\bf m}-q^{\bf m} \s)\phi^{\bf m} \big) + i e^{i\x} F^{\bf a}=0~.\qquad \label{bc_matter_2}
\eea

A special solution of these boundary conditions is obtained by imposing the Lagrangian condition 
$P_S \D\phi= P_{\widetilde{S}}\D\phiT=0$. Setting to zero the remaining terms in \eqref{bc_matter_1}, 
\eqref{bc_matter_2} we obtain 
\beq\label{bc_Dper_F}
 \D^\perp \phiT -\widetilde{\S}  ({z}-q\s) \phiT + i e^{-i\x} \FT=0~,\qquad
  \D^\perp\phi - \S(z-q \s)\phi- i e^{i\x} F=0~.
\eeq 
The term $(z-q\s)$ in \eqref{bc_Dper_F} corresponds to the standard real mass. 
By taking $S=\Id$ in flat space,
the Lagrangian condition $P_S \D\phi=0$ would become $D_+\phi=0$, with $D_+$ 
a covariantized holomorphic derivative along the coordinates of the boundary, which in that case would be a plane. 
Thus,  $P_S \D\phi= P_{\widetilde{S}}\D\phiT=0$ are the natural generalization to curved space of such 
boundary conditions. 

Comparing $P_S$ and $P_{\widetilde{S}}$ with the projector $P^{(\an,\pm)}_M$ in a non-linear sigma model, 
we observe that the boundary conditions in  \eqref{bc_matter_1} and \eqref{bc_matter_2}  are rather different 
from those describing a Lagrangian brane in the target space.  Also, the projector $P^{(\an,\pm)}_M$ was found to be 
$\an$ dependent, whereas $\an$ plays no role in $P_S$ and $P_{\widetilde{S}}$;
it only enters \eqref{bc_matter_1} and \eqref{bc_matter_2}  through $F$ and $\FT$.  

The covariant derivatives in $P_S$ and $P_{\widetilde{S}}$ contain both the 
dynamical gauge fields $\A$ and the $R$-symmetry connection $A_\mu^{(R)}$. 
As simple illustrating cases, consider the following examples. 
The covariant derivative normal to the boundary, $\D^\perp$, simplifies under
the additional assumption $A_\perp^{(R)}=0$, and reads $\D^\perp=\partial^\perp- i q \A_\perp$. 
For the covariant derivatives along the boundary, $\D_k$ and $\D_{\tilde{k}}$, 
we can borrow part of the discussion in section~\ref{sol_curv_NLSM} to understand their precise form. 
In the case of $A$-type backgrounds with twisted spinors, the twisted $R$-symmetry gauge field is such that 
$\D_k$ becomes
\bea
\label{covspecial_1}
\D_k \phi^{\bf m}- i r^{\bf m} (iH)\phi^{\bf m}&=& k^\mu\partial_\mu\phi^{\bf m} -i k^\mu(\A_\mu\phi)^{\bf m}~.
\eea
For the ellipsoid and the manifolds with $SU(2)\times U(1)$ symmetry that we introduced in 
section~\ref{SUSY_EXAMPLES}, we may also use $\tilde{k}^\mu A_\mu^{(R)}=0$ at the equator to obtain 
$\D_{\tilde{k}}$ in the simplified form
\bea
\label{covspecial_2}
\D_{\tilde{k}}\phi^{\bf a}&=& f\, k^\mu\partial_\mu\phi^{\bf m} + v^\mu\partial_\mu\phi^{\bf m} -i \tilde{k}^\mu(\A_\mu\phi)^{\bf m}~.
\eea
In that case the boundary condition $P_S\D\phi=0$ reads
\beq
\label{covspecial_3}
P_S\D\phi=\big[ f\, k^\mu\partial_\mu\phi+ v^\mu\partial_\mu\phi -i \tilde{k}^\mu(\A_\mu\phi)\big] 
+ i S  \big[  k^\mu\partial_\mu\phi - i k^\mu(\A_\mu\phi)\big] =0~.
\eeq
A similar result holds for $P_{\widetilde{S}}\D\phiT=0$.\footnote{The action of $S$ on $(\A_\mu\phi)^{\bf m}$ 
and of $\widetilde{S}$ on $(\A_\mu\phiT)^{\bf \bar{n}}$ should not be 
confused with the separate action of $U$ that was defined in the gauge sector.}

The expressions \eqref{covspecial_1}-\eqref{covspecial_3} hold under a set of simplifying assumptions for the 
background fields. For a generic $A$-type background, the full convariant derivatives, including the $R$-symmetry 
gauge fields, should be considered. 

Finally, let us obverse that the two boundary conditions $P_S\D\phi=0$ and $P_{\widetilde{S}}\D\phiT=0$ 
are genuinely complex, and the reality condition $\phiT=\phi^\star$ would impose either 
$\D_k\phi=\D_{\tilde{k}}\phi=0$ or $S^{\star}=-\widetilde{S}$. In the latter case, the boundary conditions in 
\eqref{bc_Dper_F} would decompose further into $(z-q\s)\phi=0$ and $\D^\perp\phi- i e^{i\an}F=0$.

\subsection{Closure under supersymmetry}

We conclude the analysis of the above boundary conditions, both in the gauge and the matter sector, 
with a study of their transformation under supersymmetry. We already looked at this problem when 
we discussed the boundary conditions of Lagrangian branes, 
and similar comments continue to apply here. In particular, the variation of the boundary conditions on the fermions are 
algebraic, and it is immediate to check whether they are closed under supersymmetry or not.
For CS theories, the variation of the boundary conditions on $\A_\mu$ and $\s$ are also simple and both turn out to 
be algebraic. In YM theories, the boundary conditions on the bosons are boundary conditions on the derivatives, hence 
their analysis requires specific information about the details of the background.

\subsubsection{Gauge sector}

The boundary conditions on the fermions $\psi_{\Sigma}$ and $\psiT_{\Sigma}$ are
\beq
e^{i\x}\, \zetaT\psi^{\bf a}_{\Sigma}\, =\, U_{\,\bf m}^{\bf a}\, \psi_{\Sigma}^{\bf m}\zeta~,\qquad e^{i\x}\, \zetaT\psiT^{\bf {c}}\, =\, U_{\,\bf n}^{\bf c}\, \psiT^{\bf n}\zeta~.
\eeq
The supersymmetric variation of the fermions $\d\psi_{\Sigma}$ and $\d\psiT_{\Sigma}$ under the $A$-type supersymmetry, $\theta=\widetilde{\theta}$, is
\beq\nn
\begin{array}{ccl}
\d\psi_{\Sigma}&=& \vartheta\,\Big[ \big[ D-i \s (iH + k^\mu V_\mu)- i k^\mu (j_\mu +i \partial_\mu\s)\big]\zetaT + i e^{-i\x} (n^\mu+i\tilde{k}^\mu) (a_\mu -i \D_\mu\s) \zeta  \Big]~, \\
\d\psiT_{\Sigma}&=& \vartheta\, \Big[ \big[ D-i \s (iH + k^\mu V_\mu)- i k^\mu (j_\mu -i \partial_\mu\s)\big]\zeta- i e^{+i\x} (n^\mu-i\tilde{k}^\mu) (a_\mu +i \D_\mu\s) \zetaT  \Big]~,\rule{0pt}{.7cm}\
\end{array}
\eeq
and the conditions we would like to check (assuming the matrix $U$ is invariant) are
\beq\label{check_fermion_gauge}
e^{i\x}\, \zetaT\d\psi^{\bf a}_{\Sigma}\, =\, U_{\,\bf m}^{\bf a}\, \d\psi_{\Sigma}^{\bf m}\zeta~,\qquad e^{i\x}\, \zetaT\d\psiT^{\bf {c}}\, =\, U_{\,\bf n}^{\bf c}\, \d\psiT^{\bf n}\zeta~.
\eeq
Both conditions are satisfied if
\beq
\label{BPSextra}
\begin{array}{l}
n^\mu j_\mu=\tilde{k}^\mu \D_\mu\s^{\bf a} +i U^{\bf a}_{\, \bf m} \,k^\mu \D_\mu\s^{\bf m} ~ ,\\
n^\mu \D_\mu\s^{\bf a}+ \tilde{k}^\mu ( j^{\bf a}_\mu + \s^{\bf a} V_\mu) + i U^{\bf a}_{\, \bf m}\, k^\mu j^{\bf m}_\mu =U_{\bf\ m}^{\bf a}\big[ D-i \s (iH + k^\mu V_\mu)\big]^{\bf m}~. \rule{0pt}{.6cm} 
\end{array}
\eeq
Closer inspection reveals that in YM theories, \eqref{check_fermion_gauge} reduces to a subset of 
the boundary conditions that 
we found in section \ref{sec_gaugesec_I} and \ref{sec_gaugesec_II}. Since the vector-matter 
couplings cannot appear in $\d\psi_{\Sigma}$ and $\d\psiT_{\Sigma}$, the conditions \eqref{check_fermion_gauge}
cannot lead to the most general boundary conditions \eqref{cs_cond_sg_21}, \eqref{cs_cond_sg_22}. 
Instead, they lead to the boundary conditions 
\beq\label{suppl_cond}
n^\mu j_\mu = P_U\D\s~,\qquad \D^\perp\s + \big( \tilde{k}^\mu ( j_\mu + \s V_\mu) + i U k^\mu j_\mu\big) 
= U\big[ D-i \s (iH + k^\mu V_\mu)\big]~.
\eeq

In CS theories, consider the boundary condition
$\mathcal{A}^{\bf n}_{\tilde{k}}+ U^{\bf n}_{\ \bf c}\  i\mathcal{A}^{\bf c}_k=0$. The condition on the fermions 
$\psi_{\Sigma}$ and $\psiT_{\Sigma}$ is the same as in YM, and therefore \eqref{check_fermion_gauge} leads 
to additional constraints on $D$, and on the derivatives of $\A$ and $\s$, which are precisely given by 
\eqref{suppl_cond}. On the other hand, the variation of the gauge boson at the boundary is
\beq
\d \A_\mu=
		-\theta \Big[ n_\mu ( e^{i\x} {\zetaT\psi_{\Sigma}} + e^{-i\x} {\psiT_\Sigma\zeta}) 
						- i \tilde{k^{\mu}}( e^{i\x} {\zetaT\psi_{\Sigma}} - e^{-i\x} {\psiT_\Sigma\zeta}) + k_\mu ( \psi_{\Sigma}\zeta - \zetaT\psiT_{\Sigma} ) \Big]~.
\eeq
Consequently, the special boundary condition 
$\d\mathcal{A}^{\bf n}_{\tilde{k}}+ U^{\bf n}_{\ \bf c}\  i\d\mathcal{A}^{\bf c}_k=0$ is trivially satisfied: 
\bea
\d\mathcal{A}^{\bf n}_{\tilde{k}}+ U^{\bf n}_{\ \bf c}\  i\d\mathcal{A}^{\bf c}_k &=&-i (e^{i\x}\zetaT\psi^{\bf n} - e^{-i\x}\psiT^{\bf n}\zeta)+ i U^{\bf n}_{\ \bf c}\ ( \psi^{\bf c}_{\Sigma}\zeta - \zetaT\psiT^{\bf c}_{\Sigma} ) \nn\\
						       &=& +i (e^{-i\x}\psiT^{\bf n}\zeta -U^{\bf n}_{\ \bf c} \zetaT\psiT^{\bf c}_{\Sigma}) - i (e^{i\x} \zetaT\psi^{\bf n}- U^{\bf n}_{\ \bf c}\  \psi^{\bf c}_{\Sigma}\zeta)=0~.
\eea

Finally, the scalar field $\s$, which is auxiliary in CS theories, but dynamical in YM theories, exhibits the 
supersymmetric variation
\beq
\label{deltas}
\d\s=i \theta (\zeta\psi_{\Sigma} + \zetaT\psiT_{\Sigma})~.
\eeq
We notice that the boundary conditions on the spinors relate the $\zeta$ and $\zetaT$ component of each 
fermionic field, and thus do not fix \eqref{deltas}. We could impose $\d\s=0$ by requiring
\beq\label{extra_BC_ferm}
\zetaT\psiT_{\Sigma}= \psi_{\Sigma}\zeta~.
\eeq
In that case, out of four fermionic variables (two for $\psi_{\Sigma}$ and two for $\psiT_{\Sigma}$), the boundary 
conditions \eqref{check_fermion_gauge} would fix two in terms of the rest, and by imposing \eqref{extra_BC_ferm} 
only one would remain unconstrained.

\subsubsection{Matter sector}

The supersymmetric variation of the matter fermions $\d\psi$ and $\d\psiT$ under the $A$-type supersymmetry, 
$\theta=\widetilde{\theta}$, is
\beq\nn
\begin{array}{ccl}
\d\psi_\a&=&+\,\vartheta~F\, \zeta_\a   +  \, \vartheta ~\Big[i\Big( k^\mu\D_\mu\phi - ir (iH)\phi - (z- q\s)\phi\Big)\zetaT_\a + i\Big(e^{-i\x}(n^\mu+i\tilde{k}^\mu)\D_\mu\phi\,\Big)\zeta_\a \Big]~,\\
\d\psiT_\a&=& +\,\vartheta~\FT\, \zetaT_\a   +  \, {\vartheta} ~\Big[i\Big( k^\mu\D_\mu\phiT + ir (iH)\phiT + (z- q\s)\phiT\Big)\zeta_\a - i\Big(e^{+i\x}(n^\mu-i\tilde{k}^\mu)\D_\mu\phiT\Big)\zetaT_\a \Big]~.\rule{0pt}{.6cm}
\end{array}
\eeq
The conditions we want to check are in this case 
\beq
\label{mattersusyvar}
e^{i\x}\, \zetaT\d\psi^{\bf a}_{\Sigma}\, =\, S_{\,\bf m}^{\bf a}\, \d\psi_{\Sigma}^{\bf m}\zeta~,\qquad e^{i\x}\, \zetaT\d\psiT^{\bf \bar{c}}\, =\, \widetilde{S}_{\,\bf \bar{n}}^{\bf \bar{c}}\, \d\psiT^{\bf n}\zeta
~.
\eeq
A short calculation leads to the constraints
\bea
\label{check_susy_matter1}
&& i P_S \D\phi + \big( \D^\perp\phi + S (z-q\s)\phi \big) - i e^{+i\an} F=0~, \\
&&\label{check_susy_matter2}
i P_{\widetilde{S}} \D\phiT - \big( \D^\perp\phiT - \widetilde{S} (z-q\s)\phiT \big) - i e^{-i\an} \FT =0~.
\eea
Compare these formulae with the boundary conditions \eqref{bc_matter_1} and \eqref{bc_matter_2}. The two sets 
of conditions do not coincide, because several signs do not match. Requiring that they hold simultaneously
requires
\bea
P_S \D\phi=P_{\widetilde{S}}\D\phiT=0~,\\
(z-q\s)\phi=(z-q\s)\phiT=0~,\\
\D^\perp \phi - i e^{i\an} F = \D^\perp\phiT +i e^{-i\an} \FT=0~.
\eea
This restricted set of conditions is consistent with the reality condition $\phiT=\phi^\star$ under the assumption 
$\widetilde{S}=-S^{\star}$.

\section*{Acknowledgments}

We are grateful to Francesco Benini, Lorenzo Di Pietro, Jorge Russo, Alessandro Tomasiello and Alberto Zaffaroni 
for insighful discussions. We are also grateful to CERN and the organizers of the TH-Institute on ``Recent 
Developments in M-theory" for the hospitality during the completion of this work. At the final stages of the 
preparation of the paper, FA presented the results of this work at the University of Amsterdam, and he is 
grateful to Jan de Boer, Diego Hofman, Nabil Iqbal and Joao Gomes for warm hospitality and stimulating 
discussion. This work was supported in part by European Union's Seventh Framework Programme under 
grant agreements FP7-REGPOT-2012-2013-1 no 316165, and by European Union's Horizon 2020 Programme 
under grant agreement 669288-SM-GRAV-ERC-2014-ADG. FA acknowledges support from STFC through 
Consolidated Grant ST/L000296/1.

\appendix

\section{Conventions}\label{conventions}

\paragraph{Clifford algebra.} The flat space $\g$ matrices are
\beq
\g^{1}=- \lp\begin{array}{cc} 0 & 1 \\ 1 & 0\end{array}\rp,\quad
\g^{2}=- \lp\begin{array}{cc} 0 & -i \\ i & 0\end{array}\rp,\quad 
\g^{3}=+ \lp\begin{array}{cc} 1 & 0 \\ 0 & -1\end{array}\rp.
\eeq
These $\g$ matrices satisfy the relation $\g^a\g^b=\d^{ab}+i\, \vep^{abc}\g_c$. In particular, 
$\g^{ab}\equiv \frac{1}{2}[\g^a,\g^b]=i\,\vep^{abc}\g_c$. Spinors $\chi_a$ and $\chi_b$ are contracted as follows,
\bea
\chi_a \chi_b \equiv \chi_a^\a\, C_{\a\b} \chi_b^\b\qquad \mathrm{with}\qquad C= \lp\begin{array}{cc} 0&-1\\ +1 & 0 \end{array}\rp~,
\eea
and also
\bea
\chi_a\,\g^\mu\,\chi_b  \equiv \chi_a^\a\, C_{\a\b} \lp\g^\mu\rp_{\, \s}^{\b} \chi_b^\s~.
\eea
Note the properties $C_{\a\b}=-C_{\b\a}$ and $\lp C \g  \rp_{\a\b}=\lp C \g \rp_{\b\a}$. Thus, for anticommuting spinors $\chi\g^\mu\zeta=-\zeta\g^\mu\chi$ whereas for commuting spinors $\chi\g^\mu\zeta=+\zeta\g^\mu\chi$.

The Fierz Identity for anticommuting spinors is
\beq
\lp\chi_d\g^\mu\chi_c\rp \lp \chi_b\g_\mu\chi_a\rp = -\lp\chi_d\chi_c\rp \lp\chi_b\chi_a\rp - 2 \lp \chi_d\chi_b\rp\lp\chi_c\chi_a\rp~.
\eeq
For commuting spinors we have instead
\beq
\lp\chi_d\g^\mu\chi_c\rp \lp \chi_b\g_\mu\chi_a\rp = +\lp\chi_d\chi_c\rp \lp\chi_b\chi_a\rp - 2 \lp \chi_d\chi_b\rp\lp\chi_c\chi_a\rp~.
\eeq


\paragraph{Differential geometry.}
Given an euclidean metric $ds^2=g_{\mu\nu}dx^\mu dx^\nu$, the frame fields are defined by 
$ds^2= e^a_\mu \delta_{ab} e^b_\nu$. The inverse frame fields are $e^\mu_a=g^{\mu\nu}\delta_{ab}e^b_\nu$, 
with $g^{\mu\nu}$ the inverse metric. The covariant derivative $\nabla_\mu$ acting on 1) a spinor $\chi$, 2) a 
vector field ${\bold V}^\nu$, and 3) a 1-form field ${\bold A}_\nu$ is
\beq
\begin{array}{cl}
\nabla_\mu\chi &\equiv \partial_\mu\chi +\frac{1}{4}\omega_{\mu\, ab}\g^{ab}\chi~,\\
\nabla_\mu {\bold V}^\nu &\equiv\partial_\mu {\bold V}^\nu +\G^\nu_{\mu\a} {\bold V}^\a~,\\
\nabla_\mu {\bold A}_\nu  &\equiv\partial_\mu {\bold A}_\nu -\G^\a_{\mu\nu} {\bold A}_\a~,
\end{array}
\eeq
where $\G^\nu_{\mu\a}$ is the Levi Civita connection, and we have defined the spin connection $\omega_\mu$ out of
\beq
\nabla_{\mu} e^a_\nu\equiv \partial_{\mu} e^a_\nu + \lp\omega_\mu\rp^a_b e^b_\nu -\G^\rho_{\mu\nu} e^a_\rho=0~.
\eeq


\paragraph{Supersymmetry transformations from \cite{Closset:2012ru}.}
We list the transformations rules of the components of the generic multiplet $\mathcal{S}$, 
\beq\nn
\begin{array}{ccl}
\d C &=& i\ep \chi + i \epT\widetilde{\chi}~,\\
\d \chi &=& \ep M - \epT (\s + (z-r H) C) -\g^\mu\epT\, (\D_\mu C + i a_\mu )~, \rule{0pt}{.6cm}\\
\d\widetilde{\chi} &=& \epT \widetilde{M} - \ep (\s - (z- r H) C ) - \g^\mu\ep\, (\D_\mu C - i a_\mu)~,\rule{0pt}{.6cm}\\\
\d M &=&- 2 \epT\lmb + 2i (z-(r-2) H) \epT\chi - 2i \D_\mu (\epT\g^\mu\chi)~,\rule{0pt}{.6cm}\\\
\d\widetilde{M} &=&+ 2 \ep\lm  - 2i (z-(r+2)H) \ep\widetilde{\chi} - 2i \D_\mu (\ep\g^\mu\widetilde{\chi})~, \rule{0pt}{.6cm}\\
\d a_\mu &=& -i (\ep\g_\mu\lmb + \epT\g_\mu\lm) + D_\mu (\ep\chi - \epT\widetilde{\chi})~,  \rule{0pt}{.6cm} \\
\d \s &=&-\ep\lmb + \epT\lm + i( z-r H) (\ep\chi-\epT\widetilde{\chi})~,\rule{0pt}{.6cm}\\
\d \lm &=& +i \ep (D+\s H) - i \varepsilon^{\mu\nu\rho}\g_\rho\ep\, \D_\mu a_\nu - \g^\mu\ep ( (z-rH) a_\mu + i \D_\mu\s - V_\mu \s )~,\rule{0pt}{.6cm}\\
\d\lmb &=& -i \epT (D+\s H) - i \varepsilon^{\mu\nu\rho}\g_\rho\epT\, \D_\mu a_\nu + \g^\mu\epT ((z-r H) a_\mu + i D_\mu\s + V_\mu \s)~,\rule{0pt}{.6cm}\\
\d D&=& \D_\mu (\ep\g^\mu\lmb - \epT\g^\mu\lm) - i V_\mu (\ep\g^\mu\lmb + \epT\g^\mu\lm) - H (\ep\lmb - \epT\lm)\rule{0pt}{.6cm}\\
& & \qquad + (z-rH) ( \ep\lmb + \ep \lm- i H (\ep\chi - \epT\widetilde{\chi})) + \frac{ir}{4} (R- 2 V^2 - 6 H^2) (\ep\chi-\epT\widetilde{\chi})\rule{0pt}{.6cm}~. 
\end{array}
\eeq

\section{Factorization of bilinears} \label{bilinearFactor}

In this section we explain the details of the manipulations of $\TD_{NL\s}^\perp$ that were used to obtain the 
final formula \eqref{final_result_NLSM2} in section \ref{sec_WZ_model}.
Let us recall the two basic inputs of this discussion: 
1) the main contributions to $\TD_{NL\s}^\perp$ that we want to analyze:
\bea
\TD_1+\TD_2&=& +\left[ \ep\g^\perp\g^\mu\psi^a \D_\mu\phiT^{\bar{c}} - \epT\g^\perp\g^\mu\psiT^{\bar{c}} \D_\mu\phi^a\right] K_{a\bar{c}}\\
\TD_3&=& -\left[\ep\,\g^\perp \psi^{a}\, \phiT^{\bar{c}} 
-\epT\,\g^\perp \psiT^{\bar{c}}\, \phi^a \right]  K_{a\bar{c}}\\
\TD_4&=&-\left[\ep \g^\perp\psiT^{\bar{c}} W^a + \epT\g^\perp\psi^a \Wb^{\bar{c}} \right] K_{a\bar{c}}
~,
\eea
and 2) the decomposition of the spinors with the use of the projectors $\Pr$ and $\Prb$:
\beq\label{Recap_Deco_Spinor}
\begin{array}{cclcccl}
\ep &=& \frac{1}{\Omega} (\zetaT\ep)\zeta~, &\rule{1cm}{0pt} &
\epT &= & \frac{1}{\Omega} (\epT\,\zeta)\zetaT~,\\
\psi&=& \frac{1}{\Omega} (\zetaT\psi)\zeta + \frac{1}{\Omega} (\psi\zeta)\zetaT~, & \rule{1cm}{0pt}\rule{0pt}{.5cm}&
\psiT&=&\frac{1}{\Omega} (\zetaT\psiT)\zeta+\frac{1}{\Omega} (\psiT\zeta)\zetaT~.
\end{array}
\eeq 

We begin by studying $\TD_1+\TD_2$. From \eqref{Recap_Deco_Spinor} we get
\bea
\TD_1+\TD_2
		&=&  \big[+\ep\,\g^\perp\g^\nu\psi^a \D_\nu\phiT^{\bar{c}} - \epT\,\g^\perp\g^\nu\psiT^{\bar{c}} \D_\nu\phi^a \big] \nn\\
		&=& \frac{(\zetaT\ep)}{\Omega^2} 
		\Big[ (\zeta\g^\perp\g^\nu\zeta)\, (\zetaT\psi^a) + (\zeta\g^\perp\g^\nu\zetaT)\, (\psi^a\zeta) \Big] G_{a\bar{c}}\D_\nu\phiT^{\bar{c}}+\nn \\
		& & \rule{1cm}{0pt} 
						- \frac{(\epT\,\zeta)}{\Omega^2} 
							\Big[ (\zetaT\g^\perp\g^\nu\zeta)\, (\zetaT\psiT^{\bar{c}}) + (\zetaT\g^\perp\g^\nu\zetaT)\, (\psiT^{\bar{c}}\zeta) \Big] G_{\bar{c}a}\D_\nu\phi^{a}~.\nn
\eea
By using the knowledge of the bosonic bilinears \eqref{Bilinears_2}, we obtain\footnote{In \eqref{PieceTD2gammas_2} 
and \eqref{PieceTD2gammas_3}, the sum over $\nu_{\parallel}$ is understood to run over the indices of the 
boundary $\mathcal{M}_2$.}  
\bea
\TD_1+\TD_2
				&=&  \frac{(\zetaT\ep)}{\Omega^2}\, (\zeta\zetaT)\,
					\Big[ (\psi^a\zeta)\D^\perp\phiT^{\bar{c}} + (\zetaT\psiT^{\bar{c}}) \D^\perp \phi^{a} \Big] G_{a\bar{c}}\rule{0pt}{.8cm} 
					\label{PieceTD2gammas_1}\\
				& & \rule{1cm}{0pt} 
					+ \frac{(\zetaT\ep)}{\Omega^2}\, (\zeta\g^\perp\g^{\nu\parallel}\zeta) 
							\Big[ (\zetaT\psi^a)\D_{\nu\parallel}\phiT^{\bar{c}} - e^{-2i\x}(\psiT^{\bar{c}}\zeta)\D_{\nu\parallel}\phi^{a} \Big] G_{a\bar{c}}
							\label{PieceTD2gammas_2}\\
				& &\rule{2cm}{0pt} 
					+ \frac{(\zetaT\ep)}{\Omega^2}\,  (\zeta\g^\perp\g^{\nu\parallel}\zetaT) 
							\Big[ (\psi^a\zeta)\D_{\nu\parallel}\phiT^{\bar{c}} - (\zetaT\psiT^{\bar{c}}) \D_{\nu\parallel}\phi^{a} \Big] G_{a\bar{c}}~ .
								\label{PieceTD2gammas_3}
\eea
In order to simplify our formulae, it is now convenient to use the matrix notation where 
$(\psi^a,\psiT^{\bar{c}})\rightarrow {\Psi}^I$ and $(\phi^a,\phiT^{\bar{c}})\rightarrow\Phi^I$. 
Each vector will be denoted by a corresponding bold symbol: $\bPhi$, $\bPsi$, $\bW$, and $\bK$. 
The change of variables for $(\psi^a\zeta)$ and $(\psiT^{\bar{c}}\zeta)$ results in
\beq\label{BC_ChangeofV_1}
\begin{array}{ccl}
(\psi^a\zeta)\,G_{a\bar{c}}\, \D_\nu\phiT^{\bar{c}}&=&\frac{1}{2} (\bPsi\zeta)^T (1-i J) G\, \D_\nu{\bPhi}~, \\
(\psiT^{\bar{c}}\zeta)\, G_{a\bar{c}}\, \D_\nu\phi^a &=& \frac{1}{2} (\bPsi\zeta)^T (1+iJ)G\, \D_\nu\bPhi~.\rule{0pt}{.6cm} \\
\end{array}
\eeq
For the scalar products involving $\zetaT$ we shall use the boundary condition 
$e^{i\an} \zetaT\bPsi = M \bPsi\zeta$, and write
\beq\label{BC_ChangeofV_2}
\begin{array}{ccl}
(\zetaT\psi^a)\, G_{a\bar{c}}\, \D_\nu\phiT^{\bar{c}}&=& \frac{1}{2} (\zetaT\bPsi)^T (1-iJ)G\,\D_\nu\bPhi=
										\frac{1}{2} e^{-i\x} (\bPsi\zeta)^T M^T(1-iJ)G\,\D_\nu\bPhi~,\\
(\zetaT\psiT^{\bar{c}})\, G_{a\bar{c}}\, \D_\nu\phi^a &=& \frac{1}{2} (\zetaT\bPsi)^T (1+i J) G\,\D_\nu\bPhi=
										\frac{1}{2} e^{-i\x} (\bPsi\zeta)^T M^T(1+i J) G\,\D_\nu\bPhi ~. \rule{0pt}{.6cm} 
\end{array}
\eeq
From \eqref{BC_ChangeofV_1} and \eqref{BC_ChangeofV_2}, it is a simple exercise to show that $\TD_1+\TD_2$ can be put in the following form 
\bea
\TD_1+\TD_2&=& \left[\frac{1}{\Omega} (\zetaT\ep)\,(\bPsi\zeta)^T \right]  \frac{(\zeta\zetaT)}{\Omega} 
\left[ \frac{1-iJ}{2}+ e^{-i\x} M^T \frac{1+iJ}{2} \right] G\, (\D^\perp\bPhi) \nn\\
&& \rule{1cm}{0pt} +e^{-i\x}\left[\frac{1}{\Omega} (\zetaT\ep)\,(\bPsi\zeta)^T \right]\, \left[ e^{i\x} M^T\frac{1-iJ}{2} -  \frac{1+iJ}{2} \right] G\,(k^\mu \D_\mu \bPhi) \nn\\
&&	 \rule{2cm}{0pt} -i\left[\frac{1}{\Omega} (\zetaT\ep)\,(\bPsi\zeta)^T \right]  \left[ \frac{1-iJ}{2}- e^{-i\x} M^T \frac{1+iJ}{2} \right] G\,(\tilde{k}^\mu \D_\mu \bPhi)~. \nn 
\eea
We can then introduce the projectors $P^{(\x,\pm)}_M$, and by using the properties:
\beq
\begin{array}{ccccccc}
P_{M^T}^{\lp \x ,+ \rp } J &=& J\, P_{M^T}^{\lp \x ,- \rp }\ , &\rule{1cm}{0pt}& P_{M^T}^{\lp \x ,- \rp } J &=& J\, P_{M^T}^{\lp \x ,+ \rp }~,\\
P_{M^T}^{\lp \x ,+ \rp } G &=& G\, P_{M}^{\lp \x ,+ \rp } \ ,&\rule{1cm}{0pt}\rule{0pt}{.6cm}& P_{M^T}^{\lp \x ,- \rp } G &=& G\, P_{M}^{\lp \x ,- \rp }~ ,
\end{array}
\eeq
we arrive at the final expression
\bea
\TD_1+\TD_2=& + (\ep\Psi)^T& \left[ (1-iJ) G\, P_M^{(\x,+)}\, (\D^\perp\Phi) + {i}(1-iJ) G\, P^{(\x,-)}_M\, (\tilde{k}^\mu D_\mu \Phi) \right] \\
& + (\ep\Psi)^T & \left[  e^{-i\x} (1+iJ) G\, P^{(\x,-)}_M\,(k^\mu D_\mu \Phi) \right]~.
\eea

We rearrange $\TD_3$ and $\TD_4$ with similar manipulations. In the case of $\TD_3$ we find
\bea
\TD_3 &=&  \big[-\ep\,\g^\perp\psi^a \phiT^{\bar{c}} + \epT\,\g^\perp\psiT^{\bar{c}}\phi^a\big] G_{a\bar{c}}\\
	   &=& 
	   		-\frac{(\zetaT\ep)}{\Omega^2} (\zeta\g^\perp\zeta) \left[ (\zetaT\psi^a)\phiT^{\bar{c}} + e^{-2i\x} (\psiT^{\bar{c}}\zeta)\phi^a \right]G_{\bar{c}a} \\
            &=&
         			\left[ -\frac{(\zetaT\ep)}{\Omega} (\bPsi\zeta)^T \right] \frac{(\zeta\g^\perp\zeta)}{\Omega}\ \left[ e^{-i\x}M^T\frac{1-iJ}{2} +e^{-2i\x}  \frac{1+iJ}{2} \right] G\bPhi \\ 
	   &=&     +ie^{-i\x} (\ep\bPsi)^T (1+iJ) G\, P_M^{(\x,-)}J\, \bPhi~.
\eea
In the case of $\TD_4$ we find
\bea
\TD_4  &= & \big[-\epT\,\g^\perp \psi^a \Wb^{\bar{c}} - \ep\,\g^\perp \psiT^{\bar{c}} {W}^{a} \big] G_{a\bar{c}} \\
	   &=&
	   		 -\frac{(\zetaT\ep)}{\Omega^2} (\zeta\g^\perp\zeta) \left[ -e^{-2i\x} (\psi^a\zeta)\Wb^{\bar{c}} + (\zetaT\psiT^{\bar{c}})W^a \right] G_{\bar{c}a} \\
            &=&
         			\left[ -\frac{(\zetaT\ep)}{\Omega} (\bPsi\zeta)^T \right] \frac{(\zeta\g^\perp\zeta)}{\Omega}\  \left[ -e^{-2i\x} \frac{1-iJ}{2}+ e^{-i\x} M^T\frac{1+iJ}{2} \right] G \bW \\ 
	   &=& 	-e^{-i\x} (\ep\bPsi)^T (1-iJ)\, P_{M^T}^{(-\x,-)} G\,\bW~.
\eea

\addtocontents{toc}{\protect\setcounter{tocdepth}{1}}
\addtocontents{lof}{\protect\setcounter{tocdepth}{1}}

\bibliography{Boundary_A_Biblio} 

\providecommand{\href}[2]{#2}\begingroup\raggedright\begin{thebibliography}{10}

\bibitem{Festuccia:2011ws}
G.~Festuccia and N.~Seiberg, {\it {Rigid Supersymmetric Theories in Curved
  Superspace}},  {\em JHEP} {\bf 06} (2011) 114,
  [\href{http://xxx.lanl.gov/abs/1105.0689}{{\tt arXiv:1105.0689}}].

\bibitem{Dumitrescu:2012ha}
T.~T. Dumitrescu, G.~Festuccia, and N.~Seiberg, {\it {Exploring Curved
  Superspace}},  {\em JHEP} {\bf 1208} (2012) 141,
  [\href{http://xxx.lanl.gov/abs/1205.1115}{{\tt arXiv:1205.1115}}].

\bibitem{Closset:2012ru}
C.~Closset, T.~T. Dumitrescu, G.~Festuccia, and Z.~Komargodski, {\it
  {Supersymmetric Field Theories on Three-Manifolds}},  {\em JHEP} {\bf 1305}
  (2013) 017, [\href{http://xxx.lanl.gov/abs/1212.3388}{{\tt
  arXiv:1212.3388}}].

\bibitem{Pestun:2007rz}
V.~Pestun, {\it {Localization of gauge theory on a four-sphere and
  supersymmetric Wilson loops}},  {\em Commun. Math. Phys.} {\bf 313} (2012)
  71--129, [\href{http://xxx.lanl.gov/abs/0712.2824}{{\tt arXiv:0712.2824}}].

\bibitem{Kapustin:2009kz}
A.~Kapustin, B.~Willett, and I.~Yaakov, {\it {Exact Results for Wilson Loops in
  Superconformal Chern-Simons Theories with Matter}},  {\em JHEP} {\bf 1003}
  (2010) 089, [\href{http://xxx.lanl.gov/abs/0909.4559}{{\tt
  arXiv:0909.4559}}].

\bibitem{Drukker:2012sr}
N.~Drukker, T.~Okuda, and F.~Passerini, {\it {Exact results for vortex loop
  operators in 3d supersymmetric theories}},  {\em JHEP} {\bf 07} (2014) 137,
  [\href{http://xxx.lanl.gov/abs/1211.3409}{{\tt arXiv:1211.3409}}].

\bibitem{Benini:2012ui}
F.~Benini and S.~Cremonesi, {\it {Partition Functions of ${\mathcal{N}=(2,2)}$
  Gauge Theories on S$^{2}$ and Vortices}},  {\em Commun. Math. Phys.} {\bf
  334} (2015), no.~3 1483--1527, [\href{http://xxx.lanl.gov/abs/1206.2356}{{\tt
  arXiv:1206.2356}}].

\bibitem{Doroud:2012xw}
N.~Doroud, J.~Gomis, B.~Le~Floch, and S.~Lee, {\it {Exact Results in D=2
  Supersymmetric Gauge Theories}},  {\em JHEP} {\bf 05} (2013) 093,
  [\href{http://xxx.lanl.gov/abs/1206.2606}{{\tt arXiv:1206.2606}}].

\bibitem{Kallen:2012cs}
J.~Kallen and M.~Zabzine, {\it {Twisted supersymmetric 5D Yang-Mills theory and
  contact geometry}},  {\em JHEP} {\bf 05} (2012) 125,
  [\href{http://xxx.lanl.gov/abs/1202.1956}{{\tt arXiv:1202.1956}}].

\bibitem{Hosomichi:2012ek}
K.~Hosomichi, R.-K. Seong, and S.~Terashima, {\it {Supersymmetric Gauge
  Theories on the Five-Sphere}},  {\em Nucl. Phys.} {\bf B865} (2012) 376--396,
  [\href{http://xxx.lanl.gov/abs/1203.0371}{{\tt arXiv:1203.0371}}].

\bibitem{Kim:2012ava}
H.-C. Kim and S.~Kim, {\it {M5-branes from gauge theories on the 5-sphere}},
  {\em JHEP} {\bf 05} (2013) 144,
  [\href{http://xxx.lanl.gov/abs/1206.6339}{{\tt arXiv:1206.6339}}].

\bibitem{Alday:2013lba}
L.~F. Alday, D.~Martelli, P.~Richmond, and J.~Sparks, {\it {Localization on
  Three-Manifolds}},  {\em JHEP} {\bf 10} (2013) 095,
  [\href{http://xxx.lanl.gov/abs/1307.6848}{{\tt arXiv:1307.6848}}].

\bibitem{Martelli:2015kuk}
D.~Martelli and J.~Sparks, {\it {The character of the supersymmetric Casimir
  energy}},  \href{http://xxx.lanl.gov/abs/1512.0252}{{\tt arXiv:1512.0252}}.

\bibitem{Hori:2000ck}
K.~Hori, A.~Iqbal, and C.~Vafa, {\it {D-branes and mirror symmetry}},
  \href{http://xxx.lanl.gov/abs/hep-th/0005247}{{\tt hep-th/0005247}}.

\bibitem{Okazaki:2013kaa}
T.~Okazaki and S.~Yamaguchi, {\it {Supersymmetric boundary conditions in
  three-dimensional N=2 theories}},  {\em Phys. Rev.} {\bf D87} (2013), no.~12
  125005, [\href{http://xxx.lanl.gov/abs/1302.6593}{{\tt arXiv:1302.6593}}].

\bibitem{Herbst:2008jq}
M.~Herbst, K.~Hori, and D.~Page, {\it {Phases Of N=2 Theories In 1+1 Dimensions
  With Boundary}},  \href{http://xxx.lanl.gov/abs/0803.2045}{{\tt
  arXiv:0803.2045}}.

\bibitem{Closset:2013vra}
C.~Closset, T.~T. Dumitrescu, G.~Festuccia, and Z.~Komargodski, {\it {The
  Geometry of Supersymmetric Partition Functions}},  {\em JHEP} {\bf 01} (2014)
  124, [\href{http://xxx.lanl.gov/abs/1309.5876}{{\tt arXiv:1309.5876}}].

\bibitem{Sugishita:2013jca}
S.~Sugishita and S.~Terashima, {\it {Exact Results in Supersymmetric Field
  Theories on Manifolds with Boundaries}},  {\em JHEP} {\bf 11} (2013) 021,
  [\href{http://xxx.lanl.gov/abs/1308.1973}{{\tt arXiv:1308.1973}}].

\bibitem{Yoshida:2014ssa}
Y.~Yoshida and K.~Sugiyama, {\it {Localization of 3d $\mathcal{N}=2$
  Supersymmetric Theories on $S^1 \times D^2$}},
  \href{http://xxx.lanl.gov/abs/1409.6713}{{\tt arXiv:1409.6713}}.

\bibitem{Beem:2012mb}
C.~Beem, T.~Dimofte, and S.~Pasquetti, {\it {Holomorphic Blocks in Three
  Dimensions}},  {\em JHEP} {\bf 12} (2014) 177,
  [\href{http://xxx.lanl.gov/abs/1211.1986}{{\tt arXiv:1211.1986}}].

\bibitem{Nieri:2015yia}
F.~Nieri and S.~Pasquetti, {\it {Factorisation and holomorphic blocks in 4d}},
  {\em JHEP} {\bf 11} (2015) 155,
  [\href{http://xxx.lanl.gov/abs/1507.0026}{{\tt arXiv:1507.0026}}].

\bibitem{Qiu:2014oqa}
J.~Qiu, L.~Tizzano, J.~Winding, and M.~Zabzine, {\it {Gluing Nekrasov partition
  functions}},  {\em Commun. Math. Phys.} {\bf 337} (2015), no.~2 785--816,
  [\href{http://xxx.lanl.gov/abs/1403.2945}{{\tt arXiv:1403.2945}}].

\bibitem{Hori:2013ika}
K.~Hori and M.~Romo, {\it {Exact Results In Two-Dimensional (2,2)
  Supersymmetric Gauge Theories With Boundary}},
  \href{http://xxx.lanl.gov/abs/1308.2438}{{\tt arXiv:1308.2438}}.

\bibitem{Honda:2013uca}
D.~Honda and T.~Okuda, {\it {Exact results for boundaries and domain walls in
  2d supersymmetric theories}},  {\em JHEP} {\bf 09} (2015) 140,
  [\href{http://xxx.lanl.gov/abs/1308.2217}{{\tt arXiv:1308.2217}}].

\bibitem{Gomis:2012wy}
J.~Gomis and S.~Lee, {\it {Exact Kahler Potential from Gauge Theory and Mirror
  Symmetry}},  {\em JHEP} {\bf 04} (2013) 019,
  [\href{http://xxx.lanl.gov/abs/1210.6022}{{\tt arXiv:1210.6022}}].

\bibitem{Cecotti:2013mba}
S.~Cecotti, D.~Gaiotto, and C.~Vafa, {\it {$tt^*$ geometry in 3 and 4
  dimensions}},  {\em JHEP} {\bf 05} (2014) 055,
  [\href{http://xxx.lanl.gov/abs/1312.1008}{{\tt arXiv:1312.1008}}].

\bibitem{Bullimore:2016nji}
M.~Bullimore, T.~Dimofte, D.~Gaiotto, and J.~Hilburn, {\it {Boundaries, Mirror
  Symmetry, and Symplectic Duality in 3d $\mathcal{N}=4$ Gauge Theory}},
  \href{http://xxx.lanl.gov/abs/1603.0838}{{\tt arXiv:1603.0838}}.

\bibitem{Armoni:2015jsa}
A.~Armoni and V.~Niarchos, {\it {Defects in Chern-Simons theory, gauged WZW
  models on the brane, and level-rank duality}},  {\em JHEP} {\bf 07} (2015)
  062, [\href{http://xxx.lanl.gov/abs/1505.0291}{{\tt arXiv:1505.0291}}].

\bibitem{Kapustin:2013hpk}
A.~Kapustin and B.~Willett, {\it {Wilson loops in supersymmetric
  Chern-Simons-matter theories and duality}},
  \href{http://xxx.lanl.gov/abs/1302.2164}{{\tt arXiv:1302.2164}}.

\bibitem{Assel:2015oxa}
B.~Assel and J.~Gomis, {\it {Mirror Symmetry And Loop Operators}},  {\em JHEP}
  {\bf 11} (2015) 055, [\href{http://xxx.lanl.gov/abs/1506.0171}{{\tt
  arXiv:1506.0171}}].

\bibitem{Hung:2014npa}
L.-Y. Hung, R.~C. Myers, and M.~Smolkin, {\it {Twist operators in higher
  dimensions}},  {\em JHEP} {\bf 10} (2014) 178,
  [\href{http://xxx.lanl.gov/abs/1407.6429}{{\tt arXiv:1407.6429}}].

\bibitem{Aharony:2008ug}
O.~Aharony, O.~Bergman, D.~L. Jafferis, and J.~Maldacena, {\it {N=6
  superconformal Chern-Simons-matter theories, M2-branes and their gravity
  duals}},  {\em JHEP} {\bf 10} (2008) 091,
  [\href{http://xxx.lanl.gov/abs/0806.1218}{{\tt arXiv:0806.1218}}].

\bibitem{Niarchos:2015lla}
V.~Niarchos, {\it {A Lagrangian for self-dual strings}},  {\em JHEP} {\bf 12}
  (2015) 060, [\href{http://xxx.lanl.gov/abs/1509.0767}{{\tt
  arXiv:1509.0767}}].

\bibitem{Berman:2009xd}
D.~S. Berman, M.~J. Perry, E.~Sezgin, and D.~C. Thompson, {\it {Boundary
  Conditions for Interacting Membranes}},  {\em JHEP} {\bf 04} (2010) 025,
  [\href{http://xxx.lanl.gov/abs/0912.3504}{{\tt arXiv:0912.3504}}].

\bibitem{Okazaki:2015fiq}
T.~Okazaki and D.~J. Smith, {\it {Topological M-Strings and Supergroup WZW
  Models}},  \href{http://xxx.lanl.gov/abs/1512.0664}{{\tt arXiv:1512.0664}}.

\bibitem{DiPietro:2015zia}
L.~Di~Pietro, N.~Klinghoffer, and I.~Shamir, {\it {On Supersymmetry, Boundary
  Actions and Brane Charges}},  {\em JHEP} {\bf 02} (2016) 163,
  [\href{http://xxx.lanl.gov/abs/1502.0597}{{\tt arXiv:1502.0597}}].

\bibitem{Ferrara:1988qxa}
S.~Ferrara and S.~Sabharwal, {\it {Structure of New Minimal Supergravity}},
  {\em Annals Phys.} {\bf 189} (1989) 318--351.

\bibitem{Klare:2012gn}
C.~Klare, A.~Tomasiello, and A.~Zaffaroni, {\it {Supersymmetry on Curved Spaces
  and Holography}},  {\em JHEP} {\bf 08} (2012) 061,
  [\href{http://xxx.lanl.gov/abs/1205.1062}{{\tt arXiv:1205.1062}}].

\bibitem{BlairBook}
D.~E. Blair, {\it {Riemannian Geometry of Contact and Symplectic Manifolds}},
  {\em Progress in Mathematics} (2010) Birkhauser.

\bibitem{daSilva}
A.~C. da~Silva, {\it {Lectures on Symplectic Geometry}},  {\em Lecture Notes in
  Mathematics} {\bf 1764} (2008) Springer--Verlag.

\bibitem{Isenberg}
J.~Isenberg and V.~Moncrief, {\it {On spacetimes containing Killing vector
  fields with non-closed orbits}},  {\em Classical and Quantum Gravity} (1992)
  Vol 9, Issue 7, 1683--1691.

\bibitem{LeeBook}
J.~M. Lee, {\it {Introduction to Smooth Manifolds-Second Edition}},  {\em
  Graduate Texts in Mathematics} {\bf 128} (2013) Springer--Verlag.

\bibitem{JennyHeegard}
J.~Schultens, {\it {Introduction to 3-Manifolds}},  {\em Graduate Studies in
  Mathematics} {\bf 151} (2014) American Mathematical Society.

\bibitem{Marino:2011nm}
M.~Marino, {\it {Lectures on localization and matrix models in supersymmetric
  Chern-Simons-matter theories}},  {\em J. Phys.} {\bf A44} (2011) 463001,
  [\href{http://xxx.lanl.gov/abs/1104.0783}{{\tt arXiv:1104.0783}}].

\bibitem{Hama:2011ea}
N.~Hama, K.~Hosomichi, and S.~Lee, {\it {SUSY Gauge Theories on Squashed
  Three-Spheres}},  {\em JHEP} {\bf 1105} (2011) 014,
  [\href{http://xxx.lanl.gov/abs/1102.4716}{{\tt arXiv:1102.4716}}].

\bibitem{Hama:2010av}
N.~Hama, K.~Hosomichi, and S.~Lee, {\it {Notes on SUSY Gauge Theories on
  Three-Sphere}},  {\em JHEP} {\bf 03} (2011) 127,
  [\href{http://xxx.lanl.gov/abs/1012.3512}{{\tt arXiv:1012.3512}}].

\bibitem{Imamura:2011wg}
Y.~Imamura and D.~Yokoyama, {\it {N=2 supersymmetric theories on squashed
  three-sphere}},  {\em Phys. Rev.} {\bf D85} (2012) 025015,
  [\href{http://xxx.lanl.gov/abs/1109.4734}{{\tt arXiv:1109.4734}}].

\bibitem{Martelli:2013aqa}
D.~Martelli and A.~Passias, {\it {The gravity dual of supersymmetric gauge
  theories on a two-parameter deformed three-sphere}},  {\em Nucl. Phys.} {\bf
  B877} (2013) 51--72, [\href{http://xxx.lanl.gov/abs/1306.3893}{{\tt
  arXiv:1306.3893}}].

\bibitem{Benini:2013yva}
F.~Benini and W.~Peelaers, {\it {Higgs branch localization in three
  dimensions}},  {\em JHEP} {\bf 05} (2014) 030,
  [\href{http://xxx.lanl.gov/abs/1312.6078}{{\tt arXiv:1312.6078}}].

\bibitem{Benini:2015noa}
F.~Benini and A.~Zaffaroni, {\it {A topologically twisted index for
  three-dimensional supersymmetric theories}},  {\em JHEP} {\bf 07} (2015) 127,
  [\href{http://xxx.lanl.gov/abs/1504.0369}{{\tt arXiv:1504.0369}}].

\bibitem{Closset:2014pda}
C.~Closset and S.~Cremonesi, {\it {Comments on $\mathcal{N} $ = (2, 2)
  supersymmetry on two-manifolds}},  {\em JHEP} {\bf 1407} (2014) 075,
  [\href{http://xxx.lanl.gov/abs/1404.2636}{{\tt arXiv:1404.2636}}].

\bibitem{Ooguri:1996ck}
H.~Ooguri, Y.~Oz, and Z.~Yin, {\it {D-branes on Calabi-Yau spaces and their
  mirrors}},  {\em Nucl. Phys.} {\bf B477} (1996) 407--430,
  [\href{http://xxx.lanl.gov/abs/hep-th/9606112}{{\tt hep-th/9606112}}].

\bibitem{Kapustin:2010hk}
A.~Kapustin and N.~Saulina, {\it {Topological boundary conditions in abelian
  Chern-Simons theory}},  {\em Nucl. Phys.} {\bf B845} (2011) 393--435,
  [\href{http://xxx.lanl.gov/abs/1008.0654}{{\tt arXiv:1008.0654}}].

\end{thebibliography}\endgroup
  
\end{document}